\documentclass{report}
\pdfoutput=1
  \usepackage[margin=1.3in]{geometry}

  \usepackage[T1]{fontenc} %
  \usepackage[latin2]{inputenc} %

    \usepackage[
      backend=biber,
      natbib=true,
      language=auto,
      style=ieee,
      sorting=nyt,
      maxbibnames=10,
      giveninits=true,
      backref=true,
      url=true, 
      doi=false,
      isbn=false,
      eprint=false
      ]{biblatex}
    \usepackage[colorlinks, urlcolor=RoyalBlue, linkcolor=Red, citecolor=Green]{hyperref} %

  \usepackage{graphicx} %
  \usepackage{subcaption} %
  \usepackage[dvipsnames]{xcolor} %

    \usepackage{tikz} %
    \usetikzlibrary{
      petri, %
      backgrounds, %
      arrows, %
      positioning, %
      decorations.markings, %
      calc,  %
      fit, %
    }

  \usepackage{mathtools} %
  \usepackage{amssymb} %
  \usepackage{amsthm} %

  \usepackage{listings} %

\usepackage{tabularx, cellspace} %
    \newcounter{theoremUnified} %
    \def\thetheoremUnified{\arabic{section}} %
    \numberwithin{theoremUnified}{section} %
    \numberwithin{theoremUnified}{section} %

    \newtheoremstyle{plainStyle} %
    {2mm} %
    {2mm} %
    {} %
    {} %
    {\bfseries} %
    {.} %
    {.5em} %
    {} %

    \newtheoremstyle{italicStyle} %
    {2mm} %
    {2mm} %
    {\itshape} %
    {} %
    {\bfseries} %
    {.} %
    {.5em} %
    {} %

    \theoremstyle{plainStyle} %
      \newtheorem{example}[theoremUnified]{Example}
        \AfterEndEnvironment{example}{\noindent\ignorespaces}
      \newtheorem{examplehard}[theoremUnified]{Example*}
        \AfterEndEnvironment{examplehard}{\noindent\ignorespaces}
      \newtheorem{remark}[theoremUnified]{Remark}
        \AfterEndEnvironment{remark}{\noindent\ignorespaces}
      \newtheorem{remarkhard}[theoremUnified]{Remark*}
      \AfterEndEnvironment{remarkhard}{\noindent\ignorespaces}
    \theoremstyle{italicStyle} %
      \newtheorem{definition}[theoremUnified]{Definition}
        \AfterEndEnvironment{definition}{\noindent\ignorespaces}
      \newtheorem{proposition}[theoremUnified]{Proposition}
        \AfterEndEnvironment{proposition}{\noindent\ignorespaces}
      \newtheorem{lemma}[theoremUnified]{Lemma}
        \AfterEndEnvironment{lemma}{\noindent\ignorespaces}
      
        \AfterEndEnvironment{theorem}{\noindent\ignorespaces}
      
        \AfterEndEnvironment{corollary}{\noindent\ignorespaces}

\newcommand{\Naturals}{\mathbb{N}} %
\newcommand{\Integers}{\mathbb{Z}} %

  \def\backgrnd{black!10}	%

	\tikzstyle{place}=
		[circle,thick,draw=blue!75,fill=blue!20,minimum size=6mm]
	\tikzstyle{transition}=
		[rectangle,thick,draw=black!75,fill=black!20,minimum size=4mm]

	\newcommand{\Mset}[1]{{#1}^{\Naturals}} %
	\newcommand{\MsetBase}[2]{{#1}^{\Naturals}_{#2}} %
	\newcommand{\Msets}[1]{{#1}^{\oplus}} %

		\newcommand{\Disjoint}{\sqcup} %
		\newcommand{\Zeromset}[1]{\emptyset_{#1}} %
		\newcommand{\Inject}{\hookrightarrow} %

	\newcommand{\Net}[1]{(\Pl{#1},\Tr{#1},\Pin{-}{#1},\Pout{-}{#1})} %
	\newcommand{\Pl}[1]{P_{#1}} %
	\newcommand{\Tr}[1]{T_{#1}} %
	\newcommand{\Pin}[2]{{^\circ}(#1)_{#2}} %
	\newcommand{\Pout}[2]{{(#1)_{#2}^\circ}} %
	\newcommand{\Marking}[1]{\Mset{#1}} %
  \newcommand{\Obj}[1]{\operatorname{Obj} \, #1} %
  \newcommand{\Homtotal}[1]{\operatorname{Hom}_{\,#1}} %
  \newcommand{\Source}[1]{\operatorname{s}(#1)} %
  \newcommand{\Target}[1]{\operatorname{t}(#1)} %
  \newcommand{\Id}[1]{id_{#1}} %

  \newcommand{\CategoryC}{\mathcal{C}}
  \newcommand{\CategoryD}{\mathcal{D}}
  \newcommand{\CategoryE}{\mathcal{E}}

  \newcommand{\Set}{\textbf{Set}} %
  \newcommand{\Hask}{\textbf{Hask}} %
  \newcommand{\Group}{\textbf{Group}} %
  \newcommand{\Top}{\textbf{Top}} %
  \newcommand{\hTop}{\textbf{hTop}} %
  \newcommand{\Cat}{\textbf{Cat}} %

  \newcommand{\Tensor}{\otimes} %
  \newcommand{\TensorUnit}{I} %

  \newcommand{\Suchthat}[2]{\left\{#1 \: \middle\vert \: #2\right\}} %

  \tikzset{ %
		oriented WD/.style={%
			every to/.style={
        out=0,in=180,draw
      },
			label/.style={
				font=\everymath\expandafter{\the\everymath\scriptstyle},
				inner sep=0pt,
        node distance=2pt and -2pt
      },
			semithick,
			node distance=1 and 1,
			decoration={
        markings, mark=at position \stringdecpos with \stringdec
      },
			ar/.style={
        postaction={decorate}
      },
			execute at begin picture={
        \tikzset{
					x=\bbx, y=\bby,
					every fit/.style={
            inner xsep=\bbx, inner ysep=\bby
          }
        }
      }
		},
		string decoration/.store in=\stringdec,
		string decoration={
      \arrow{stealth};
    },
		string decoration pos/.store in=\stringdecpos,
		string decoration pos=.7,
		bbx/.store in=\bbx,
		bbx = 1.5cm,
		bby/.store in=\bby,
		bby = 1.5ex,
		bb port sep/.store in=\bbportsep,
		bb port sep=1.5,
		bb port length/.store in=\bbportlen,
		bb port length=4pt,
		bb penetrate/.store in=\bbpenetrate,
		bb penetrate=0,
		bb min width/.store in=\bbminwidth,
		bb min width=1cm,
		bb rounded corners/.store in=\bbcorners,
		bb rounded corners=2pt,
		bb small/.style={
      bb port sep=1, 
      bb port length=2.5pt, 
      bbx=.4cm, bb min width=.4cm, 
      bby=.7ex
    },
		bb medium/.style={
      bb port sep=1, 
      bb port length=2.5pt, 
      bbx=.4cm, 
      bb min width=.4cm, 
      bby=.9ex
    },
		bb/.code 2 args={%
			\pgfmathsetlengthmacro{\bbheight}{\bbportsep * (max(#1,#2)+1) * \bby}
			\pgfkeysalso{
        draw,
        minimum height=\bbheight,
        minimum width=\bbminwidth,
        outer sep=0pt,
        rounded corners=\bbcorners,
        thick,
				prefix after command={
          \pgfextra{\let\fixname\tikzlastnode}
        },
				append after command={
          \pgfextra{
            \draw
            \ifnum #1=0
              {} 
            \else 
              foreach \i in {1,...,#1} {
						  	($(\fixname.north west)!{\i/(#1+1)}!(\fixname.south west)$) +(-
							\bbportlen,0) 
              coordinate (\fixname_in\i) -- +(\bbpenetrate,0) coordinate (\fixname_in\i')
              }
            \fi 
            \ifnum 
              #2=0{} 
            \else 
              foreach \i in {1,...,#2} {
							($(\fixname.north east)!{\i/(#2+1)}!(\fixname.south east)$) +(-
							\bbpenetrate,0) 
              coordinate (\fixname_out\i') -- +(\bbportlen,0) coordinate (\fixname_out\i)
              }
            \fi;
          }
        }
      }
		},
		bb name/.style={
      append after command={
        \pgfextra{
          \node[anchor=north] at (\fixname.north) {#1}
        ;}
      }
    }
	}
  \newcommand{\Type}[1]{\texttt{#1}} %
  \newcommand{\Bool}{\Type{Bool}} %
	\newcommand{\True}{\Type{True}} %
  \newcommand{\False}{\Type{False}} %
  \newcommand{\Int}{\Type{Int}} %

  \newcommand{\Petri}{\textbf{Petri}} %
  \newcommand{\PetriGrounded}{\textbf{Petri}_G} %
  \newcommand{\FSSMC}{\textbf{FSSMC}} %
  \newcommand{\FSSMCGrounded}{\textbf{FSSMC}_G} %

  \newcommand{\PetriZ}{\textbf{Petri}^\Integers} %
  \newcommand{\FSCCC}{\textbf{FSCCC}} %

  \newcommand{\Sym}[1]{\mathcal{S}_{#1}} %

  \newcommand{\Fold}[1]{\mathfrak{F}(#1)} %
  \newcommand{\UnFold}[1]{\mathfrak{U}(#1)} %

  \newcommand{\Strings}[1]{#1^{\otimes}} %
	\newcommand{\Multiplicity}[1]{\mathfrak{M}_{#1}} %
	\newcommand{\Ordering}[1]{\mathfrak{O}_{#1}} %

  \newcommand{\ListInt}{\Type{List[\Int]}} %

  \newcommand{\Semantics}{\mathcal{S}} %
\title{{\Huge\bfseries The Mathematical Specification\\
		 of the\\
		 Statebox Language}}
\author{
	\includegraphics[width = 65mm]{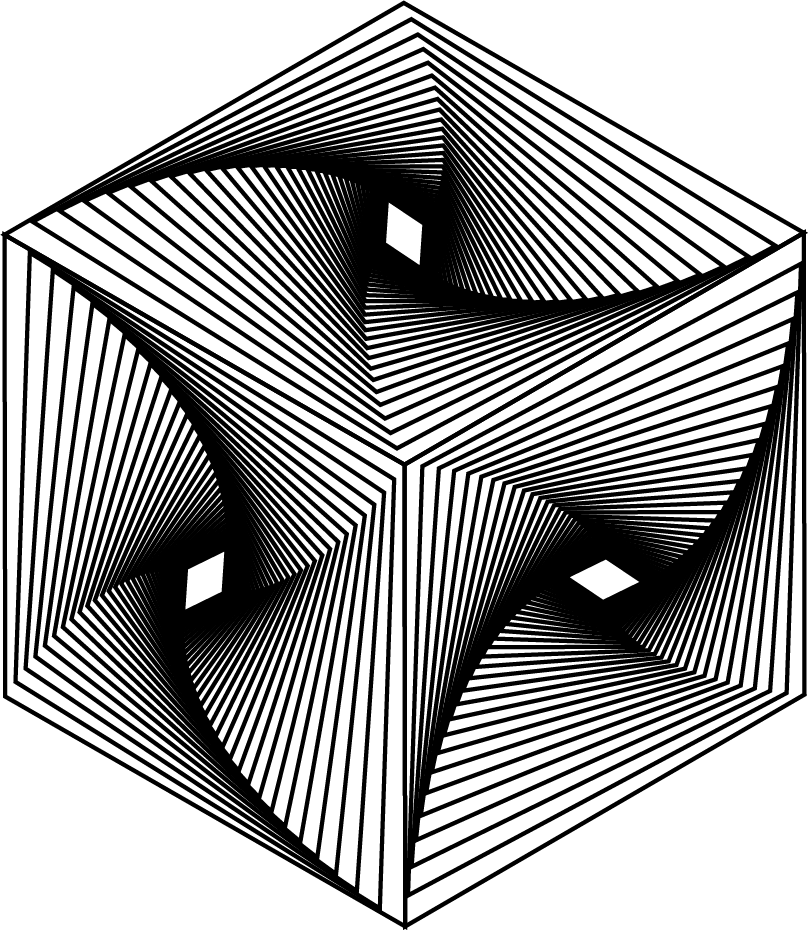}\\[4ex] 
	{
    \huge \bfseries Statebox Team\footnote{
      The list of people that contributed to this document is contained in~\nameref{ch: contributors}.
    }
  }
  \\[4ex]
  {
    \LARGE 
    \href{https://statebox.org}{\texttt{statebox.org}}
  }
}
\begin{document}
\maketitle
{
  \hypersetup{hidelinks} %
  \tableofcontents
}
\chapter*{Contributors}
  \label{ch: contributors}
  \addcontentsline{toc}{chapter}{Contributors}
This document is the result of years of discussion, joint work and 
development by different members of the Statebox team. Ideas, help and 
feedback from our advisors and many other people met in many different 
circumstances (at conferences, on the internet, etc.) have also been 
invaluable and fundamental. 

Jelle Herold is to be credited with the original idea of building a 
programming language based on Petri nets and category theory.
Fabrizio Genovese took care of formalizing this idea into a mathematically 
precise framework, and materially wrote the majority of this document. 
He is to blame for any typo or inaccuracy in what follows.
Many other people contributed in laying down these mathematical 
foundations, either by proving results or by suggesting central ideas,
most notably: 
Jelle Herold, 
David Spivak,
Neil Ghani,
Dani\"el van Dijk and 
Stefano Gogioso. 

We also want to explicitly thank former and current team members
Alex Gryzlov,
Fredrik Nordvall-Forsberg,
Jack Ek,
Marco Perone,
Andr\'e Videla,
Andre Knispel,
Erik Post, 
Anton Livaja, 
Bert Span,
Ryan Wisnesky and
Anthony Di Franco for the useful technical discussions and material 
they provided, which made this document better.

Emi Gheorghe, Anton Livaja and Erik Post have to be credited for 
having done the majority of proofreading of this text.
\section*{Acknowledgements}
  \addcontentsline{toc}{section}{Acknowledgements}
In addition to this, we want to thank all the researchers working in 
areas related to what we do. They all contributed, either directly or 
indirectly, by making this document more mathematically grounded.
Many of them also dedicated time to our project
 by taking part in our research meetings and summits, by hosting 
us at their research institutions and homes or by providing opportunities for us to join 
community conferences and workshops.
This includes many people in the Applied Category Theory community,
in particular
Pawel Sobocinski, 
Neil Ghani,
Robin Piedeleu, 
Jules Hedges,
David Spivak,
Brendan Fong, 
Jade Master,
Fabio Gadducci,
Philipp Zahn, 
Viktor Winschel,
Bob Coecke, 
John Baez, 
Bas Spitters,
Helle Hvid Hansen,
Christina Vasilakopoulou,
Fabio Zanasi,
Bartosz Milewski,
Dan Ghica,
Christian Williams,
David Reutter,
Michael Robinson,
Francisco Rios,
Blake Pollard,
Daniel Cicala,
 and
Dusko Pavlovic,
as well as people coming from different research fields, 
such as 
Andrew Polonsky, 
Yoichi Hirai, 
Arian van Putten,
Jason Teutsch,
Aron Fischer,
Jon Paprocki and
Martin Lundfall.

Furthermore, we need to thank many of the active members in our 
online communities (Telegram, Twitter, etc.), which provided 
conceptual insights, spotted typos, and gave any sort of feedback. 
We know some of these people only by their digital handles, so we will 
refer to them in this way when no alternatives are 
possible: 
Zans, 
@no\_identd,
Hj\"orvar,
Matthew York, 
Dotrego, 
Nikolaj-K, 
Arseniy Klempner, 
Herve Moal.
Special thanks go to Kasper Keunen and Josh Harvey,
which provided early feedback and insights and to Roy Blackstone and 
Greedy Ferengi, which proofread our document and spotted errors.

As one can see this document is the result of many different, 
entangled contributions. We are sure we forgot to mention some people, 
and we apologize in advance for this. 

In this setting, talking about contribution and ownership in a traditional 
sense is difficult. For this reason, we opted to use the wording ``Statebox team'' to broadly  
refer to the authors and contributors of this paper.
\section*{How to cite this document}
  \addcontentsline{toc}{section}{How to cite this document}
Statebox Team. \emph{The Mathematical Specification 
of the Statebox Language}, 2018. ArXiv: TBD.
\chapter{Introduction}\label{ch: introduction}
This document defines the mathematical backbone of the Statebox 
language. In the simplest way possible, Statebox can be seen as a 
clever way to tie together different theoretical structures to maximize 
their benefits and limit their downsides. Since consistency and 
correctness are central requisites for our language, it became 
clear from the beginning that such tying could not be achieved by just 
hacking together different pieces of code representing implementations 
of the structures we wanted to leverage: Rigorous mathematics is employed 
to ensure both conceptual consistency of the language 
and reliability of the code itself. The mathematics presented here is what guided 
the implementation process, and we deemed very useful to release it 
to the public to help people wanting to audit our work to better 
understand the code itself.
\section{What to expect}
This document is a work in progress, and will be released together 
with each version of the Statebox language, suitably expanded to 
cover the new features we will gradually implement. Each version 
of it will contain more theoretical material than what will actually 
be implemented in the Statebox version it comes together with. This 
serves the purpose of helping the audience 
understand what we are working on, and what to expect from 
the upcoming releases.

In this document there is very little code involved, and quite 
a lot of mathematics. The maths will always be introduced together 
with intuitive explanations meant to clarify the ideas we are trying 
to formalize. Notice that here we care more about giving the bigger 
picture of the language itself and will focus on technical details only 
when strictly needed. There are a number of seminal papers that 
explain, with a much greater deal of precision, some of the theoretical 
material that we are employing to implement the Statebox language, 
and we will constantly refer the reader to them for details. On the other hand,
sometimes the material covered here is genuinely new,  in which case
details can be found in papers we published ourselves in peer 
reviewed venues, as in~\cite{Genovese2018}. 
Again, in this case we will reference 
the audience to our own contributions for a thorough presentation of 
the concepts covered. 

All in all, the reader should consider this document as a 
high-level presentation of how concepts we are using
interact together, and should follow the references provided to 
understand the technicalities.
\begin{itemize}
	\item The audience with a strong background in theoretical 
	computer science can use this document to understand how we 
	plan to use results in different research fields to create a new 
	programming language, and how we achieve consistent 
	interaction between them, especially when they are expressed 
	using very different formalisms. An exhaustive explanation of the 
	concepts presented, if needed, will be found in the bibliographic 
	references;

	\item The inexperienced reader will be able to understand the 
	content of cutting-edge research that would be otherwise difficult 
	or impossible to access directly. Hopefully, reading this document 
	will make the reader's attempt to read the papers firsthand 
	easier -- if they choose to do so.
\end{itemize}
It is also worth stressing that we did our best to keep the bibliography 
to a bare minimum, to help the people willing to dig deeper 
focus on a few, selected resources. In particular, when possible, 
we relied on works which are considered the standard reference in 
their field, as in the case of~\cite{MacLane1978} for category theory.
\section{Prerequisites}\label{sec: prerequisites}
We did our best to make this document as accessible 
as possible. This clearly required a trade-off between 
exhaustive presentation and conceptual accessibility. In general, we 
assume very little previous knowledge. Our ideal reader knows some 
basic set theory, knows how to manipulate equalities and, at least in 
principle, understands how coding works. This does not mean that it 
is necessary to be a programmer to understand this document. 
What we require is having a vague idea of how, 
conceptually, humans instruct machines on how to perform tasks. 
This said, an inclination toward logical thinking and approaching 
problems rationally and in a pragmatic way is surely needed to 
understand this work properly.

Throughout this document, we will often make remarks and examples 
intended for a more experienced audience. These are marked with an 
\emph{asterism superscript} (like this\textsuperscript{$\star$}) and 
can be safely ignored without undermining the general comprehension 
of the concepts exposed if too difficult to grasp.

We moreover tried as much as possible to stick to common 
mathematical notation to avoid any kind of discomfort, making 
exceptions only when ambiguity could arise.
\section{Synopsis}
We conclude this short introduction by presenting a synopsis of 
what we are going to do in each Chapter of this document. As we 
already mentioned, this document is a work in progress, and its synopsis 
will be changed accordingly as the amount of released material grows.
\begin{itemize}
	\item This document is divided into parts. Part I is named 
	``first concepts'' and introduces the basic ideas behind Statebox;
	\begin{itemize}
		\item In Chapter~\ref{ch: Petri nets} we will introduce Petri nets, 
		one of the fundamental ingredients in our language. 
		The emphasis in this Chapter falls on why Petri nets make a 
		great graphical tool to reason about complex infrastructure.
		We will also describe some of the most interesting properties 
		that nets can have, and why it is important to study them;
		
		\item In Chapter~\ref{ch: introduction to category theory} we 
		will introduce category theory, the mathematical framework that 
		will allow us to find a common ground to tie Petri nets with 
		other theoretical structures. This will ultimately enable us to 
		export Petri nets from the realm of theoretical research to true 
		software engineering, turning them into a great way of 
		designing complex code while guaranteeing consistency and 
		reliability. The categories we use come endowed with a diagrammatic
		formalism which we will explain in detail. It will serve the purpose 
		of backing up the strength of mathematical reasoning with a
		visual, intuitive representation of concepts;
		\item In Chapter~\ref{ch: executions} we will give a first 
		"categorification" of nets, expressing some of the concepts 
		covered in Chapter~\ref{ch: Petri nets} using category theory. 
		We will show how this allows us to use Petri nets in a much more 
		powerful way and to fine-tune our reasoning about them, for 
		instance by allowing us to track the whole history of a token in a 
		net. This will give us the needed tools to see nets as deterministic 
		objects by defining their categories of executions, which is a 
		fundamental step to make the implementation of Petri nets useful;
		\item In Chapter~\ref{ch: folds} we will elaborate on the results 
		of Chapter~\ref{ch: executions}, showing how we can map Petri 
		nets to other programming languages to produce actual software 
		in a conceptually layered fashion. This is achieved by a functorial 
		mapping from net executions to semantic categories of functional 
		programming languages, allowing us to achieve a separation 
	        between software topology and software meaning;
	\end{itemize}
	\item More parts will follow in the upcoming months, as our research 
	becomes stable enough to be added to this document.
\end{itemize}
Throughout the document, often at the end of a chapter, 
we will make direct reference to our codebase to point out how we 
implemented in practice a mathematical concept. We hope this will 
help the reader to establish links between the theory presented here 
and the codebase hosted on Github~\cite{StateboxTeam2017a}.

\newtoggle{togglePetriNets}\toggletrue{togglePetriNets}
	\chapter{Petri nets}\label{ch: Petri nets}
\emph{Petri nets} were invented by Carl Adam Petri in 1939 
to model chemical reactions~\cite{Petri2008}. In the subsequent 
years, they have met incredible success, especially in computer 
science, to study and model distributed/concurrent 
systems~\cite{Nielsen1991, Riemann1999}. In this Chapter, 
we will start explaining what a Petri net is, and why we 
chose this structure to be at the very core of Statebox.

We will start with an informal introduction, relying on the graphical 
formalism of nets to present concepts in an intuitive way. Then 
we will proceed by formalizing everything in mathematical terms. 
Finally, we will define some useful properties of nets which we 
will be interested in studying later on.
\section{Petri nets, informally}\label{sec: Petri nets informally}
\begin{figure}[!ht]
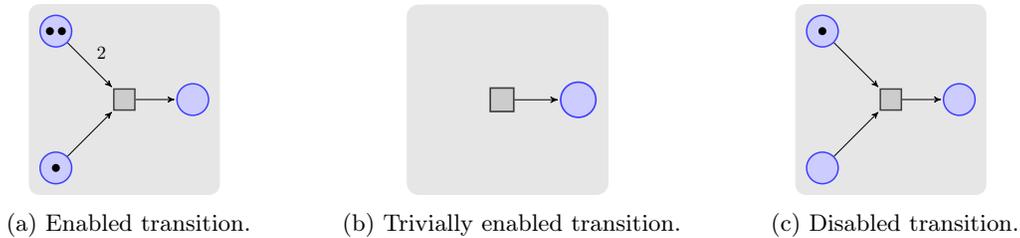

	\centering
	\begin{subfigure}[t]{0.32\textwidth}\centering
		\input{Pictures/PetriNets/01-TransitionEnabled.tex}
		\caption{Enabled transition.}
		\label{fig: Petri net enabled}
	\end{subfigure}
	\hfill
	\begin{subfigure}[t]{0.32\textwidth}\centering
		\input{Pictures/PetriNets/01-TransitionTriviallyEnabled.tex}
		\caption{Trivially enabled transition.}
		\label{fig: Petri net trivially enabled}
	\end{subfigure}
	\hfill
	\begin{subfigure}[t]{0.32\textwidth}\centering
		\input{Pictures/PetriNets/01-TransitionDisabled.tex}
		\caption{Disabled transition.}
		\label{fig: Petri net disabled}
	\end{subfigure}
	\caption{Example of enabled and disabled Petri nets.}
\end{figure}

\begin{figure}[!ht]
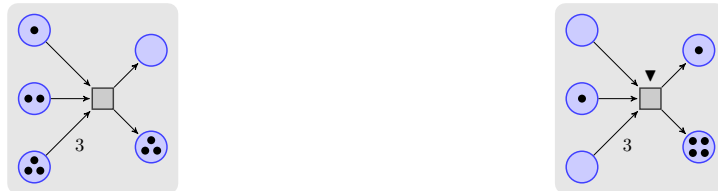

	\centering
	\begin{subfigure}[t]{0.45\textwidth}\centering
		\input{Pictures/PetriNets/02-FiringBefore.tex}
	\end{subfigure}
	\hspace{2ex}
	\begin{subfigure}[t]{0.45\textwidth}\centering
		\input{Pictures/PetriNets/02-FiringAfter.tex}
	\end{subfigure}
	\caption{An enabled transition before (left) and after (right) firing.}
	\label{fig: Petri net firing}
\end{figure}

A \emph{Petri net} is composed of \emph{places}, 
\emph{transitions} and \emph{arcs weighted on the natural numbers}. 
Any place contains a given number of \emph{tokens}, which 
represent resources. Transitions are connected to places through 
the arcs, and can turn resources into other resources: A transition 
can \emph{fire}, consuming tokens living in places connected to 
its input, and producing tokens living in places connected to its 
output. An example of a Petri net is shown in 
Figure~\ref{fig: Petri net example}, where:
\begin{figure}[!ht]
	\centering
	\input{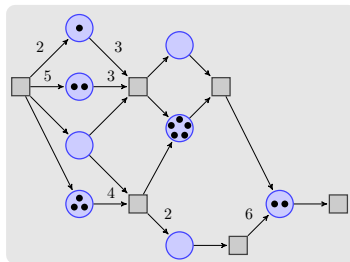}
	\caption{Example of a Petri net.}
	\label{fig: Petri net example}
\end{figure}
\begin{itemize}
	\item Places are represented by blue circles;
	\item Tokens are represented by black dots in each circle;
	\item Transitions are represented by gray rectangles;
	\item A weighted directed arc going \emph{from a place to a 
	transition} represents the transition input; the weight signifies 
	the number of consumed tokens. To avoid clutter, we omit the 
	weights when they are equal to 1;
	\item A weighted directed arc going \emph{from a transition to a 
	place} represents the transition output; the weight signifies the 
	number of produced tokens. To avoid clutter, we omit the weights 
	when they are equal to 1.
\end{itemize}
A Petri net should be thought of as representing some sort of system. 
Tokens are resources, and places are containers that hold resources 
of a given type. Transitions are processes that convert resources from 
one type to another. Weights on the arcs identify how many resources 
of some kind a process needs to be executed, and how many 
resources of some other kind will be produced when the process 
finishes. With respect to this, we say that a transition can be in two 
states:
\begin{description}
	\item[Enabled,] if, in \emph{all} the places having edges towards 
	the transition, there is a number of tokens at least equal to the 
	weight of the edge itself (see Figure~\ref{fig: Petri net enabled}). 
	Note that if a transition has no inbound edges 
	(as in Figure~\ref{fig: Petri net trivially enabled}), then it is 
	always considered enabled;
	\item[Disabled,] otherwise (see Figure~\ref{fig: Petri net disabled}).
\end{description}
When a transition is enabled, then we say that it may \emph{fire}. 
Firing represents the act of executing the process the transition 
represents. When a transition fires, a number of tokens are 
\emph{removed} from each input place, according to the arc 
weight, and similarly a number of tokens are \emph{added} to each 
output place, again according to the arc weight. 
Figure~\ref{fig: Petri net firing} shows an enabled transition 
before (left) and after (right) firing. As you can see, we 
highlight firing transitions with a black triangle.
\begin{remark}[Generalized nets]
	Note that the behavior of Petri nets can be generalized 
	much further than this, for example by annotating the 
	arcs with logical conditions that have to be satisfied to 
	consider a transition enabled, or by introducing transitions 
	that -- a bit counterintuitively -- fire only when there are 
	no tokens in one of their input places. Working in a greater 
	degree of generality, though, can make much more 
	difficult -- or even impossible -- to answer questions pertaining 
	reachability and absence/presence of deadlocks, which are 
	important concepts that will be formally introduced later. 
	In Statebox, the fundamental requirement is that we should 
	always be able to tell what is going on in our processes. For 
	this reason we do prefer working with a restricted set of rules 
	and to be very careful in adopting any generalization. The 
	study of how suitably extend the expressivity of the nets 
	considered here will be the focus of the second part of this 
	document. 
\end{remark}
\section{Multisets}\label{sec: multisets}
The first concrete goal of this Chapter is to state the intuitive 
concepts presented above in mathematical terms. Before we 
can introduce Petri nets formally, we need a way to 
formalize \emph{multisets}. Intuitively, a multiset is just 
a \emph{set with repetition}, meaning that each element is 
allowed to occur multiple times in the same set. To make things 
easier to understand, consider the following writings:
\begin{equation}\label{eq: multiset example}
	\{a,b,d,e,k\} \qquad \{a,b,b,d,e,e,e,k\} \qquad \{a,b,b,d,e,e,k,k\}
\end{equation}
When seen as sets, the ones above denote the same thing, since 
sets ignore repeated elements. The 
reason why we are interested in the concept of a multiset is 
precisely because, in our case, we want to be able to consider 
the three sets above as distinct. The experienced reader will 
have already noted how the need for multisets naturally arises 
when dealing with Petri nets. Specifically, multisets will be 
useful in:
\begin{itemize}
	\item Describing the transitions of a Petri net, since we can 
	represent how many tokens a transition consumes (produces) 
	from (in) a place as the number of occurrences of that place in 
	a multiset;
	\item Describing the state of a Petri net, since we can represent 
	the number of tokens in each place as the number of occurrences 
	of that place in a multiset.
\end{itemize}
Without further ado, let us introduce the first mathematical definition 
of this document.
\begin{definition}[Multiset]\label{def: multiset}
	A \emph{multiset on $S$} is a function 
	$\MsetBase{X}{S}: S \to \Naturals$, where $S$ is a set. 
	A multiset is called \emph{finite} when there is only a 
	finite number of $s \in S$ such that $\MsetBase{X}{S}(s)> 0$. 
	Finite multisets will usually be denoted with a $\Naturals$ used 
	as superscript. For instance, $\MsetBase{X}{S}$ represents a 
	finite multiset on $S$.
\end{definition}
\begin{remark}[Non-finite multisets]
	In this work, we are only interested in finite multisets. 
	To avoid clutter, \emph{we will refer to finite multisets just as 
	multisets}.
\end{remark}
\begin{example}[Multisets are functions]
	As we said, multisets have to be interpreted as sets where 
	the same element can be repeated a finite number of times. 
	If we go back to the sets displayed in 
	Equation~\ref{eq: multiset example}, we readily see how 
	these can indeed be expressed as functions 
	$f,g,h:\{a,b,d,e,k\} \to \Naturals$, taking values:
	\begin{align*}
		f(a) &= 1 &\qquad f(b) &= 1 &\qquad f(d) &= 1 &
			\qquad f(e) = &1 &\qquad f(k) &= 1\\
		g(a) &= 1 &\qquad g(b) &= 2 &\qquad g(d) &= 1 &
			\qquad g(e) = &3 &\qquad g(k) &= 1\\
		h(a) &= 1 &\qquad h(b) &= 2 &\qquad h(d) &= 1 &
			\qquad h(e) = &2 &\qquad h(k) &= 2
	\end{align*}
	Where $f$ is the function representing the first multiset, 
	$g$ the function representing the second, and $h$ the function 
	representing the third, respectively.
\end{example}
\begin{remark}[Same multiset, different functions]
	Note that our definition of $\Naturals$ includes $0$, and hence 
	if we have a function $g': \{a,b,c,d,e,k\} \to \Naturals$ defined as:
	\begin{align*}
		g'(a) &= 1 &\quad g'(b) &= 2 &\quad g'(c) &= 0 &
			\quad g'(d) &= 1 &\quad g'(e) = &3 &\quad g'(k) &= 1
	\end{align*}
	This also defines the multiset $\{a,b,b,d,e,e,e,k\}$, like $g$. This can be a 
	source of confusion, and hence in the notation for 
	multiset -- namely $\MsetBase{X}{S}$ -- we make the base 
	set explicit. Also note that subsets of a set $S$ correspond to 
	functions $f:S \to \{0,1\}$, and can thus be seen as particular 
	multisets on $S$ where each element is mapped to $0$ or $1$.
\end{remark}
\begin{definition}[Set of multisets over $S$]\label{def: set of multisets}
	$\Msets{S}$ denotes the set of all possible finite multisets 
	over $S$, that is,
	\begin{equation*}
		\Msets{S}:= \{\MsetBase{X}{S}:S \to \Naturals, \mid 
			\text{$\MsetBase{X}{S}(s)>0$ for a finite number of $s \in S$}\}
	\end{equation*}
\end{definition}
\subsection{Operations on multisets}\label{subsec: operations on multisets}
To be able to proficiently use multisets to formalize Petri nets, 
we need to understand what we can do with them. Given two 
multisets $\MsetBase{X}{S}, \MsetBase{Y}{S}$ on $S$, we can 
generalize many operations from sets to multisets, as inclusion, 
union and difference using point-wise definitions.
\begin{definition}[Operations on multisets]\label{def: operations on multisets}
	Let $\MsetBase{X}{S}, \MsetBase{Y}{S} \in \Msets{S}$. 
	Set \emph{inclusion} generalizes easily setting, for all $s \in S$,
	\begin{equation*}
		\MsetBase{X}{S} \subseteq \MsetBase{Y}{S} := 
			\MsetBase{X}{S}(s) \leq \MsetBase{Y}{S}(s)
	\end{equation*}
	Similarly, \emph{union} can be generalized to 
	multisets $\MsetBase{X}{S}$ and $\MsetBase{Y}{S}$, setting:
	\begin{align}
		\cup:\Msets{S} \times \Msets{S} & \to \Msets{S} \\
		(\MsetBase{X}{S} \cup \MsetBase{Y}{S})(s) & := 
			\MsetBase{X}{S}(s) + \MsetBase{Y}{S}(s)\nonumber
	\end{align}
	When $\MsetBase{X}{S} \subseteq \MsetBase{Y}{S}$, 
	we can moreover define their \emph{multiset difference}, 
	that unsurprisingly is just:
	\begin{align*}
		-:\Msets{S} \times \Msets{S} &\to \Msets{S}\\
		(\MsetBase{Y}{S} - \MsetBase{X}{S})(s) &:= 
			\MsetBase{Y}{S}(s) - \MsetBase{X}{S}(s)
	\end{align*}
	Another intuitive operation that can be defined on multisets 
	is the one of \emph{scalar multiplication}, that is similar 
	in concept to scalar products for vector spaces. 
	For each $n \in \Naturals$ and $s \in S$, we set:
	\begin{align*}
		\cdot :\Naturals \times \Msets{S} &\to \Msets{S}\\
		(n \cdot \MsetBase{X}{S})(s) &:= n\, \MsetBase{X}{S}(s)
	\end{align*}
	Denoting with $S_1 \Disjoint S_2$ the disjoint union of sets, 
	that we recall being defined as:
	\begin{equation*}
		S_1 \Disjoint S_2 := \{(s_1,0) \mid s_1 \in S_1\}
			 \cup \{(s_2,1) \mid s_2 \in S_2\}
	\end{equation*}
	We can moreover define the analogous 
	\emph{disjoint union of multisets}, setting for 
	all $s \in S \Disjoint S'$:
	\begin{align*}
		\Disjoint:\Msets{S} \times \Msets{S'} &\to \Msets{S \Disjoint S'}\\
		(\MsetBase{X}{S_1} \Disjoint \MsetBase{Y}{S_2})(s) &:=
		\begin{cases}
			\MsetBase{X}{S_1}(s_1) \text{ iff } s = (s_1,0) \\
			\MsetBase{Y}{S_2}(s_2) \text{ iff } s = (s_2,1)
		\end{cases}
	\end{align*}
	For each set $S$ we denote with $\Zeromset{S}$ the multiset 
	in $\Msets{S}$ with the following property:
	\begin{equation*}
		\forall s \in S,~\Zeromset{S}(s) = 0
	\end{equation*}
	Finally, we define the \emph{cardinality} of a 
	multiset $\MsetBase{X}{S}$ as:
	\begin{equation*}
		\left|\MsetBase{X}{S} \right| := \sum_{s \in S} \MsetBase{X}{S}(s)
	\end{equation*}
\end{definition}
\begin{remarkhard}[Multisets are free commutative monoids]\label{rem: multisets are free commutative monoids}
	The reader fluent in algebra will have noted that multiset 
	union defines an operation in the algebraic sense, 
	that makes $\Msets{S}$, for each $S$, 
	\emph{the free commutative monoid generated by $S$}, 
	where the unit is the multiset $\Zeromset{S}$.
\end{remarkhard}
\begin{remarkhard}[Injections of multisets]\label{rem: injection}
	The multiset $\Zeromset{S_1}$ can be cleverly used to 
	inject $\Msets{S_2}$ into $\Msets{S_1 \Disjoint S_2}$, as follows:
	\begin{align*}
		\Msets{S_2} &\Inject \Msets{S_1 \Disjoint S_2}\\
		\MsetBase{Y}{S_2} &\mapsto \Zeromset{S_1} \Disjoint \MsetBase{Y}{S_2}
	\end{align*}
	The set $S$ may be embedded into $\Msets{S}$ via a 
	function $\delta: S\to\Msets{S}$, defined as:
	\begin{equation*}
		\delta(s)(s') := \begin{cases}
			1, \text{ iff }s=s' \\
			0, \text{ iff }s\neq s' \\
		\end{cases}
	\end{equation*}
	Finally, given a function $f:S_1 \to \Msets{S_2}$, we can 
	abuse notation and consider $f$ as a function of multisets 
	$\Msets{S_1}\to\Msets{S_2}$, by defining
	\begin{equation*}
		f: \MsetBase{X}{S_1} \in \Msets{S_1} \mapsto 
			\bigcup_{s_1 \in S_1} \Mset{X}(s_1)\cdot f(s_1) \in \Msets{S_2}
	\end{equation*}
\end{remarkhard}
In the remainder of this document, we will just 
write $\Mset{X}$ instead of $\MsetBase{X}{S}$ when the 
base set $S$ is clear from the context.
\section{Petri nets, formally}\label{sec: Petri nets, formally}
\begin{figure}[!ht]
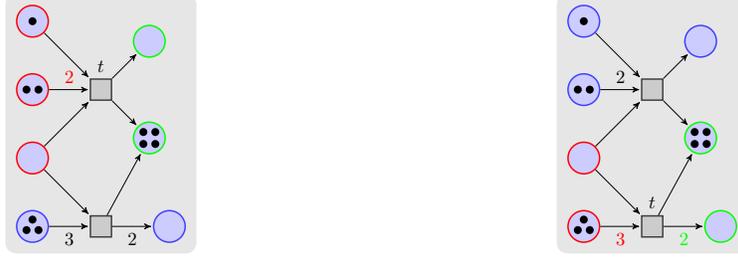

	\centering
	\begin{subfigure}[t]{0.45\textwidth}\centering
		\input{Pictures/PetriNets/03-TransitionInputOutput1.tex}
	\end{subfigure}
	\hspace{1em}
	\begin{subfigure}[t]{0.45\textwidth}\centering
		\input{Pictures/PetriNets/03-TransitionInputOutput2.tex}
	\end{subfigure}
	\caption{Examples of input/output places of transitions.}
	\label{fig: transition example}
\end{figure}

Now that we have some intuition about how Petri nets work 
and have introduced multisets, it is time to define Petri nets formally.
\begin{definition}[Petri net]\label{def: Petri net}
	A \emph{Petri net} is a quadruple
	\begin{equation*}
		N := \Net{N}
	\end{equation*}
	Where:
	\begin{itemize}
		\item $\Pl{N}$ is a \emph{finite} set, representing places;
		\item $\Tr{N}$ is a \emph{finite} set, representing transitions;
		\item $\Pl{N}$ and $\Tr{N}$ are disjoint: Nothing can be a 
		transition and a place at the same time;
		\item $\Pin{-}{N}:\Tr{N} \to \Msets{\Pl{N}}$ is a function 
		assigning to each transition the multiset of $\Pl{N}$ 
		representing its \emph{input} places;
		\item $\Pout{-}{N}:\Tr{N} \to \Msets{\Pl{N}}$ is a function 
		assigning to each transition the multiset of $\Pl{N}$ 
		representing its \emph{output} places.
	\end{itemize}
	We will often denote with $\Tr{N}, \Pl{N}, \Pin{-}{N}, \Pout{-}{N}$ 
	the set of places, transitions and input/output functions 
	of the net $N$, respectively.
\end{definition}
\begin{example}[Input and output places]
	In Figure~\ref{fig: transition example} we highlighted 
	the action of $\Pin{t}{N}$ in red for two different transitions, 
	denoted with $t$. We did the same for $\Pout{t}{N}$, 
	highlighted in green.
\end{example}
\begin{remarkhard}[Generalized input and output]
	Given a Petri net $N$, we can generalize 
	$\Pin{-}{N}$ and $\Pout{-}{N}$ to functions of 
	multisets $\Msets{\Tr{N}} \to \Msets{\Pl{N}}$ using the 
	procedure explained in Remark~\ref{rem: injection}, that is, 
	we can extend them so that they act on multisets of transitions, 
	as follows:
	\begin{gather*}
		\Pin{-}{N}: \Mset{U} \in \Msets{\Tr{N}} \mapsto 
			\bigcup_{t \in \Tr{N}} \Mset{U}(t)\cdot \Pin{t}{N} \in \Msets{\Pl{N}}\\
		\Pout{-}{N}: \Mset{U} \in \Msets{\Tr{N}} \mapsto 
			\bigcup_{t \in \Tr{N}} \Mset{U}(t)\cdot \Pout{t}{N} \in \Msets{\Pl{N}}
	\end{gather*}
\end{remarkhard}
\subsection{Markings, enabled transitions}\label{subsec: markings, enabled transitions}
Up to now, we still did not formalize the concept of a marking. 
At the moment, our Petri nets are empty, meaning that we 
do not have a way to populate places with tokens. This can be 
readily expressed using multisets again.
\begin{definition}[Marking]\label{def: marking}
	Given a Petri net $N$, a \emph{marking} (also called a 
	\emph{state}) for $N$ is a multiset 
	on $\Pl{N}$, $\Marking{X}:\Pl{N} \to \mathbb{N}$.
\end{definition}
The interpretation is that the marking assigns a finite, positive 
or zero number of tokens to each place of $N$. We denote 
that a net $N$ comes endowed with a marking $\Marking{X}$ using 
the notation $N_\Marking{X}$. Equivalently, we can also say 
that $N$ is in the \emph{state} $\Marking{X}$ to refer 
to $N_\Marking{X}$.

Having formalized the concept of a marking, we can now 
take care of defining the dynamics of a Petri net.
\begin{definition}[Enabled transition]
	Given a Petri net $N$ in the state $\Marking{X}$, we say that a 
	transition $t \in \Tr{N}$ is \emph{enabled} if:
	\begin{equation*}
		\Pin{t}{N} \subseteq \Marking{X}
	\end{equation*}
\end{definition}
Note that since we are working with multisets, this is equivalent to
\begin{equation*}
	\forall p \in \Pl{N},\, \Pin{t}{N}(p) \leq \Marking{X}(p)
\end{equation*}
meaning, as we would expect, that a transition is enabled if 
and only if in any input place for $t$ there are at least as many 
tokens available as $t$ will have to consume.
\begin{remarkhard}[Enabled check]
	$\Msets{\Pl{N}}$ denotes the set of all possible multisets 
	over $\Pl{N}$. For a net $N$ we can define a function
	\begin{equation*}
		\overline{(-)}_{(-)}: \Tr{N} \times \Msets{\Pl{N}} \to \{\top, \bot\}
	\end{equation*}
	that takes a transition $t$ and a marking $\Marking{X}$ as input 
	and returns $\top$ if $t$ is enabled in $\Marking{X}$, and $\bot$ 
	otherwise. This function can be generalized to sets of 
	transitions $U \subseteq \Tr{N}$ by setting 
	$\overline{U}_{\Marking{X}}:=\bigwedge\limits_{t \in U} \overline{t}_M$, 
	where $\bigwedge$ denotes the usual logical conjunction 
	of predicates. This function is important from an implementation 
	point of view as it allows for an efficient way to determine 
	if a given transition can fire in a given state.
\end{remarkhard}
\subsection{Firing semantics for Petri nets}\label{subsec: firing semantics for Petri nets}
Now, we have to define a \emph{firing policy} -- also called 
\emph{firing semantics} -- by mathematically formalizing what 
happens when a transition fires. Given a Petri net $P$ in the state 
$\Marking{X}$, the firing of a transition $t$ should have two properties:
\begin{itemize}
	\item $t$ should be able to fire only when enabled;
	\item Firing $t$ should consume some tokens and produce others, 
	thus changing the state of $P$ from $\Marking{X}$ to some other 
	marking $\Marking{Y}$. We will indicate this using the notation 
	$N_\Marking{X} \xrightarrow{t} N_\Marking{Y}$.
\end{itemize}
These two requirements can be captured by the following definition.
\begin{definition}[Firing rule]\label{def: Petri firing policy}
	Let $N$ be a Petri net in a state $\Marking{X}$, and let $t \in \Tr{N}$. 
	We define:
	\begin{equation*}
		N_\Marking{X} \xrightarrow{t} N_\Marking{Y} := 
			\left(\Pin{t}{N} \subseteq \Marking{X}\right) \,\wedge\, 
			\left(\Pout{t}{N} \subseteq \Marking{Y}\right) \,\wedge\, 
			\left(\Marking{X} - \Pin{t}{N} = \Marking{Y} - \Pout{t}{N}\right)
	\end{equation*}
	and say that \emph{$t$ fires, carrying $N$ from $\Marking{X}$ to 
	$\Marking{Y}$}, if it is $N_\Marking{X} \xrightarrow{t} N_\Marking{Y}$.
\end{definition}
Note that in Definition~\ref{def: Petri firing policy} the requirements 
$\Pin{t}{N} \subseteq \Marking{X}$ and $\Pout{t}{N} \subseteq \Marking{Y}$ 
are redundant, since $\Marking{X} - \Pin{t}{N}$ and 
$\Marking{Y} - \Pout{t}{N}$ are defined only under such assumption. 
We decided to list them explicitly to elucidate the fact that for 
$N_\Marking{X} \xrightarrow{t} N_\Marking{Y}$ to be true, $t$ has to 
be enabled in $N_\Marking{X}$.

Our firing policy says that, given a place $p \in \Pl{N}$, when a 
transition fires, \emph{exactly} $\Pin{t}{N}(p)$ tokens are consumed 
from $p$, and exactly $\Pout{t}{N}(p)$ tokens are produced in $p$. This 
causes the net to go from the state $\Marking{X}$ to the state $\Marking{Y}$, 
where $\Marking{Y}$ is obtained from $\Marking{X}$ by adding/subtracting 
the relevant number of tokens as prescribed by the input and output 
functions evaluated on $t$.

\begin{figure}[!ht]
	\centering
	\input{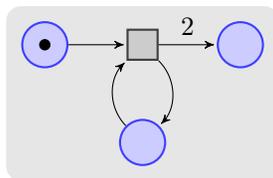}
	\caption{An example of a transition with intersecting input and output places.}
	\label{fig: Petri same input output}
\end{figure}

Note that the same transition can produce and consume tokens from 
the same places, that is, a place can act both as an input and an 
output -- they are not mutually exclusive. The net in 
Figure~\ref{fig: Petri same input output} is an example of this, 
where $\Pin{t}{N} \cap \Pout{t}{N}$ is non-empty.
\begin{remarkhard}[Generalized firing policy]
	Our firing policy can of course be generalized to arbitrary 
	sets of transitions $U \subseteq \Tr{N}$, defining things in the obvious way:
	\begin{equation*}
		N_\Marking{X} \xrightarrow{U} N_\Marking{Y} := \Pin{U}{N} 
			\subseteq \Marking{X} \,\wedge\, \Pout{U}{N} \subseteq \Marking{Y}
			\,\wedge\, \Marking{X} - \Pin{U}{N} = \Marking{Y} - \Pout{U}{N}
	\end{equation*}
\end{remarkhard}
\section{Examples}\label{sec: examples of Petri nets}
\begin{figure}[!ht]
	\centering
	\input{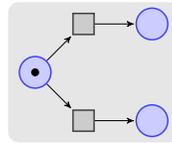}
	\caption{Petri net modeling exclusive choice.}
	\label{fig: Petri transition choice}
\end{figure}

\begin{figure}[!ht]
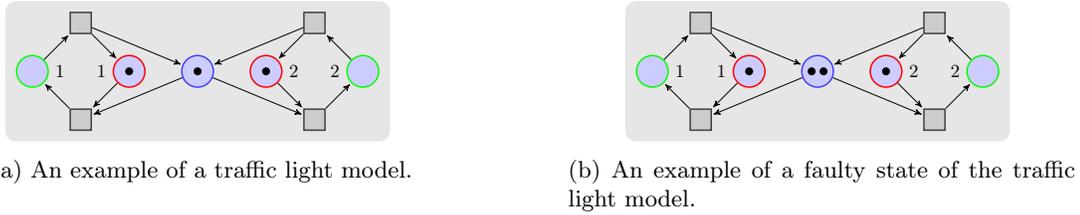

	\centering
	\begin{subfigure}[t]{0.45\textwidth}\centering
		\input{Pictures/PetriNets/07-TrafficLight.tex}
		\caption{An example of a traffic light model.}
		\label{fig: Petri traffic light good}
	\end{subfigure}
	\hfill
	\begin{subfigure}[t]{0.45\textwidth}\centering
		\input{Pictures/PetriNets/07-TrafficLightBad.tex}
		\caption{An example of a faulty state of the traffic light model.}
		\label{fig: Petri traffic light bad}
	\end{subfigure}
	\caption{Traffic light models.}
	\label{fig: Petri traffic lights}
\end{figure}

Petri nets are good for representing the development stage of a product,
 and concurrent behavior. We will show this using examples.
\begin{figure}[!ht]
	\centering
	\input{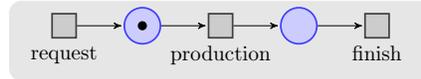}
	\caption{Product development example.}
	\label{fig: Petri process example}
\end{figure}
\begin{example}[Product development]\label{ex: product development}
	We can describe the life stages of a product, from order to production, using 
	Petri nets. The simplest case we can think of is the one in 
	Figure~\ref{fig: Petri process example}. In this case, transitions correspond to 
	different processing stages for a product. Clearly, we can design processes that 
	are much more complicated than this, for instance introducing \emph{exclusive choices} 
	as in the Petri net in Figure~\ref{fig: Petri transition choice}, where the two 
	transitions must compete to fire. Here we can imagine that a user can decide 
	which transition fires, maybe by pressing a button or by filling in a form. And 
	with this model we can represent the fact that once one decision is taken the other 
	one is automatically disabled.
\end{example}
\begin{example}[Traffic Light]\label{ex: traffic light}
	Concurrent behavior models situations where two or more systems have to 
	compete to get the needed resources to run. One typical example is given by 
	a couple of traffic lights (denoted $1$ and $2$, respectively) at a crossing: For 
	simplicity, each traffic light can be green or red, but they cannot be both green 
	at the same time, otherwise cars might crash. We can model this using Petri nets 
	(see Figure~\ref{fig: Petri traffic light good}), where two systems -- representing 
	the traffic lights -- have to compete for the token in the middle to turn the light to green.

	\noindent
	In Figure~\ref{fig: Petri traffic light good}, the places have been colored in red and 
	green, representing ``the colour a traffic light is in''. Transitions represent the 
	switches that change a given traffic light's color. The numbers labeling places 
	represent the traffic light that each place refers to.

	Note that with the marking provided as in the figure above, it can never happen 
	that both lights are green at the same time, thanks to the token in the center place: 
	One light, say $1$, could always be ``better'' at becoming green, thus preventing 
	the second one to ever fire, but situations causing crashes would never happen. Note 
	moreover that since there could be more than one token in each place, there are other 
	markings that do not prevent this situation from happening, such as the one 
	in Figure~\ref{fig: Petri traffic light bad}.
\end{example}
\section{Further properties of Petri nets}\label{sec: further properties of Petri nets}
Now that we have defined Petri nets formally and clarified 
why we deem them useful, it is time to explore the properties 
that a Petri net can have, and to state them formally.
\subsection{Reachability, safeness and deadlocks}\label{subsec: reachability, safeness and deadlocks}
\begin{figure}[!ht]
	\centering
	\input{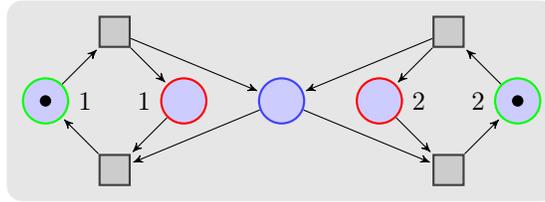}
	\caption{The evolution of a faulty state of the traffic light model.}
	\label{fig: Petri traffic light bad 2}
\end{figure}

\begin{figure}[!ht]
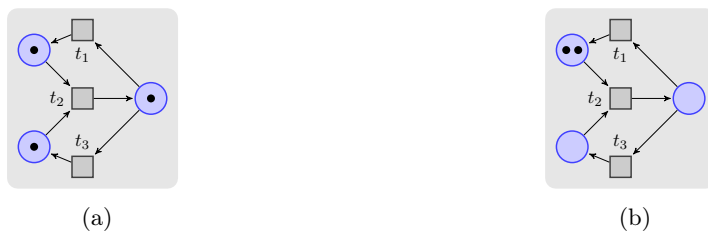

	\centering
	\begin{subfigure}[t]{0.45\textwidth}\centering
		\input{Pictures/PetriNets/09-Deadlock.tex}
		\caption{}
		\label{fig: Petri deadlock}
	\end{subfigure}
	\hspace{0.5em}
	\begin{subfigure}[t]{0.45\textwidth}\centering
		\input{Pictures/PetriNets/09-DeadlockTrivial.tex}
		\caption{}
		\label{fig: Petri deadlock evident}
	\end{subfigure}
	\caption{An example of a deadlocked Petri net.}
\end{figure}

Let us go back to Example~\ref{ex: traffic light}. We already 
saw two different markings, generating two completely 
different behaviors, in Figure~\ref{fig: Petri traffic lights}. 
We recognize that, in Figure~\ref{fig: Petri traffic light bad}, 
firing both transitions at the bottom leads us to the marking 
in Figure~\ref{fig: Petri traffic light bad 2}. ...But this is exactly 
the situation we wanted to avoid, since now cars might start crashing! 
This prompts a question: Given some marking $\Marking{X}$, 
is it possible to \emph{reach} a marking $\Marking{Y}$ with a 
sequence of transition firings?
The traffic light example should clarify how important answering 
this question is. As usual, something important deserves a definition.
\begin{definition}[Reachability]\label{def: reachability}
	Given a Petri net $N_\Marking{X}$ we say that a marking 
	$\Marking{Y}$ is \emph{reachable} from $\Marking{X}$ if there 
	is a \emph{finite} sequence of transitions $t_0, ..., t_n$ such that
	\begin{equation*}
		\Marking{X} \xrightarrow{t_0} \Marking{X}_1 \xrightarrow{t_1} 
			\dots \xrightarrow{t_{n-1}} \Marking{X}_n \xrightarrow{t_n} \Marking{Y}
	\end{equation*}
	If $s$ is a finite sequence of transitions $(t_0, ..., t_n)$, we express 
	the statement above by simply writing $\Marking{X} \xrightarrow{s} \Marking{Y}$.
\end{definition}
\begin{remark}[Studying reachability]
	In the traffic light example our Petri net is simple enough to 
	allow us to manually deduce if some marking can be reached 
	from its initial state by writing out all the possible states of the 
	net. Clearly, when we start designing complex systems, we want 
	to develop formal tools to automatically provide answers to 
	this question. This will be covered later on.
\end{remark}
Another important concept is the one of \emph{deadlock}. The idea 
behind this is that a net is deadlocked if ``it is going to jam'', 
meaning that at some point nothing will be able to fire anymore. 
This can be formalized as follows:
\begin{definition}[Deadlock]\label{def: deadlock}
	Given a Petri net $N$ in the state $\Marking{X}$, we say 
	that $N_\Marking{X}$ is \emph{deadlocked} if there is some 
	marking $\Marking{Y}$, reachable from $\Marking{X}$, in 
	which no transition can fire.
\end{definition}
\begin{example}[Deadlocked net]
	In Figure~\ref{fig: Petri net disabled} it is very easy to see 
	that the net is deadlocked, but things are not always so clear. 
	Consider, for instance, Figure~\ref{fig: Petri deadlock} on the 
	left: Here everything seems fine, but firing $t_2$ and then $t_1$ 
	two times gets us to the state in Figure~\ref{fig: Petri deadlock evident} 
	on the right, which is deadlocked. Deadlock is undesirable because 
	it means that our process cannot progress in any way. As in the 
	case of reachability, we want to develop higher order tools to study 
	if a given Petri net is deadlocked or not.
\end{example}
On the other end of the spectrum, opposed to the 
concept of deadlock, we have the concept of \emph{liveness}. 
Liveness means, in short, absence of deadlocks, as it can be 
easily seen from the following definition:
\begin{definition}[Liveness]\label{def: liveness}
	Given a Petri net $N_\Marking{X}$, we say that a transition $t \in \Tr{N}$ is
	\begin{itemize}
		\item \emph{Dead} if it can never fire;
		\item \emph{Alive} if, for any marking $\Marking{Y}$ 
		reachable from $\Marking{X}$, there is a firing sequence 
		that, from $\Marking{Y}$, leads to a marking $\Marking{Z}$ 
		in which $t$ can be fired.
	\end{itemize}
	We say that $N_\Marking{X}$ is \emph{alive} if all of its transitions are alive.
\end{definition}
This in particular means that, starting from $N_\Marking{X}$, we 
can apply any firing sequence and be sure that, if we keep going, 
we will always end up in a situation in which $t$ can fire.
\begin{remark}[Dead and alive are not incompatible]
	Note that a transition can be both not dead and not alive 
	at the same time: It may be fireable in the state $\Marking{X}$ 
	but become dead later on. The definition above can be 
	generalized, introducing intermediate degrees of liveness 
	between the ones we gave above, but we are not interested 
	in this right now. What is interesting for us is that it is trivial 
	to prove that an alive Petri net is not deadlocked, and that 
	every transition will always be enabled in the future, no 
	matter what we do.
\end{remark}
\begin{example}[Traffic light nets are alive]
	The traffic light nets provided in Figure~\ref{fig: Petri traffic lights} 
	are both alive, while the net in Figure~\ref{fig: Petri net disabled} 
	is not, consisting only of a dead transition.
\end{example}
Being deadlocked is considered a bad quality for a Petri net 
to have. Another property, called \emph{boundedness}, is 
instead considered good:
\begin{definition}[Boundedness and safeness]\label{def: boundedness and safeness}
	Given a Petri net $N_\Marking{X}$, we say that a 
	place $p \in \Pl{N}$ is \emph{$k$-bounded} if it 
	never contains more than $k$ tokens in any reachable 
	marking. We also say that a place is \emph{bounded} 
	if it is $k$-bounded for some $k$.

	We can extend these definitions to the whole net 
	saying that $N_\Marking{X}$ is $k$-bounded (bounded) 
	if all its places are $k$-bounded (bounded) in any state 
	reachable from $\Marking{X}$. Finally, we say that a 
	Petri net is \emph{safe} if it is $1$-bounded.
\end{definition}
\begin{example}[Traffic light nets are bounded]
	The nets in Figure~\ref{fig: Petri traffic lights} are both 
	bounded. The one in Figure~\ref{fig: Petri traffic light good} is also safe.
\end{example}
If we think about Petri nets as modeling process behavior, 
boundedness is a desirable quality because it means that 
at any stage tokens do not accumulate. For example, consider 
the net in Figure~\ref{fig: Petri process example}: This net is 
not bounded, because the leftmost transition could keep firing 
accumulating tokens in the leftmost place. This would happen 
if, for instance, the firing rate of the leftmost transition exceeds 
the firing rate of the middle one, meaning that the demand 
exceeds the production capability.
\begin{figure}[!ht]
	\centering
	\input{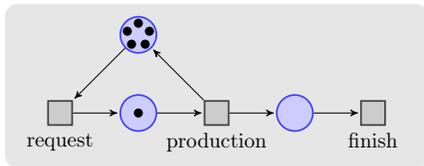}
	\caption{An example of a bounded Petri net modeling production.}
	\label{fig: Petri process example bounded}
\end{figure}
\begin{remark}[Making a net bounded]
	Note that in a situation like the one in Figure~\ref{fig: Petri process example} 
	we can make the net $k$-bounded artificially by 
	adding a place with $k$ tokens in it, as in 
	Figure~\ref{fig: Petri process example bounded}, 
	where we just made the net above $6$-bounded. The 
	interpretation of the added place is that it represents 
	the \emph{maximum production capabilities of the process}. 
	In this case the ``request'' transition is \emph{automatically 
	disabled} (i.e. it cannot fire) when there are six or more 
	tokens waiting for production.
\end{remark}
As usual, we would like a theoretical framework to establish 
when a net is bounded, or when strategies like the one above 
can work to make it such. Many of the questions asked here 
will be answered formally with the categorification of 
Petri nets, that will unveil their compositional nature.
\begin{figure}[!ht]
	\centering
	\includegraphics[height=100px]{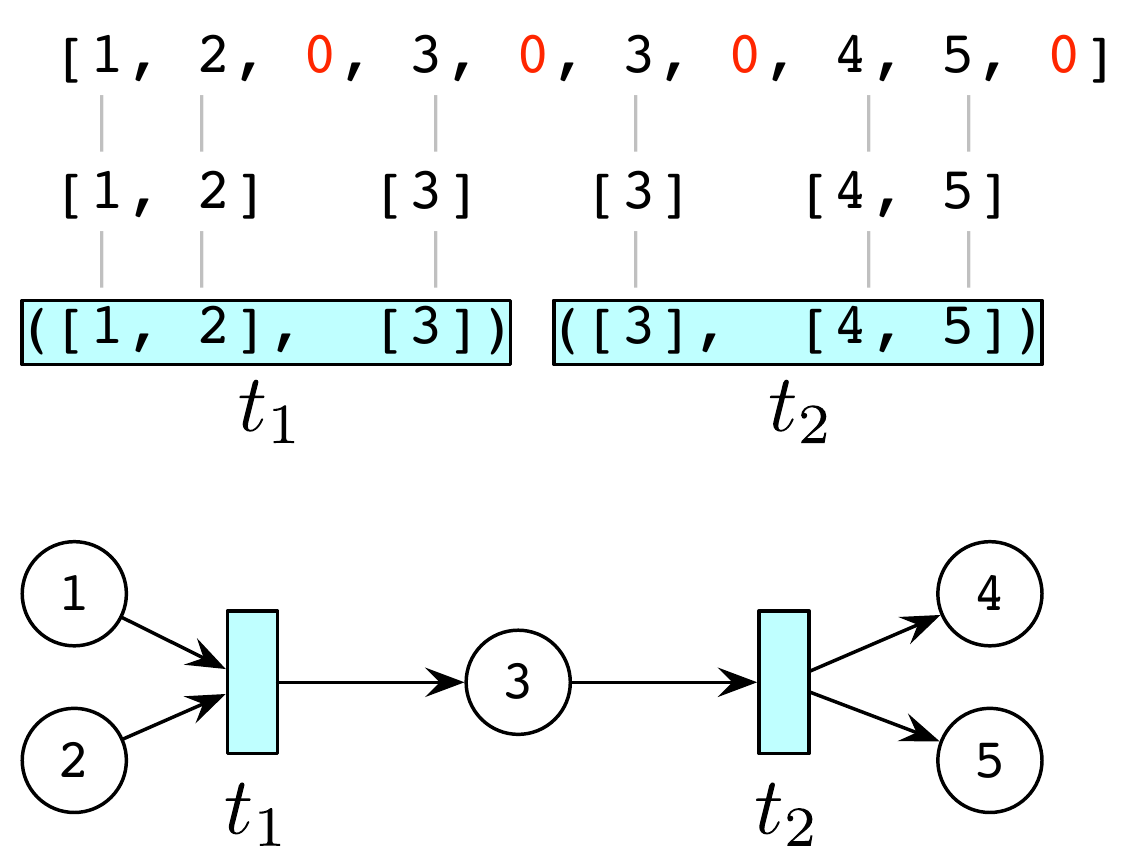}
  \caption{A way to convert a string to a Petri net, and vice-versa.}
  \label{fig: string to net}
\end{figure}
\section{Implementation}\label{sec: Petri nets implementation}
At the moment, we are implementing Petri nets on different parts of our stack, from 
frontend to core, using a plethora of different languages. We are also building 
parsers that allow us to import Petri nets designed in widespread editors such as 
GreatSPN~\cite{UniversityofTorino2018}. 
This is important since it allows users to design nets in the 
editor they like the most, and also to use whatever model checking features 
such editors provide. 

Since Petri nets are a bit all over the place in our 
codebase, and many of the repositories where we are carrying this work 
are yet to be made open, it is probably better to focus on how we manage 
to ``pass Petri nets around'' between different components of our 
stack. We obtain this by means of \emph{serializing/deserializing a net}.

With serializing/deserializing a Petri net 
we mean that we need a way to pass around the information needed 
to define a Petri net between machines, and to do so we need a procedure 
to convert this information into an actual net and vice-versa. Considering 
the very nature of computer networking, this means that we need a way 
to convert a net to a \emph{string}, and back. This is shown in 
Figure~\ref{fig: string to net}.

The procedure is quite self-explanatory, but we will try 
to comment on that nevertheless: We start with a string of numbers,
where $0$ is treated as a special character. Scanning the string, we chop 
it every time we encounter a zero. What we are left with now is a bunch 
of substrings, which we sequentially group into couples. Each of these 
couples defines input and output of a transition, and as we see this is 
enough information to build a Petri net.

With this procedure we are able to convert Petri nets to strings and 
viceversa, and exchange them between components such as, say, 
the frontend codebase displaying the net to the user and the core 
codebase dealing with processing net firings in a formally 
consistent way.  As we will see in 
Section~\ref{sec: lack of functoriality is not the end of the world}, 
this exchange format has the advantage of being able to exchange 
not only nets, but the categories defining their histories, which 
we will introduce in Chapter~\ref{ch: executions}.
\section{Why is this useful?}\label{sec: why is this useful - Petri}
This is a perfectly legitimate question, that deserves a 
prompt answer. We will proceed by analytically listing 
the ways in which Petri nets can be useful for software-design 
purposes. We hope to give the reader reason to believe 
that Petri nets are, in fact, a good formalism to base a 
programming language on. In introducing Petri nets, we 
pointed out the following characteristics that make them 
very appetible for software-design purposes:
\begin{itemize}
	\item Petri nets are inherently graphical. Since the very 
	beginning, we were able to introduce and manipulate Petri 
	nets diagrammatically. The pictorial representation of Petri 
	nets is intuitive, and allows us to quickly draft how a 
	complex system is supposed to work. This makes designing 
	infrastructure with Petri nets much easier than by, say, using 
	traditional code;
	\item Petri nets represent concurrency well. The idea of 
	transitions having to ``fight'' for resources is very useful in 
	representing processes that could be run independently on 
	those resources. Again, this can be easily represented 
	graphically, giving us a neat, intuitive explanation of what is 
	going on. Such a feature is of great value in modeling 
	complex systems, often consisting of multiple, independent 
	parties performing concurrent operations on different machines;
	\item Petri nets can be studied formally. The graphical 
	formalism we rely on to model systems is backed up by a 
	sound mathematical model. This guarantees that our 
	drawings are not just drawings, but that 
	computers can ``understand'' our drawings by means 
	of the corresponding mathematics;
	\item Interesting properties of nets can be expressed in 
	terms of reachability. Since reachability is formally 
	defined, we can develop technical tools to infer if a 
	given condition holds or not for a net. This, in particular, 
	means that we can ask a computer to answer such questions 
	for us. If the possibility of algorithmically deciding if 
	some property holds or not for a given net may not seem 
	very important, it is because all the examples provided 
	up to now consisted of very small nets. The reader 
	should be aware that in production applications, Petri 
	nets can easily have many hundreds of places/transitions, 
	and answering reachability questions without the aid of a 
	computer is basically impossible. Clearly, up to now we 
	do not know how efficient algorithms can be 
	in solving such problems, and indeed verifying some 
	properties can take an exponential time (or even worse) in 
	the size of the net. This means that as our net grows in size 
	the time needed to know if some property holds or not for 
	it will increase exponentially. This prompts for the development 
	of efficient methods, such that when an efficient solution to 
	answer a question exists, it is attained.
\end{itemize}
We decided to use Petri nets as the language 
that Statebox uses to design code at the highest level of abstraction. 
More precisely, the programmer will be able to use Petri nets to 
draft how the software should behave by modeling it as a process, 
and then dive into details by filling in all the remaining information 
by means of a well defined procedure, backed up by sound 
mathematics to ensure consistency of such method. We call 
this way of writing code \emph{behavioral programming}. A 
tutorial about how to employ this technique to write programs can 
be found in~\cite{Genovese2018a}.

As this is the direction we want to take, we need a way to recast 
the Petri nets formalism in a way that makes it compatible with 
other mathematical gadgets we want to use for the ``filling the blanks'' 
stage we mentioned above. This will be done with the aid 
of \emph{category theory}, that we will introduce in the next Chapter.

\newtoggle{toggleCategoryTheory}\toggletrue{toggleCategoryTheory}
  \chapter{Introduction to category theory}
  \label{ch: introduction to category theory}
In Chapter~\ref{ch: Petri nets} we introduced Petri nets, 
and definined some of their properties. Now we proceed 
by introducing the other main actor in the Statebox project, 
category theory. Category theory is a relatively young 
branch of mathematics that originated during the second 
half of the last century~\cite{Eilenberg1945}, and since 
then it has had an increasingly pervasive influence in the 
way modern mathematicians and computer scientists think. 
Category theory can be seen as ``the glue of mathematics'' 
and has the marvelous ability of making different 
theories interact consistently with each other. Set theory 
is also a universal language for mathematics, with the 
difference that while sets focus on defining a structure 
``imperatively'' -- e.g. by specifying which properties the 
elements of a structure need to satisfy -- category theory 
defines mathematical structures behaviorally, that is, by 
specifying patterns and interactions of a structure with 
structures of similar kind. In this sense, it is clear how 
working from a categorical perspective makes studying 
the interaction of different theories easier.

Since one of the main characteristics of the Statebox 
project is unification of advancements in very different 
fields of computer science, the reader can already appreciate 
why category theory will end up being very relevant for 
us. Indeed, the standard modus operandi of this document 
will most often reduce to the following pattern:
\begin{itemize}
  \item Introduce a new idea;
	\item Find a mathematical theory that captures the idea well;
	\item Categorify it, that is, translate it to the language 
	of category theory;
	\item Study how what we obtained interacts with what 
	we already had. One of the main advantages of category 
	theory is that its extensive toolbox makes this step much easier.
\end{itemize}
Albeit having just sketched out why category theory will 
have a central role in the development of our theoretical 
framework, we already have what we need to introduce it 
in all of its glory. The 
reader should employ this Chapter as a reference, and not 
worry too much if something explained here initially does not 
seem very ``useful''. Eventually, every detail will find its 
place in the environment we are building up.
\section{What is category theory?}
  \label{sec: what is category theory}
That is a great question -- in many ways the answer 
deepens every day. Category theory is primarily a 
way of thinking, more than just a theory in the usual 
sense of the term. Probably the simplest idea of category 
theory is that everything is interrelated. This applies 
not only to mathematics, but also computation, physics, 
and other sciences which are just beginning to be elucidated 
and unified via the use of categories -- and is precisely the 
reason why category theory has such natural real-world 
applications. Of course the pertinent application here is in 
the context of computer science, and the mission of 
Statebox is to make programming concrete, principled, 
and universal. First, we begin with a simple mathematical 
overview of category theory.

According to~\cite{MacLane1986}, a nice way to describe 
category theory is as \emph{the language for describing 
and observing patterns in mathematics}. Every \emph{object} 
is of a certain kind, which is interrelated by a \emph{morphism} 
intrinsic to the kind. For example, a morphism of 
\emph{sets} is simply a \emph{function}, while a 
morphism of structured objects cooperates with the 
structure, e.g. an algebraic operation. Taken together, the 
objects and morphisms form a \emph{category}, which 
encapsulates the particular notion, and more so, connects 
it to all of mathematics -- the category is \emph{itself} a 
kind of object, and we can consider the category of 
categories! The morphisms between categories, called 
\emph{functors}, respect the \emph{composition} of 
morphisms in the related categories, providing a fundamental 
connection between distinct concepts. A functor witnesses 
how the reasoning patterns found in a certain theory are 
``compatible'' with the patterns found in another. This entails 
a well-behaved notion of compatibility between different 
theories, an essential aspect of principled theoretical modeling 
called \emph{compositionality}. In a way, this perspective 
already empowers us when thinking about mathematics as 
a whole. But let us slow down and see the basic definitions.
\begin{definition}[Category]
	A \emph{category} $\CategoryC$ consists of
	\begin{itemize}
		\item A collection of \emph{objects}, denoted as 
		$\Obj{\CategoryC}$;
		\item A collection of \emph{morphisms}, denoted 
		as $\Homtotal{\CategoryC}$;
		\item Two functions 
		$\Source{-},\Target{-}:
			\Homtotal{\CategoryC} \to \Obj{\CategoryC}$ 
		called \emph{source} (or \emph{domain}) and 
		\emph{target} (or \emph{codomain}), respectively;
		\item A partial function 
		$(-);(-): \Homtotal{\CategoryC} \times 
			\Homtotal{\CategoryC} \to \Homtotal{\CategoryC}$, 
		called \emph{composition}, that assigns to every 
		pair $f,g \in \Homtotal{\CategoryC}$, such 
		that $s(g) = t(f)$, the arrow $f;g$;
		\item An \emph{identity} function 
		$\Obj{\CategoryC} \to \Homtotal{\CategoryC}$, that 
		assigns to every object $A$ an arrow $\Id{A}$.
	\end{itemize}
	Moreover, we require that the following 
	axioms have to be satisfied:
	\begin{itemize}
		\item $\Source{f} = \Source{f;g}$ and 
		$\Target{f;g} = \Target{g}$;
		\item $\Source{\Id{A}} = A = \Target{\Id{A}}$;
		\item $f ; (g ; h) = (f ; g) ; h$ for each $f,g,h$ arrows 
		such that composition is defined;
		\item $f ; \Id{\Target{f}} = f = \Id{\Source{f}} ; f$ 
		for each arrow $f$.
	\end{itemize}
	An arrow $f$ such that $\Source{f}=A, \Target{f}=B$ is
	 often denoted with $f:A \to B$ or $A \xrightarrow{f} B$.
\end{definition}\label{def: category}
The concept of category is a very powerful one, 
and we redirect the reader who wants to know 
more to~\cite{MacLane1978}: Category theory 
can indeed become very difficult to grasp only with 
the introduction of its simplest concepts, and this 
document is not the right place for an in-dept exposition.
Nevertheless, it is worth to give an intuitive explanation 
of the definition provided above: \emph{Objects} can be 
thought of as representing systems, resources, or states 
of a machine. \emph{Arrows} represent transformations 
between them, that is, processes that turn a given 
system (or resource, or state) into another according 
to some rules. Moreover, \emph{composition} tells 
us that transformations can be serialized: 
Transforming $A$ into $B$ using $f$ and 
then $B$ into $C$ using $g$ is the same as 
transforming $A$ into $C$ using $f;g$. The axioms 
tell us that composing transformations is \emph{associative}, 
and moreover that for each system $A$ ``doing nothing'' can 
be regarded as an \emph{identity transformation} $\Id{A}$.
\begin{remark}
In interpreting objects as states of a system and morphisms 
as transformations between them, we already see some 
similarity with the interpretation we gave of Petri nets in 
Chapter~\ref{ch: Petri nets}. This similarity will be 
described in depth in Chapter~\ref{ch: executions}.
\end{remark}
\begin{example}[Sets and functions]
  \label{ex: category of sets and functions}
	There is a category, denoted with $\Set$, whose 
	objects are sets and whose morphisms are functions 
	between them. It is easy to see that composition of 
	functions is a function, composition is associative, 
	and that every set has an identity function carrying 
	every element into itself. Hence $\Set$ is indeed a 
	well-defined category.
\end{example}
\begin{remark}[Notation]
	From now on, we will stick to the convention of 
	indicating generic categories with curly letters, 
	like $\CategoryC, \CategoryD, \CategoryE$. Objects 
	will be denoted with capital Latin letters, preferably 
	from the beginning of the alphabet, $A, B, C$ etc. 
	Morphisms will be denoted with lower-case Latin letters, 
	preferably from the middle of the alphabet, $f,g,h$ etc. 
	Categories that deserve a name of their own, like the 
	one in Example~\ref{ex: category of sets and functions}, 
	will have the name denoted in bold letters, as in $\Set$.
\end{remark}
\begin{examplehard}[Functional programming]
  \label{rem: Haskell category}
	We can build a category $\Hask$ where objects 
	are data types and morphisms are Haskell~\cite{HaskellWiki} 
	functions from one type to another. Associativity is 
	composition of functions, and identity morphisms 
	are the algorithms sending terms to themselves. 
	Defining the category $\Hask$ is actually not as 
	easy as it seems and we will discuss more about this 
	issue in Remark~\ref{rem: Haskell not a category}.
\end{examplehard}
\begin{examplehard}[Groups, topological spaces]
	Groups and homomorphisms between them form 
	a category, called $\Group$. So do topological 
	spaces and continuous functions, forming the category $\Top$.
\end{examplehard}
\begin{remarkhard}[Free categories from graphs]
	There is an evident connection between the 
	definition of a category and the definition of 
	a graph. A category just looks like ``the transitive 
	closure of a graph, with loops added at every 
	vertex''. This connection between categories and 
	graphs is indeed real, and one can always generate 
	a free category from a directed 
	graph~\cite[Ch. 2, Sec. 7]{MacLane1978}.
\end{remarkhard}
\begin{remarkhard}[Size issues]
	The reader with experience in mathematics will 
	have noted how we have been vague in saying 
	what we mean by ``a collection of objects'' in the 
	definition of a category. Indeed, note how the objects 
	of the category $\Set$, $\Group$ and $\Top$ do not 
	form a set, but a \emph{proper class}. All these 
	\emph{size issues} are deeply covered in any 
	comprehensive book about category theory, and 
	we refer the reader to~\cite[Ch. 1, Sec. 6]{MacLane1978} 
	for details.
\end{remarkhard}
\begin{figure}[!ht]
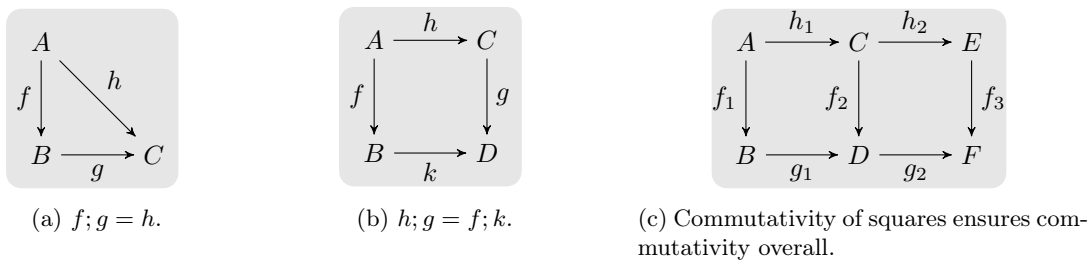

	\centering
	\begin{subfigure}[t]{0.24\textwidth}\centering
		\input{Pictures/CategoryTheoryIntroduction/00-CommutativeTriangle.tex}
		\caption{$f;g = h$.}
		\label{fig: commutative triangle example}
	\end{subfigure}
	\hfill
	\begin{subfigure}[t]{0.24\textwidth}\centering
		\input{Pictures/CategoryTheoryIntroduction/00-CommutativeSquare.tex}
		\caption{$h;g = f;k$.}
		\label{fig: commutative square example}
	\end{subfigure}
	\hfill
	\begin{subfigure}[t]{0.4\textwidth}\centering
		\input{Pictures/CategoryTheoryIntroduction/00-CommutativeRectangle.tex}
		\caption{Commutativity of squares ensures commutativity overall.}
		\label{fig: commutative rectangle example}
	\end{subfigure}
  \caption{Examples of commutative diagrams.}
  \label{fig:examples of commutative diagrams}
\end{figure}
\begin{remark}[Commutative diagram]
  \label{rem: commutative diagram}
	A neat way to express equations between morphisms 
	in a category is via \emph{commutative diagrams}. 
	A commutative diagram is just a picture that shows 
	us how morphisms compose with each other. 
	Commutative diagrams are interpreted as follows: 
	Vertexes are objects in a category. Paths between 
	objects are compositions of morphisms. If there are 
	multiple paths from one object to another, this means 
	that the corresponding morphisms are equal. For 
	example, the diagram in 
	Figure~\ref{fig: commutative triangle example} states 
	that  $f;g = h$, while the diagram in 
	Figure~\ref{fig: commutative square example} states 
	that $h;g = f;k$.
\end{remark}
Commutative diagrams are a fundamental tool in category 
theory, and are routinely used to prove things. The standard 
way to prove something in category theory is to draw a 
diagram representing our thesis, and then try to prove 
that the diagram commutes. A way to do this is by 
dividing the diagram into multiple sub-diagrams and 
proving the commutativity of each of them separately. 
The commutativity of the overall diagram can then be 
inferred by the commutativity of its components. To see 
how this works, consider 
Figure~\ref{fig: commutative rectangle example}: If we 
know that the left and right squares commute, then so 
does the rectangle obtained from their composition, in fact:
\begin{equation*}
  f_1;g_1;g_2 = (f_1;g_1);g_2 = (h_1;f_2);g_2 =
     h_1;(f_2;g_2) = h_1;(h_2;f_3) = h_1;h_2;f_3
\end{equation*}
where the second equality follows from the commutativity 
of the left square, while the fourth follows from the 
commutativity of the right one.

To conclude this Section, we introduce the concept 
of \emph{isomorphism}. Intuitively, an isomorphism 
in a category is a morphism that allows us ``to go back 
and forth between two objects''. This is easily defined as follows:
\begin{definition}[Isomorphism]\label{def: isomorphsims}
  Given a category $\CategoryC$ we say that a 
  morphism of $\CategoryC$ $f:A \to B$ is an 
  \emph{isomorphism} (or just an \emph{iso}) if 
  there is a morphism $f^{-1} : B \to A$ such that
	\begin{equation*}
		f;f^{-1} = \Id{A} \qquad f^{-1};f = \Id{B}
	\end{equation*}
\end{definition}
The definition of isomorphism is nothing new, and 
captures the idea of a ``reversible process''. We already 
know examples of this:
\begin{example}[Isos in $\Set$]
	In $\Set$, the isomorphisms are exactly the bijective functions.
\end{example}
\begin{examplehard}[Isos in $\Group$ and $\Top$]
  In $\Group$, the isomorphisms are exactly the bijective 
  homomorphisms. In $\Top$, the isomorphisms are exactly 
  the homeomorphisms.
\end{examplehard}
\section{Functors, natural transformations, natural isomorphisms}
  \label{sec: functors}
We mentioned functors en passant in the introduction 
of this Chapter, when we said that \emph{a functor is 
a morphism between categories}. We moreover added 
that a morphism in a category can be thought of as a 
transformation that preserves \emph{all the relevant structure} 
from its domain to its codomain. So, if a functor is a morphism 
of categories, which is the relevant structure it has to preserve?

Well, in a general category the only things that we have are 
identities for each object and composition of morphisms, so 
it seems reasonable to require these to be preserved by a 
functor. This is, indeed, enough:
\begin{definition}[Functor]
  A \emph{functor} $F$ from a category $\CategoryC$ to a 
  category $\CategoryD$, often denoted with 
  $F:\CategoryC \to \CategoryD$ or 
  $\CategoryC \xrightarrow{F} \CategoryD$, consists of the 
  following:
	\begin{itemize}
    \item A map from $\Obj{\CategoryC}$ to $\Obj{\CategoryD}$, 
    that associates to the object $A$ of $\CategoryC$ the 
    object $FA$ of $\CategoryD$;
    \item A map from $\Homtotal{\CategoryC}$ to 
    $\Homtotal{\CategoryD}$, that associates to the 
    morphism $f:A \to B$ of $\CategoryC$ the morphism 
    $Ff:FA \to FB$ of $\CategoryD$.
		\item We moreover require that the following equalities hold:
		\begin{equation}\label{eq: functor condition}
			F_{\Id{A}} = \Id{FA} \qquad F(f;g) = Ff ; Fg
		\end{equation}
	\end{itemize}
\end{definition}
In particular, Equations~\ref{eq: functor condition} mean 
that identities get carried to identities and compositions 
to compositions, as we would have expected. Note how 
\emph{this is enough to guarantee that $F$ sends any 
commutative diagram in $\CategoryC$ to a commutative 
diagram in $\CategoryD$}. This is the whole point about 
functors: If the main way to prove facts in category theory 
is by using commutative diagrams, a functor is basically 
sending facts about $\CategoryC$ to facts about $\CategoryD$. 
This allows us to ``export'' theorems from one category to 
another, and is a tremendously powerful feature to carry 
results across mathematical theories.
\begin{example}[Identity functor]\label{rem: identity functor}
  For each category $\CategoryC$ there is a functor 
  $\Id{\CategoryC}$ that sends each object and each 
  morphism of $\CategoryC$ to itself, respectively.
\end{example}
\begin{examplehard}[Homotopy groups]
  There is a functor from the category of pointed topological 
  spaces and homotopy classes of continuous functions, 
  $\hTop\star$, to the category $\Group$. This is exactly 
  what makes it possible to deduce if a given topological 
  space is connected or not -- studying its homotopy group.
\end{examplehard}
\begin{remark}[Functor composition]
  \label{rem: functor composition}
  Given two functors $\CategoryC \xrightarrow{F} 
  \CategoryD \xrightarrow{G} \CategoryE$ we can compose 
  them by composing their maps on objects and morphisms. 
  The composition sends an object $A$ of $\CategoryC$ to 
  an object $FGA$ of $\CategoryE$, and a morphism 
  $f:A \to B$ in $\CategoryC$ to $FGf: FGA \to FGB$ 
  in $\CategoryE$.
\end{remark}
\begin{remark}[Notation]
  It is commonplace to denote functors using capital Latin 
  letters from the middle of the alphabet, $F, G, H$ etc. 
  Also, the application of a functor to an object or a morphism 
  is usually written without using parentheses, as in $FA, Ff$.
\end{remark}
We can now start playing with the definition of functor a bit more. 
First, something simple:
\begin{definition}[Isomorphism of categories]
  Using Remarks~\ref{rem: identity functor} 
  and~\ref{rem: functor composition} it is not difficult to 
  convince ourselves that categories and functors form the 
  objects and morphisms, respectively, of a category, 
  called $\Cat$. Then we can apply 
  Defintion~\ref{def: isomorphsims} in this context and obtain 
  that the two categories $\CategoryC$ and $\CategoryD$ are 
  \emph{isomorphic} when there are functors 
  $F:\CategoryC \to \CategoryD$ and 
  $F^{-1}:\CategoryD \to \CategoryC$ such that 
  $F;F^{-1} = \Id{\CategoryC}$ and 
  $F^{-1};F = \Id{\CategoryD}$.
\end{definition}
The definition of isomorphism between categories is not really 
the interesting one for us. This is because it is too restrictive. 
However, we can relax it a little to make it more manageable. 
To do this, we first need to introduce some properties.
\begin{definition}[Full and faithful functors]
  \label{def: full and faithful functors}
  A functor $F: \CategoryC \to \CategoryD$ is called 
  \emph{full} if, for any objects $A, B$ in $\CategoryC$ and 
  any morphism $f: FA \to FB$ in $\CategoryD$, there is 
  always a morphism $g$ in $\CategoryC$ such that $Fg = f$.

  On the other hand, $F$ is called \emph{faithful} if given 
  morphisms $f,g:A \to B$ in $\CategoryC$, $f \neq g$ implies 
  $Ff \neq Fg$ in $\CategoryD$.

  When a functor is full and faithful, we sometimes say that it 
  is \emph{fully faithful}.
\end{definition}
In essence, a functor $F: \CategoryC \to \CategoryD$ is full 
when every morphism between objects of the form 
$FA$, $FB$ -- that is, objects that are hit by $F$ -- comes
from $\CategoryC$. This means that the morphisms 
$A \to B$ are \emph{at least as many} as the morphisms 
$FA \to FB$. Similarly, faithfulness implies that the morphisms 
$FA \to FB$ are \emph{at least as many} as the ones $A \to B$, 
since different morphisms from $A$ to $B$ go to different 
morphisms from $FA$ to $FB$. When a functor is fully faithful, 
then, all the morphisms between objects of $\CategoryC$ are 
carried to $\CategoryD$ exactly as they are, and all the objects 
of the form $FA$ for some $A$ in $\CategoryC$, together with 
their morphisms, form ``a copy'' of $\CategoryC$ in $\CategoryD$.

This is pretty close to an equivalence of categories, 
but in $\CategoryD$ there could be other objects that are 
not hit by $F$, viz. objects that cannot be written as $FA$ for 
some $A$ in $\CategoryC$. Since these objects are not hit by 
$F$ they could behave as they want to, while the structure of 
objects of type $FA$ and their morphisms is completely 
determined by $\CategoryC$ and the full faithfulness of $F$. 
To rule out this eventuality, we give the following definition:
\begin{definition}[Equivalence of categories]
  \label{def: equivalence of categories}
  Two categories $\CategoryC$ and $\CategoryD$ are said to be 
  \emph{equivalent} when there is a functor 
  $F:\CategoryC \to \CategoryD$ that is fully faithful and 
  \emph{essentially surjective}, meaning that each object 
  of $\CategoryD$ is isomorphic to an object of the form $FA$ 
  for some $A$ in $\CategoryC$.
\end{definition}
Now we see that Definition~\ref{def: equivalence of categories} 
is the right one to describe categories that are, structurally 
speaking, the same: All the objects and morphisms in 
$\CategoryD$ are forced to behave like objects and morphisms 
in $\CategoryC$, since either they are hit by the functor $F$, and 
then are taken care of by the full faithfulness of $F$, or they are 
not, in which case they are isomorphic to some object which is. 
Equivalence of categories will have a big role in 
Chapter~\ref{ch: executions}, and we postpone any meaningful 
example until then.

\noindent
Now that functors have been introduced, it is legitimate to ask 
if there is a notion of ``morphism between functors'': Suppose 
we have categories $\CategoryC$ and $\CategoryD$, and 
functors $F,G:\CategoryC \to \CategoryD$. We know that 
if $\mathfrak{D}$ stands for a commutative diagram in 
$\CategoryC$, the functors $F, G$ carry it to a couple of 
commutative diagrams in $\CategoryD$. The question, then, is: 
Is it possible to establish a relationship between the diagrams 
in $\CategoryD$ to which $\mathfrak{D}$ is carried to by $F$ 
and $G$, respectively? The answer to this question is \emph{yes}, 
and the notion we are looking for is called a \emph{natural 
transformation}.
\begin{definition}[Natural transformation]
  \label{def: natural transformation}
  Given functors $F,G: \CategoryC \to \CategoryD$, a 
  \emph{natural transformation from $F$ to $G$}, denoted 
  with $\eta:F \to G$, consists of a collection of morpshisms 
  of $\CategoryD$
	\begin{equation*}
		\{\eta_A:FA \to GA\}_{A \in \Obj{\CategoryC}}
	\end{equation*}
  such that, for every morphism of $\CategoryC$ $f:A \to B$, the 
  diagram in Figure~\ref{fig: natural transformation} commutes.
\end{definition}
\begin{figure}[!ht]
	\centering
	\input{Pictures/CategoryTheoryIntroduction/01-NaturalTransformationDiagram.tex}
	\caption{Commutativity for a natural transformation.}
	\label{fig: natural transformation}
\end{figure}
Definition~\ref{def: natural transformation} is slightly tricky. 
The morphisms defining $\eta$, also called its 
\emph{components}, live in $\CategoryD$, but are 
indexed by objects of $\CategoryC$. This is for the 
following reason: We want to find a procedure to turn 
every diagram where each vertex and edge is an application 
of $F$  to an object or morphism of $\CategoryC$, respectively, 
into a diagram where each vertex and edge is an application 
of $G$ to the same object or morphism of $\CategoryC$. In 
practice, this means looking for a rewriting procedure that 
strips all the occurences of $F$ from the diagram and replaces 
them with $G$. To do this, what we need to do is establish a 
correspondence between vertexes, that is, a correspondence 
$FA \to GA$ for each object $A$. Since $FA, GA$ are objects 
of $\CategoryD$, this correspondence will have to be a 
morphism of $\CategoryD$. Clearly, we need as many of these 
correspondences as there are $FA$ and $GA$, so one for each 
object $A$ of $\CategoryC$. Moreover, it is easy to prove that 
the commutativity of the square in 
Definition~\ref{def: natural transformation} is everything we 
need so that the correspondence between the diagrams does not 
break their commutativity.

Finally, we can combine 
Definitions~\ref{def: natural transformation} 
and~\ref{def: isomorphsims} to capture the concept of 
``going back and forth between diagrams only made of 
applications of $F$ and diagrams made only by applications 
of $G$'', as follows:
\begin{definition}[Natural isomorphism]
  Given functors $F,G:\CategoryC \to \CategoryD$, a 
  natural transformation $\eta:F \to G$ is called a 
  \emph{natural isomorphism} if each component of $\eta$ is 
  an isomorphism in $\CategoryD$.
\end{definition}
We see that the concept of a natural isomorphism is a 
very strong one. It consists of a number of 
isomorphisms $\eta_A: FA \to GA$ which are 
\emph{consistently connected with each other}. 
Moreover, it is easy to say that if $\eta: F \to G$ is a 
natural isomorphism, then there is a natural 
transformation $G \to F$ defined by taking the 
inverse of each $\eta_A$, and that these two natural 
transformations are each other's inverses. The concept 
of natural isomorphism is very useful to express all sorts 
of ``coherence conditions'' which are the categorical tools 
capturing the idea of ``it does not matter in which way you 
stack up these commutative diagrams, the result will be the 
same''. We will see an example of this in the next Section.
\begin{remark}[Notation]
  It is commonplace to denote natural transformations 
  using Greek letters, $\eta, \mu, \tau$ etc. The component 
  of natural transformation $\eta$ on object $A$ is usually 
  denoted with subscripts, $\eta_A, \mu_A, \tau_A$ etc.
\end{remark}
\section{Monoidal categories}\label{sec: monoidal categories}
\begin{figure}[!ht]
	\centering
	\begin{subfigure}[t]{1\textwidth}\centering
		\input{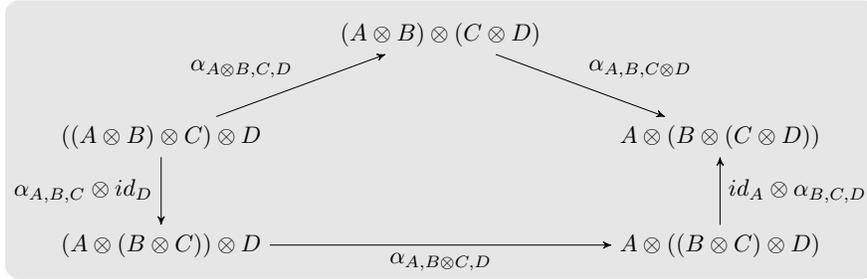}
		\caption{Coherence condition for the associator.}
		\label{fig: coherence condition pentagon}
	\end{subfigure}
	\hfill
	\begin{subfigure}[t]{1\textwidth}\centering
		\input{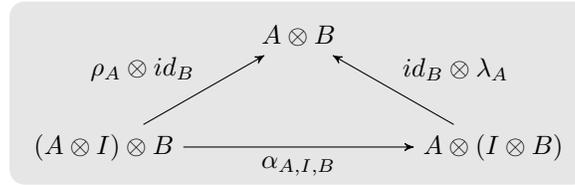}
		\caption{Coherence condition for unitors.}
		\label{fig: coherence condition triangle}
	\end{subfigure}
  \caption{Coherence conditions for monoidal categories.}
  \label{fig: coherence conditions for mc}
\end{figure}

Up to now, we have only had one operation between 
morphisms in a category, composition. Composition 
has a very clear time-like interpretation, especially if 
we interpret objects as states of a system, and morphisms 
between them as processes. In fact, we can clearly 
read $f;g$ as ``apply $f$ and \emph{then} apply $g$''. 
The question, then, is if there is a categorical notion that 
captures the idea of ``things happening in parallel''. The 
answer to this question is positive, and is provided by the 
following definition.
\begin{definition}[Monoidal category]
  \label{def: monoidal category}
  A \emph{monoidal structure} for a category $\CategoryC$ 
  consists of:
	\begin{itemize}
    \item A functor 
    $\Tensor: \CategoryC \times \CategoryC \to \CategoryC$, 
    called the \emph{monoidal product} or, sometimes, 
    the \emph{tensor product} (because traditionally the 
    symbol used to denote it, $\Tensor$, denotes tensor 
    products in linear algebra).
    Note that in this case $\CategoryC \times \CategoryC$ is 
    a product of categories, which will be formally introduced 
    in Definition~\ref{def: products} and 
    Example~\ref{ex: products of categories}. 
    Intuitively, the functor $\Tensor$ can be interpreted 
    as having two arguments: It associates an object 
    (a morphism, respectively ) of $\CategoryC$ to each 
    couple of objects (morphisms, respectively) 
    of $\CategoryC$ such that the functor laws hold for 
    both components:
    \begin{equation*}
      \Id{A} \Tensor \Id{B} = \Id{A \Tensor B} 
      \qquad (f;g) \Tensor (h;k) = (f \Tensor h) ; (g \Tensor k)
    \end{equation*}
    \item A selected object $\TensorUnit$ of $\CategoryC$, 
    called the \emph{monoidal unit};
		\item A natural isomorphism
		\begin{equation*}
      \alpha:((-) \Tensor (-)) \Tensor (-) 
        \xrightarrow{\simeq} (-) \Tensor ((-) \Tensor (-))
		\end{equation*}
		called \emph{associator}, with components in the form
		\begin{equation*}
      \alpha_{A,B,C}: (A \Tensor B) \Tensor 
        C \xrightarrow{\simeq} A \Tensor (B \Tensor C)
		\end{equation*}
    that expresses the fact that the tensor operation is associative;
		\item Natural isomorphisms
		\begin{equation*}
      \lambda: \TensorUnit \Tensor (-) 
        \xrightarrow{\simeq} (-) \qquad 
        \rho (-) \Tensor \TensorUnit \xrightarrow{\simeq} (-)
		\end{equation*}
    called \emph{left and right unitors}, respectively, 
    with components in the form:
		\begin{equation*}
      \lambda_A: \TensorUnit \Tensor A 
      \xrightarrow{\simeq} A \qquad \rho_A: A \Tensor 
      \TensorUnit \xrightarrow{\simeq} A
		\end{equation*}
    that express the fact that $\TensorUnit$ behaves as a unit;
    \item These natural isomorphisms have to respect 
    the so called \emph{coherence conditions}, that imply 
    that associator and unitors are well behaved, and can 
    thus be used in full generality. Coherence conditions are 
    expressed in the form of \emph{commutative diagrams}, 
    as in Figure~\ref{fig: coherence conditions for mc}.
	\end{itemize}
  A category $\CategoryC$, together with a monoidal structure, 
  is called a \emph{monoidal category}.
\end{definition}
Let us try to make this definition more explicit: 
The monoidal product $\Tensor$ captures the idea of 
parallel composition. $A \Tensor B$ represents two 
systems existing at the same time. $f \Tensor g$ represents 
two processes $f,g$ being applied at the same time on different 
systems.

The associator captures the idea that the monoidal product 
is associative: We can always go from 
$(A \Tensor B) \Tensor C$ to $A \Tensor (B \Tensor C)$ 
and vice-versa without destroying any fact proven by 
commutative diagrams (that is why we need associators 
to be \emph{natural} isomorphisms!).

The monoidal unit represents the \emph{trivial system}. 
This behavior is enforced by left and right unitors, which  
tell us that we can go from $A \Tensor \TensorUnit$ to $A$ 
to $\TensorUnit \Tensor A$ in any way we want, without 
losing information. We can deduce that the system 
$\TensorUnit$ does not add or remove any information 
when composed with $A$.

Coherence conditions require a few more words. They are 
what make associators behave like associators and unitors 
behave like unitors, and are expressed by two commutative 
diagrams. These two diagrams are very important, because 
it can be proved (see~\cite[Ch. 7, Sec.2]{MacLane1978}) that when they commute 
\emph{any other diagram} made uniquely of associators, 
monoidal products and unitors commutes, effectively meaning 
that adding $I$ to a monoidal product or changing the bracketing 
in any possible way does not change anything, as we would 
expect.
\begin{remarkhard}[Monoidal categories as higher categories]
  The reader versed in higher category theory can equivalently see a
  monoidal category as a bicategory with one 0-cell. 1-cells represent
  the objects of the monoidal category, with 1-cell composition as monoidal
  product. The identity on the unique 0-cell stands for the monoidal unit.
  2-cells represent the morphisms of the monoidal category. Vertical 
  composition of 2-cells represents the usual morphism composition, while
  horizontal composition of 2-cells represents the monoidal product on 
  morphisms. Coherence conditions follow directly from the coherence 
  conditions of horizontal and vertical composition of 2-cells.
\end{remarkhard}
\begin{remark}[Notation]
  When we want to make explicit that $\CategoryC$ is a 
  monoidal category, we use the notation 
  $(\CategoryC,\Tensor, \TensorUnit)$, where $\TensorUnit$ 
  represents the monoidal unit and $\Tensor$, the monoidal 
  product. For instance, if we say that 
  $(\CategoryC,\Tensor,\TensorUnit)$ and 
  $(\CategoryD, \square, \TensorUnit')$ are monoidal categories, 
  we are denoting the tensor product as $\Tensor$ in 
  $\CategoryC$ and as $\square$ in $\CategoryD$, and their 
  monoidal units as $\TensorUnit, \TensorUnit'$, respectively.
\end{remark}
\begin{example}[Products of sets]\label{ex: set product}
  The category $\Set$ can be made into a monoidal category 
  $(\Set, \times, \{\star\})$, where $\times$ is the 
  cartesian product of sets and $\{ \star \}$ is the one element set. 
  The associator is the usual rebracketing for tuples, while 
  unitors are the isomorphisms sending both $(a, \star)$ and 
  $(\star, a)$ to $a$.
\end{example}
\begin{example}[Coproducts of sets]\label{ex: set coproduct}
  The category $\Set$ admits another monoidal structure, 
  and can thus also be turned into a monoidal category 
  $(\Set, \sqcup, \emptyset)$, where $\sqcup$ denotes 
  the disjoint union of sets and $\emptyset$ denotes the usual 
  empty set. The associator is the usual rebracketing of disjoint 
  unions, while unitors are the identities expressing the fact 
  that taking the disjoint union of a set $A$ with the empty 
  set gives back $A$.
\end{example}
\begin{remark}[Monoidal structures are not unique]
  Examples~\ref{ex: set product} and~\ref{ex: set coproduct} 
  show that the category $\Set$ admits two different monoidal 
  structures. It is very easy to see that 
  $(\Set, \times, \{\star\})$ and 
  $(\Set, \sqcup, \emptyset)$ are different monoidal 
  categories, since in general $A \times B \neq A \sqcup B$. 
  This proves that it is often incorrect to refer to a category as 
  monoidal without explicitly stating what the monoidal 
  structure is, unless it is clear from the context. If we say 
  that $\Set$ is a monoidal category, to which monoidal 
  structure are we referring to?
\end{remark}
\begin{examplehard}[Monoidal structures for $\Group$ and $\Top$]
  The cartesian product of groups, with operations defined 
  component-wise, defines a monoidal structure on $\Group$. 
  Similarly, the product of topological spaces defines a monoidal 
  structure on $\Top$.
\end{examplehard}
Note that, in a monoidal category, $A \Tensor B$ is not the 
same object as $B \Tensor A$, and there is no general way 
to go from one to the other. This can be a useful feature if 
we want to model a notion of parallel composition which is 
``position-sensitive'', but in other situations it can be a blocker. 
For instance, it conflicts with the idea of systems that can be 
\emph{swapped}, meaning that it does not matter which system 
is on the left and which system is on the right, since we can 
always exchange their places.

\noindent
If we want to describe entities that can be composed in 
parallel where swapping is permitted, we have to require 
this explicitly, imposing more properties that our monoidal 
category has to satisfy.
\begin{definition}[Symmetric monoidal category]
  \label{def: symmetric monoidal category}
  A \emph{symmetric monoidal category} is a monoidal 
  category $(\CategoryC,\Tensor,\TensorUnit)$ together 
  with a natural isomorphism
	\begin{equation*}
		\sigma: (-) \Tensor (-) \xrightarrow{\simeq} (-) \Tensor (-)
	\end{equation*}
  called \emph{symmetry} (or \emph{swap}), with 
  components in the form
	\begin{equation*}
		\sigma_{A,B}: A \Tensor B \xrightarrow{\simeq} B \Tensor A
	\end{equation*}
  such that the diagram in 
  Figure~\ref{fig: coherence condition symmetry} commutes 
  and, moreover,
	\begin{equation}\label{eq: symmetry equation}
		\sigma_{A,B};\sigma_{B,A} = id_{A \Tensor B}
	\end{equation}
\end{definition}
\begin{figure}[!ht]
	\centering
	\input{Pictures/CategoryTheoryIntroduction/03-SymmetryCoherence.tex}
	\caption{Additional coherence condition for symmetric monoidal categories.}
	\label{fig: coherence condition symmetry}
\end{figure}
Notice how Equation~\ref{eq: symmetry equation} suffices to state that $\sigma$ is its own inverse 
(consistent with the idea that swapping $A$ for $B$ 
and then $B$ for $A$ amounts to do nothing), while 
the diagram in Figure~\ref{fig: coherence condition symmetry} 
guarantees that the order in which we swap more than two 
objects does not matter.
\begin{example}(Symmetric monoidal categories in $\Set$)
  Both $(\Set, \times, \{\star\})$ and 
  $(\Set, \sqcup, \emptyset)$ are symmetric 
  monoidal categories. In the first case, $\sigma$ is just 
  the natural isomorphism that swaps terms in a couple:
	\begin{equation*}
	(a,b) \mapsto (b,a)
	\end{equation*}
  In the second, remembering that $A \sqcup B$ can be 
  represented as couples $(x,y)$ where $y = 0$ if 
  $x \in A$ and $y=1$ if $x \in B$, then $\sigma$ is the 
  natural isomorphism
	\begin{equation*}
	(x,y) \mapsto (x,y+1 \mod 2)
	\end{equation*}
\end{example}
\begin{examplehard}[Non-symmetric monoidal category]
  Left modules over a ring $R$ and their module 
  homomorphisms form a category. The usual tensor 
  product of modules defines a monoidal category, with 
  the trivial left module $R$ serving as unit. If $R$ is not 
  commutative, this monoidal category is not symmetric.
\end{examplehard}
We conclude this Section with a last definition, that 
is just a strengthening of Definition~\ref{def: monoidal category}.
\begin{definition}[Strict monoidal category]
  \label{def: strict monoidal category}
  We say that a category is \emph{strict monoidal} 
  when \emph{associators and unitors are identities}. 
  This means that, in a strict monoidal category,
	\begin{equation*}
  (A \Tensor B) \Tensor C = A \Tensor (B \Tensor C) 
    \qquad \TensorUnit \Tensor A = A = A \Tensor \TensorUnit
	\end{equation*}
\end{definition}
Note how in a strict monoidal category the coherence 
conditions for associators and unitors become trivial, since 
all the morphisms are equalities.
\begin{examplehard}[The category of endofunctors is strict monoidal]
  Given a category $\CategoryC$ we can consider the 
  category $[\CategoryC, \CategoryC]$ that has functors 
  of the form $F:\CategoryC \to \CategoryC$ as objects and 
  natural transformations between them as arrows. Maybe 
  counterintuitively, functor composition defines a monoidal 
  structure on $[\CategoryC, \CategoryC]$, with monoidal unit 
  being the identity functor $\Id{\CategoryC}$. Strictness 
  of $[\CategoryC, \CategoryC]$ follows immediately from 
  associativity of composition and identity laws between 
  functors, that hold with equality.
\end{examplehard}
\begin{example}[Products of sets are not strict]
  $(\Set, \times, \{\star\})$ is not a strict monoidal 
  category. This is because a generic couple $((a,b),c)$ is not 
  equal to the couple $(a,(b,c))$, albeit one can be mapped 
  into the other and vice-versa. While mathematicians often 
  ignore this phenomenon, functional programmers are 
  particularly sensitive to this sort of nuance, which often 
  prevents a program from correctly type-checking.
\end{example}
\begin{remark}[Monoidal categories are equivalent to strict ones.]
    \label{rem: monoidal equivalence}
  A quite useful result, that can be found 
  in~\cite[Ch. 11, Sec. 3, Thm. 1]{MacLane1978}, proves 
  that every monoidal category is \emph{monoidally equivalent} to a 
  strict monoidal one. In this document, we did not formally define 
  what an monoidal equivalence of monoidal categories 
  is, but you can guess it by massaging the 
  Definition~\ref{def: equivalence of categories}: It is just a 
  normal equivalence where our functor is monoidal 
  (monoidal functors will be defined in 
  Section~\ref{sec: monoidal functors})! This is useful since 
  it means that every time we are working with a monoidal 
  category we can also work with a strict version of it, 
  where many of the important properties stay the same 
  but life is easier. This scales to symmetric monoidal 
  categories in the obvious way.
\end{remark}
\section{String diagrams}\label{sec: string diagrams}
One of the most striking features of strict 
monoidal categories is that they admit a 
convenient \emph{graphical calculus} that allows 
us to forget the mathematical notation altogether and 
work just using pictures - string diagrams. The best 
thing about this approach is that these pictures are formally 
defined, ensuring that if we manipulate our drawings 
following some basic rules we are correctly manipulating 
morphisms in the underlying category.
\begin{figure}[!h]
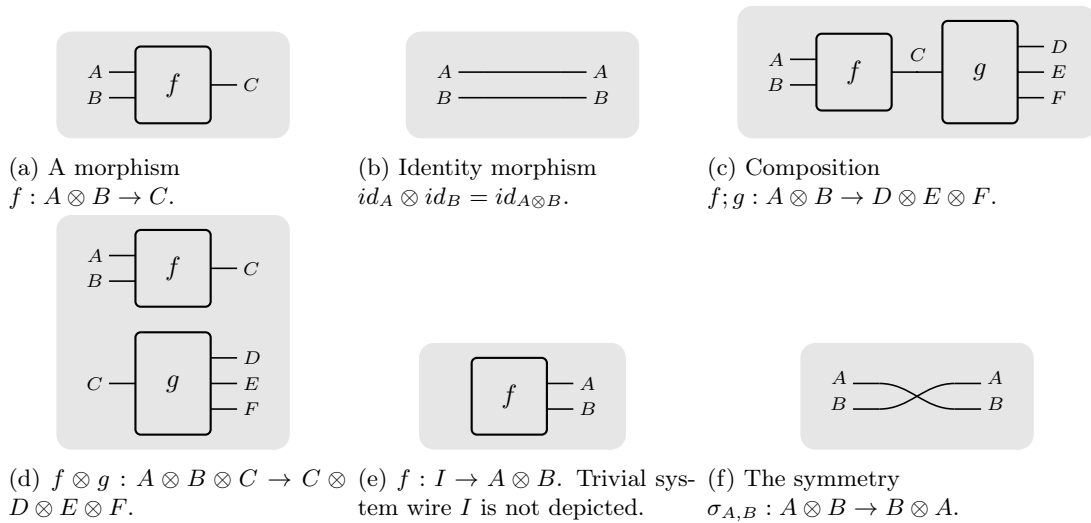

	\centering
	\begin{subfigure}[b]{0.3\textwidth}\centering
		\input{Pictures/CategoryTheoryIntroduction/04-Box.tex}
		\caption{A morphism\\ $f:A \Tensor B \to C$.}
		\label{fig: box}
	\end{subfigure}
	\hfill
	\begin{subfigure}[b]{0.3\textwidth}\centering
		\input{Pictures/CategoryTheoryIntroduction/04-Identity.tex}
		\caption{Identity morphism\\ $\Id{A} \Tensor \Id{B} = \Id{A \Tensor B}$.}
		\label{fig: identity}
	\end{subfigure}
	\hfill
	\begin{subfigure}[b]{0.38\textwidth}\centering
		\input{Pictures/CategoryTheoryIntroduction/04-Composition.tex}
		\caption{Composition\\ $f;g:A \Tensor B \to D \Tensor E \Tensor F$.}
		\label{fig: composition}
	\end{subfigure}
	\hfill
	\begin{subfigure}[b]{0.3\textwidth}\centering
		\input{Pictures/CategoryTheoryIntroduction/04-Tensor.tex}
		\caption{$f \Tensor g:A \Tensor B \Tensor C \to C \Tensor D \Tensor E \Tensor F$.}
		\label{fig: tensor}
	\end{subfigure}
		\hfill
	\begin{subfigure}[b]{0.3\textwidth}\centering
		\input{Pictures/CategoryTheoryIntroduction/04-State.tex}
		\caption{$f:\TensorUnit\to A \Tensor B$. Trivial system wire $\TensorUnit$ is not depicted.}
		\label{fig: state}
	\end{subfigure}
	\hfill
	\begin{subfigure}[b]{0.38\textwidth}\centering
		\input{Pictures/CategoryTheoryIntroduction/04-Swap.tex}
		\caption{The symmetry\\ $\sigma_{A,B}:A \Tensor B \to B \Tensor A$.}
		\label{fig: symmetry}
	\end{subfigure}
  \caption{Graphical calculus for symmetric monoidal categories.}
    \label{fig: graphical calculus for symmetric monoidal categories}
\end{figure}
\noindent
In the graphical formalism, to be read left to right, 
objects are represented as \emph{typed wires}, and 
morphisms as \emph{boxes}, as in Figure~\ref{fig: box}. 
Identity morphisms are just represented as wires 
(see Figure~\ref{fig: identity}), which is clearly consistent 
with the idea of identity morphisms ``doing nothing''. As 
we already noted, composition of morphisms can express 
the idea of sequential composition, and is thus represented 
by \emph{connecting the output wire of a box with the input 
wire of another when the wire types match}, as in 
Figure~\ref{fig: composition}.

The monoidal product, representing the idea of parallel 
composition, is depicted by \emph{placing boxes and 
wires next to each other}, as shown in Figure~\ref{fig: tensor}. 
This is consistent with the idea that 
$(f \Tensor g) \Tensor h \simeq f \Tensor (g \Tensor h)$ via the 
associator, hence we do not need to represent any bracketing. 
The unit wire $\TensorUnit$ represents the trivial system, 
and is thus \emph{not drawn}, see Figure~\ref{fig: state}. 
This again backs up the intuition that 
$\TensorUnit \Tensor A$, $A$ and $A \Tensor \TensorUnit$ 
are morally the same. Finally, symmetry is represented by 
just \emph{swapping wires}, as in Figure~\ref{fig: symmetry}.
\begin{remark}[Equivalence to a strict category is necessary 
  for diagrammatics]
  Note that in depicting monoidal products without brackets, 
  and in choosing not to draw the monoidal unit, we are implicitly 
  making use of the result mentioned in 
  Remark~\ref{rem: monoidal equivalence}. Working in the 
  graphical formalism means exactly working in the strict 
  symmetric monoidal category equivalent to the monoidal 
  category we want to study.
\end{remark}
\begin{figure}[!ht]
	\centering
	\input{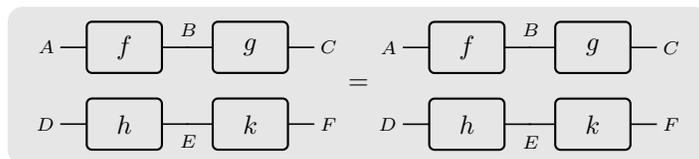}
	\caption{Graphical proof of the Eckmann-Hilton argument.}
	\label{fig: eckmann-hilton}
\end{figure}
\begin{example}[Eckmann-Hilton argument]
  \label{ex: eckmann-hilton argument}
  To point out how powerful the diagrammatic formalism is, 
  note that results such as the Eckmann-Hilton argument for 
  monoidal categories, that is, one of the equations expressing 
  the functoriality of the monoidal product:
	\begin{equation*}
		(f ; g) \Tensor (h ; k) = (f \Tensor h) ; (g \Tensor k)
	\end{equation*}
  reduce to tautologies, making proofs much easier 
  (see Figure~\ref{fig: eckmann-hilton}). This is very 
  interesting considering that the equation above does not 
  look trivial, while the corresponding diagrams surely do: 
  The graphical formalism helps by stripping away many of 
  the irrelevant details when we work with monoidal categories.
\end{example}
\begin{remark}[References for string diagrams]
  The study of string diagrams goes often under the name of 
  \emph{process theory}, of which~\cite{Coecke2017} 
  is one of the most complete references.
\end{remark}
\begin{figure}[!ht]
	\centering
	\input{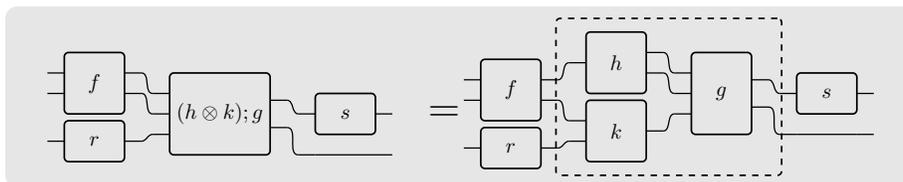}
	\caption{The graphical calculus allows us to explode boxes.}
	\label{fig: box explosion}
\end{figure}

\noindent
The analogy between process theories and programming 
is more than evident: A box can be thought of as a piece of 
software that performs some operations on data having 
certain types, and functional programming can be entirely 
formalized using these diagrams. Moreover, note that we can 
\emph{explode} a box, that is, boxing more components into 
a unique one. For instance, in Figure~\ref{fig: box explosion} 
(where the wire types have been omitted to avoid clutter), we 
are considering the morphism $(h \Tensor k) ; g$ as a unique 
box (dashed). This allows us to zoom in/out our processes and 
hide the features that are irrelevant at a given level of generality. 
We can then form new boxes by just stacking up some other 
processes and considering them as one.

\begin{remark}[Completeness of graphical calculi]
  The kind of string diagrams covered here is one of the most 
  simple graphical formalisms studied in process theories, but 
  it is good to unveil how category theory can provide nice tools 
  to reason about compositionality without having to learn 
  difficult maths. The reason why it works, viz. why categorical 
  proofs can be carried out graphically, relies on a 
  \emph{completeness theorem} which results from linking things 
  that are graphically provable to things that are provable in 
  monoidal categories. Details about this can be found 
  in~\cite{Selinger2010}.
\end{remark}
\begin{remark}[String diagrams and commutative diagrams 
  are different things]
  Often pictures in the graphical calculus are referred to as 
  \emph{diagrams}. Do not confuse these diagrams with the 
  \emph{commutative diagrams} introduced in 
  Remark~\ref{rem: commutative diagram}!
\end{remark}
As we hinted in the beginning of this Chapter, category theory 
together with its links to graphical calculi will act as ``deus 
ex machina'' in the formalization of Statebox: All the theories 
presented in the remainder of this document will admit a strong 
categorical formalization, from which it is possible to 
create an equivalent graphical formalization in a safe way. ``Pure'' 
category theory is used to ``sew together'' all these different 
theories, and obtain a formally organic and satisfying foundation 
on which Statebox is implemented.

This is exactly what backs up our claim that, in Statebox, it is 
possible to do software engineering in a way that is at the same 
time~\emph{purely graphical and purely correct}.
\section{Monoidal functors}\label{sec: monoidal functors}
What happens to our functors if our categories are monoidal? 
If $(\CategoryC, \Tensor, \TensorUnit)$ and 
$(\CategoryD, \square, \TensorUnit')$ are monoidal categories 
and there is a functor $F:\CategoryC \to \CategoryD$, there is 
nothing in principle that says that monoidal products will be 
preserved. For instance, if we consider $A \Tensor B$, then 
$F(A \Tensor B)$ and $FA \square FB$ may be totally unrelated. 
Embracing the idea that ``a functor is a morphism between 
categories'', we see that in restricting to monoidal categories 
there is some additional, relevant structure that functors are 
not preserving. We deduce, then, that the notion of a functor 
is not the correct one to model morphisms between monoidal 
categories. Here, we want to find conditions for $F$ to 
\emph{preserve the monoidal structure}. This idea prompts 
various different definitions, that are nevertheless related to 
each other.
\begin{figure}[!ht]
	\centering
	\begin{subfigure}[b]{1\textwidth}\centering
		\input{Pictures/CategoryTheoryIntroduction/07-LaxMonoidalAssociator.tex}
		\caption{Associator condition for lax monoidal functors.}
		\label{fig: lax monoidal associator}
	\end{subfigure}
	\hfill
	\begin{subfigure}[b]{0.45\textwidth}\centering
		\input{Pictures/CategoryTheoryIntroduction/07-LaxLeftUnitor.tex}
		\caption{Left unitor condition for lax monoidal functors.}
		\label{fig: lax monoidal left unitor}
	\end{subfigure}
	\hfill
	\begin{subfigure}[b]{0.45\textwidth}\centering
		\input{Pictures/CategoryTheoryIntroduction/07-LaxRightUnitor.tex}
		\caption{Right unitor condition for lax monoidal functors.}
		\label{fig: lax monoidal right unitor}
	\end{subfigure}
  \caption{Coherence conditions for a lax monoidal functor.}
    \label{fig: lax monoidal conditions}
\end{figure}
\begin{definition}[Lax monoidal functor]
  \label{def: lax monoidal functor}
  A \emph{lax monoidal functor} between two monoidal 
  categories $F: (\CategoryC, \Tensor, \TensorUnit) \to 
  (\CategoryD, \square, \TensorUnit')$ is specified by the 
  following infomation:
	\begin{itemize}
		\item A functor $F:\CategoryC \to \CategoryD$;
		\item A morphism in $\CategoryD$
		\begin{equation*}
			\epsilon:\TensorUnit' \longrightarrow F\TensorUnit
		\end{equation*}
		\item A natural transformation
		\begin{equation*}
			\varphi: F(-) \square F(-) \longrightarrow F((-)\Tensor(-))
		\end{equation*}
		with components in the form
		\begin{equation*}
			\varphi_{A,B}: FA \square FB \longrightarrow F(A \Tensor B)
		\end{equation*}
	\end{itemize}
  Such that the diagrams in 
  Figure~\ref{fig: lax monoidal conditions} commute, 
  where superscripts $\CategoryC,\CategoryD$ denote 
  if the associator/left unitor/right unitor are the ones 
  in $\CategoryC$ or the ones in $\CategoryD$, respectively.
\end{definition}
The diagram in Figure~\ref{fig: lax monoidal associator} 
expresses the idea that the monoidal functor respects associators: 
It says that there is no real difference in applying the 
associator in $\CategoryC$ and then applying $F$ to the 
result or applying the associator in $\CategoryD$ to the images 
through $F$ of the objects in $\CategoryC$. Same reasoning 
applies for left and right unitors, as depicted in 
Figures~\ref{fig: lax monoidal left unitor} 
and~\ref{fig: lax monoidal right unitor}.

The concept of a lax monoidal functor is one of the weakest 
ways to relate monoidal categories. In the following definition, 
we will refine this concept requiring more properties to be 
satisfied, making the way monoidal categories are related to 
each other increasingly stronger.
\begin{definition}[Symmetric, strong, strict monoidal functors]
  \label{def: symmetric strong strict monoidal functors}
  A lax monoidal functor 
  $F: (\CategoryC, \Tensor, \TensorUnit) \to 
  (\CategoryD, \square, \TensorUnit')$ is said to be:
	\begin{itemize}
    \item \emph{Symmetric} if $F$ preserves symmetries, 
    meaning that the diagram in 
    Figure~\ref{fig: lax monoidal symmetry} also commutes;
    \item \emph{Strong} if both $\varphi$ and $\epsilon$ are 
    natural isomorphisms;
    \item \emph{Strict} if both $\varphi$ and $\epsilon$ are 
    equalities. In this case we have
		\begin{equation*}
      F\TensorUnit = \TensorUnit' \qquad F(A \Tensor B) = 
      FA \square FB
		\end{equation*}
	\end{itemize}
\end{definition}
\begin{figure}[!ht]
	\centering
	\input{Pictures/CategoryTheoryIntroduction/08-LaxSymmetry.tex}
  \caption{Additional coherence condition for lax symmetric 
  monoidal functor.}
	\label{fig: lax monoidal symmetry}
\end{figure}
\begin{remark}[strictness of symmetries follows from strictness.]
  Note that, if $F$ is symmetric and strict, strictness and the 
  diagram in Figure~\ref{fig: lax monoidal symmetry} 
  automatically imply
	\begin{equation*}
		F\sigma_{A,B} = \sigma_{FA, FB}
	\end{equation*}
\end{remark}
\section{Products, coproducs, pushouts}
\label{sec: products, coproducts, pushouts}
Now we introduce another well known concept in 
category theory. The arguments covered here are just a 
small fragment of a much more developed theory, and are 
particular instances of \emph{limits} and \emph{colimits}. 
Due to the risk of losing the reader's attention, we refer one 
to~\cite[Ch. 2]{Borceux1994} and~\cite[Ch. 3]{MacLane1978} 
for a fully-detailed coverage of the story.

Let us think about the category of sets and functions, $\Set$. 
In Examples~\ref{ex: set product} and~\ref{ex: set coproduct} 
we already mentioned the concepts of a \emph{cartesian product} 
and \emph{disjoint union} of sets, and we highlighted how these 
constructions can be used to define different symmetric 
monoidal structures. But what is a cartesian product? 
And a disjoint union? Do we have a way to capture these notions 
purely categorically, that is, without making any explicit 
reference to elements?

In principle, we would be tempted to say ``no''. The main idea 
when dealing with cartesian products is that if we have two 
sets $A,B$ then we are able to consider \emph{the set of couples}:
\begin{equation*}
	A \times B := \Suchthat{(a,b)}{a \in A, b \in B}
\end{equation*}
This definition makes explicit use of elements, so how can 
we restate it just in terms of sets and functions? Surprisingly, 
it turns out that there is a way, as we are about to show.

\begin{figure}[!ht]
	\centering
	\input{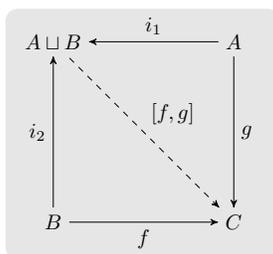}
	\caption{Universal property for coproducts.}
	\label{fig: universal property for coproducts}
\end{figure}

First things first, we note that if we have a cartesian product 
$A \times B$ then we have a couple of functions, usually 
called \emph{projections}, that ``forget'' about one side of the 
product:
\begin{align*}
	\pi_1: A \times B &\to A &\qquad  \pi_2: A \times B &\to B\\
	(a,b) &\mapsto a &\qquad (a,b) &\mapsto b
\end{align*}
Moreover, we also note that every time we have two functions 
$f:C \to A$ and $g: C \to B$, we can construct a function 
$f \times g$ pairwise, setting
\begin{align*}
	\langle f, g \rangle: C &\to A \times B\\
	c &\mapsto (f(c),g(c))
\end{align*}
We also see quite easily that, by definition,
\begin{equation*}
	\langle f, g \rangle;\pi_1 = f \qquad \langle f, g \rangle;\pi_2 = g
\end{equation*}
All this information is indeed enough to capture the idea 
of cartesian product of sets, and can be presented as follows:
\begin{figure}[!ht]
	\centering
	\input{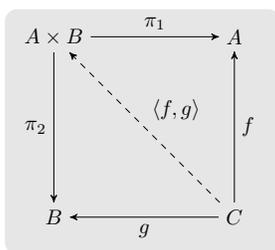}
	\caption{Universal property for products.}
	\label{fig: universal property for products}
\end{figure}
\begin{example}[Products in $\Set$]\label{ex: products of sets}
  For any two sets $A,B$, there exists a set $A \times B$, 
  together with functions 
  $\pi_1:A \times B \to A$, $\pi_2: A \times B \to B$, such that 
  every time we have another set $C$ and a couple of functions 
  $f:C \to A$, $g:C \to B$, there is 
  \emph{one and only one function}, denoted with $\langle f, g \rangle$, 
  that makes the diagram in 
  Figure~\ref{fig: universal property for products} commute.
\end{example}
But now this definition of product does not make use of elements 
at all, and we can use it for any category!
\begin{definition}[Products]\label{def: products}
  Let $\CategoryC$ be a category. We say that $\CategoryC$ has 
  \emph{products} when the condition stated in 
  Example~\ref{ex: products of sets} holds for any couple of 
  objects $A,B$ and for any couple of morphisms 
  $C \to A$, $C \to B$.
\end{definition}
\begin{example}[Product of categories]
  \label{ex: products of categories}
  It is not hard to see that \Cat, the category of all 
  small\footnote{There are issues in considering the category 
  of all categories that make the theory inconsistent, exactly as 
  it happens in set theory. To solve this, we have to restrict 
  ourselves to particular types of categories, called \emph{small} 
  categories. All the categories usually considered in ordinary 
  mathematics are small, so this is not a big deal for us!} 
  categories and functors between them, admits a product 
  structure. Given categories $\CategoryC, \CategoryD$, 
  their product $\CategoryC \times \CategoryD$ can be defined 
  as just
	\begin{equation*}
    \Obj{\CategoryC \times \CategoryD} := 
      \Obj{\CategoryC} \times \Obj{\CategoryD} \qquad 
      \Homtotal{\CategoryC \times \CategoryD} := 
      \Homtotal{\CategoryC} \times \Homtotal{\CategoryD}
	\end{equation*}
	With source and target defined component-wise as
	\begin{equation*}
    \Source{(f,g)} := (\Source{f}, \Source{g}) \qquad 
      \Target{(f,g)} := (\Target{f}, \Target{g})
	\end{equation*}
  Note how we used this product in 
  Definition~\ref{def: monoidal category} to define the 
  functor $\Tensor$.
\end{example}
\begin{figure}[!ht]
	\centering
	\input{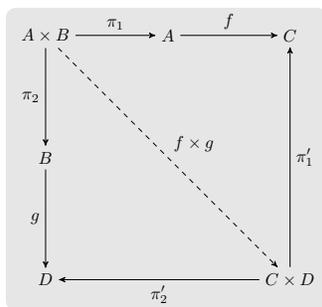}
	\caption{Product of morphisms $f,g$.}
	\label{fig: product of morphisms}
\end{figure}
\begin{example}[Product of morphisms]
  We can make immediate use of the property of products, 
  as follows: Suppose that we have morphisms 
  $f:A \to C$, $g:B \to D$. \emph{Thanks to the property 
  of $C \times D$}, we can obtain a unique morphism 
  $f \times g: A \times B \to C \times D$ setting
	\begin{equation*}
		f \times g := \langle \pi_1;f, \pi_2;g \rangle
	\end{equation*}
  where $\pi_1, \pi_2$ are the projections from 
  $A \times B$ to $A, B$, respectively, as in 
  Figure~\ref{fig: product of morphisms}. This is exactly 
  what allowed us to use the cartesian product to define a 
  monoidal structure in Example~\ref{ex: set product}, but 
  holds in general: In any category with products, the product 
  defines a monoidal structure.
\end{example}
Similarly, we can characterize the idea of ``disjoint union" 
categorically, as follows:
\begin{definition}[Coproducts]
  A category $\CategoryC$ has \emph{coproducts} if, for 
  every couple of objects $A,B$, there is an object 
  $A \sqcup B$ together with morphisms 
  (called \emph{injections}) $i_1: A \to A \sqcup B$ and 
  $i_2:B \to A \sqcup B$ such that, for each couple of 
  morphisms $f:A \to C$ and $g:B \to C$, there is 
  \emph{one and only one} morphism 
  $[f, g]: A \sqcup B \to C$ that makes the diagram 
  in Figure~\ref{fig: universal property for coproducts} commute.

  Given two morphisms $f: A \to C$ and $g: B \to D$ we 
  can, as for products, obtain a morphism 
  $f \sqcup g: A \sqcup B \to C \sqcup D$ by setting:
	\begin{equation}\label{eq: coproduct is a monoidal functor}
		f \sqcup g := [f;i_1, g;i_2]
	\end{equation}
  Where $i_1, i_2$ are the injections from $C, D$ to 
  $C \sqcup D$, respectively.
\end{definition}
Note that the usual disjoint union of sets respects the 
condition given above. Moreover, we are now able to see 
how disjoint union (categorically known as \emph{coproduct}) 
and the cartesian product (categorically just known as 
\emph{product}) are somehow connected: The definition 
of coproduct is the same as the one of product, but 
\emph{with all the arrows reversed!}
\begin{remark}[Coproducts induce monoidal structures]
  It is again true that, in any category with coproducts, the 
  coproduct can be used to define a monoidal structure. As 
  in the case of products, this is implied by 
  Equation~\ref{eq: coproduct is a monoidal functor}. We saw 
  this explicitly with the category $\Set$, in 
  Example~\ref{ex: set coproduct}.
\end{remark}
The product and coproduct construction, respectively, are said 
to be built by means of \emph{universal properties}. Intuitively, 
the idea of universal property is that for each set of 
``preconditions'' -- whatever this means depends on context -- 
there is \emph{exactly one morphism} that makes some 
diagram commute.

\begin{figure}[!ht]
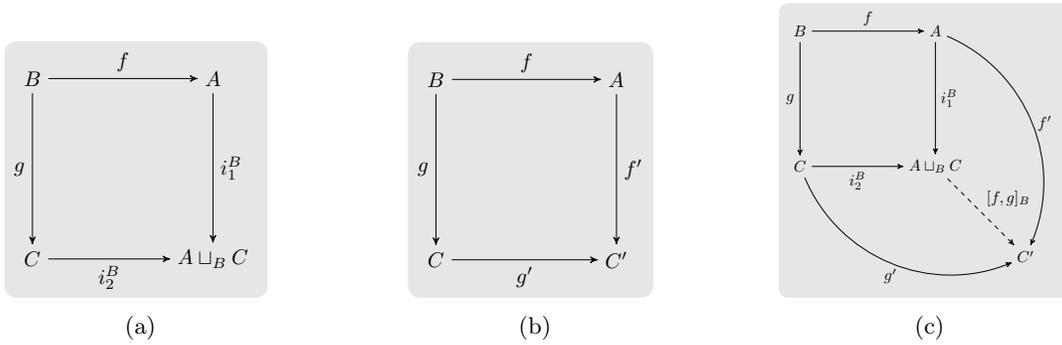

	\centering
	\begin{subfigure}[b]{0.30\textwidth}\centering
		\input{Pictures/CategoryTheoryIntroduction/12-Pushout1.tex}
		\caption{}
		\label{fig: pushout 1}
	\end{subfigure}
	\hfill
	\begin{subfigure}[b]{0.30\textwidth}\centering
		\input{Pictures/CategoryTheoryIntroduction/12-Pushout2.tex}
		\caption{}
		\label{fig: pushout 2}
	\end{subfigure}
	\hfill
	\begin{subfigure}[b]{0.30\textwidth}\centering
		\input{Pictures/CategoryTheoryIntroduction/12-Pushout3.tex}
		\caption{}
		\label{fig: pushout 3}
	\end{subfigure}
  \caption{Universal property for pushouts.}
    \label{fig: universal property for pushouts}
\end{figure}

There is another universal construction (that is, a categorical 
construction made by means of universal properties) that is 
worth mentioning. This 
construction is called \emph{pushout}:
\begin{definition}[Pushout]
  A category $\CategoryC$ has \emph{pushouts} if, for each 
  couple of morphisms $f: C \to A$ and $g: C \to B$, there is 
  an object $A \sqcup_B C$ and morphisms $i_1^B: A 
  \to A \sqcup_B C$, $i_2^B: C \to A \sqcup_B C$ that make 
  the diagram in Figure~\ref{fig: pushout 1} commute.

  Moreover, if we have morphisms $f':A \to C'$ and 
  $g': C \to C'$ such that the diagram in Figure~\ref{fig: pushout 2} 
  commutes, then there is a unique morphism (here is where the 
  universal property kicks in) ${[f,g]}_B: A \sqcup_B C \to C'$ 
  such that the diagram in Figure~\ref{fig: pushout 3} commutes 
  too.
\end{definition}
Note how in the pushout case we are conceptually going 
backwards: Before, we took set-theoretic concepts and we 
generalized them to arbitrary categories. Now, instead, we 
are giving a categorical definition, and in principle we do not 
even know if there are categories that satisfy it or, more 
specifically, if the category $\Set$, on which we 
based many examples, does.

\begin{figure}[!ht]
	\centering
	\input{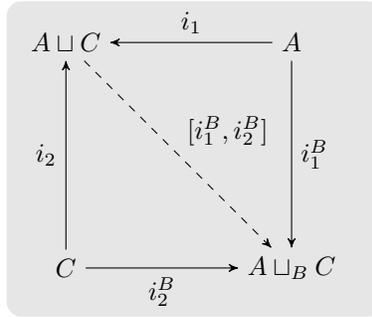}
	\caption{Interaction between coproducts and pushouts.}
	\label{fig: coproduct pushout interaction}
\end{figure}

\noindent
This is a very important point when working abstractly, 
viz. while giving categorical definitions that are not based 
on specific examples we already know. Every time we give 
a categorical property we have to check:
\begin{itemize}
	\item If the category we are interested in satisfies that property;
	\item How can that property be explicitly described in the category.
\end{itemize}
In the case of $\Set$, we are indeed lucky.
\begin{examplehard}[Pushouts in $\Set$]\label{ex: pushouts in set}
  The category $\Set$ indeed has pushouts. 
  The pushout of morphisms $f:B \to A$,and $g:B \to C$ 
  can be characterized as follows:
	\begin{equation*}
		A \sqcup_B C := A \sqcup C/\sim
	\end{equation*}
  where $\sim$ is the smallest equivalence relation that 
  identifies $a \in A$ with $c \in C$ if there is some $b\in B$ 
  such that $a = f(b)$ and $c = g(b)$. $i^B_1$ 
  ($i^B_2$, respectively) sends every element of $A$ (of $C$, 
  respectively) to its equivalence class. It is easy to see that, 
  for each $b \in B$, it is true by definition that 
  $f;i_1(b) = g;i_2(b)$.
\end{examplehard}
To conclude, we note that in making the definition of a pushout 
explicit in $\Set$ we used the disjoint union, which 
we know is the coproduct in $\Set$. We may 
hypothesize that these two things are somehow connected 
and this is indeed true in any category that has both pushouts 
and coproducts.
\begin{remark}[Coproducts and pushouts are connected]
  The pushout of morphisms $f:B \to A$ and $g:B \to C$ gives 
  us morphisms 
  $i^B_1: A \to A \sqcup_B C$, $i_2^B: C \to A \sqcup_B C$. 
  In a category that has both pushouts and coproducts, we 
  can then consider the situation in 
  Figure~\ref{fig: coproduct pushout interaction}, where the 
  existence and uniqueness of the diagonal arrow is guaranteed 
  by the universal property defining coproducts.

  So, every time we have pushouts of some morphisms $f,g$ and 
  coproducts in a category, we always have a unique morphism 
  connecting the two.
\end{remark}
\begin{examplehard}
  In the category $\Set$, as we already saw, the unique 
  morphism in Figure~\ref{fig: coproduct pushout interaction} 
  is the one sending each element of the disjoint union to its 
  equivalence class with respect to the relation $\sim$ defined 
  in Example~\ref{ex: pushouts in set}.
\end{examplehard}
\section{Implementation}
Some of the concepts covered in this Chapter have been 
implemented by our team in a Idris~\cite{Brady2013, Brady2017} 
library, called \texttt{idris-ct}~\cite{StateboxTeam2019}.
Idris is a \emph{dependently-typed}, functional programming language. 
Dependent types are a very expressive typesystem that allow us to 
implement mathematical proofs in our code. In our case, this 
means that categorical concepts can be implemented in a way 
which is precisely equivalent to their mathematical counterpart.

For instance, our implementation of the concept of \texttt{Category} consists
of types representing objects and morphisms, function types representing 
identities and morphism composition and, most notably, three 
types whose terms are proofs 
that the left, right identity and associative laws, respectively, hold. 
This means that in defining a category the user does not just have to 
specify morphisms and objects, but \emph{must also provide proofs 
that such definition is correct}. 

Moreover, our code is written in \emph{literate Idris} and can be compiled 
down to \LaTeX: The source code itself can either be compiled into 
an executable or be compiled into its own \LaTeX-typeset instruction manual!
More details about this can be found in~\cite{StateboxTeam2019}. 

Among the categorical constructions we implemented there are:
\begin{itemize}
  \item The definitions of category, functor and natural transformation;
  \item The definitions of monoidal category, symmetric monoidal 
  category, monoidal functors and their strict counterparts;
  \item The definition of product of categories;
  \item Proofs that $\Cat$ is a category, and that Idris types and functions 
  item between them form a category as well.
\end{itemize}
\section{Why is this useful?}
We admit that it is difficult to answer this question at this 
stage. This Chapter has been very dense in terms of mathematical 
definitions and results, and quite poor in terms of applicative 
purposes and examples. We could not do much better than this, 
since the learning curve for category theory is steep and one 
needs to build quite a bit of machinery to successfully employ 
it in modeling real-world problems.

The best we can do for now is reassure the audience that the 
fruits of such an involved reading will be reaped very soon, 
and limit ourselves to the following considerations:
\begin{itemize}
  \item Categories are ubiquitous, and we can do all the known 
  mathematics with them. The concepts of functor and natural 
  transformation are very powerful, and allow us to establish 
  formally consistent links between mathematical theories. We 
  understand that if we have a categorical definition of Petri nets -- 
  that we will work out in Chapter~\ref{ch: executions} -- 
  and a categorical definition of some other meaningful tool 
  we want to use, then we can employ our categorical techniques 
  to combine these two together, as we will do in 
  Chapter~\ref{ch: folds};

  \item Categories have a clear operative interpretation, allowing 
  us to talk about processes happening in series or in parallel. 
  This makes category theory readily applicable in the context 
  of software design, as we will see in Chapter~\ref{ch: folds}. 
  What we lack is the idea of processes competing for resources, 
  as we had for Petri nets, and this is exactly why we want to 
  categorize them, bringing together the best of both worlds;

  \item As in the case of Petri nets, monoidal categories admit a 
  neat graphical formalism. This means once more that in 
  implementing a programming language based on monoidal 
  categories we get a visual way to code/debug for free. The 
  debugging functionality is particularly advantageous, since code 
  that may be hard to read is often translated to 
  straightforward images, as we saw for the Eckmann-Hilton 
  argument in Example~\ref{ex: eckmann-hilton argument}.
\end{itemize}
How proficuous category theory will be for us will already 
become clear in the next Chapter, where we use category 
theory to turn Petri nets into fully deterministic structures.

\newtoggle{toggleExecutions}\toggletrue{toggleExecutions}
  \chapter{Executions of Petri nets}\label{ch: executions}
Up to now, we introduced two main concepts: Petri nets 
-- in Chapter~\ref{ch: Petri nets} -- and category theory -- 
in Chapter~\ref{ch: introduction to category theory}. We 
moreover promised that the two things are related, 
and that we use the second one to help us reason 
about the first. In this Chapter we start honoring this promise, 
modeling the executions of a Petri net categorically.
\section{Problem overview}
\begin{figure}[!ht]
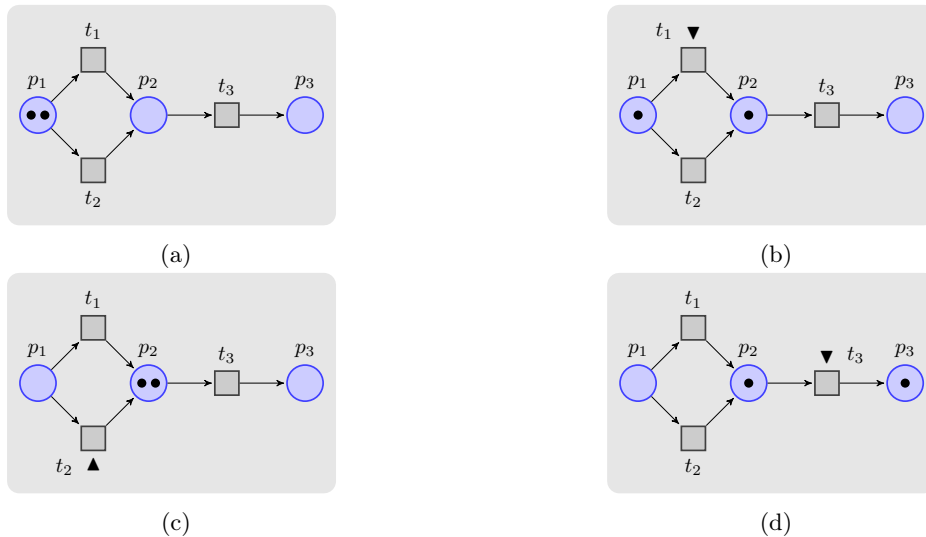

	\centering
	\begin{subfigure}[t]{0.45\textwidth}\centering
		\input{Pictures/ExecutionsOfPetriNets/00-Evolution1.tex}
		\caption{}
		\label{fig: petri evolution 1}
	\end{subfigure}
	\hspace{1cm}
	\begin{subfigure}[t]{0.45\textwidth}\centering
		\input{Pictures/ExecutionsOfPetriNets/00-Evolution2.tex}
		\caption{}
		\label{fig: petri evolution 2}
	\end{subfigure}
	\hspace{1cm}
	\begin{subfigure}[t]{0.45\textwidth}\centering
		\input{Pictures/ExecutionsOfPetriNets/00-Evolution3.tex}
		\caption{}
		\label{fig: petri evolution 3}
	\end{subfigure}
	\hspace{1cm}
	\begin{subfigure}[t]{0.45\textwidth}\centering
		\input{Pictures/ExecutionsOfPetriNets/00-Evolution4.tex}
		\caption{}
		\label{fig: petri evolution 4}
	\end{subfigure}
	\caption{Evolution of a net.}
	\label{fig: petri evolution}
\end{figure}

Consider the images in Figure~\ref{fig: petri evolution}, describing 
the evolution of a Petri net.

\noindent
In Figure~\ref{fig: petri evolution 2} transition $t_1$ has fired. 
If we name the markings in Figures~\ref{fig: petri evolution 1}, 
\ref{fig: petri evolution 2},~\ref{fig: petri evolution 3} 
and~\ref{fig: petri evolution 4} respectively as
$\Marking{X},\Marking{Y},\Marking{Z},\Marking{W}$, the notation
\begin{equation*}
  \Marking{X} \xrightarrow{t_1} \Marking{Y} \xrightarrow{t_2} 
    \Marking{Z} \xrightarrow{t_3} \Marking{W}
\end{equation*}
does not help us understand which one of the two tokens in 
$p_2$ the transition $t_3$ is consuming: It is impossible to say 
if the token $t_3$ is consuming has been previously produced by 
transition $t_1$ or by transition $t_2$. This is problematic if we 
think of transitions as processes that consume and produce 
resources: A token represents a resource of the type of the place 
it is in, but these resources are not necessarily all the same. If a 
place in a net holds resources of type $\Bool$, for instance, then 
tokens in that place will be boolean entities, meaning that they can 
either represent the value $\True$ or the value $\False$. Similarly, 
if a place holds resources of type $\Int$ then each token represents 
an integer number, and two tokens in the same place may be 
different from each other. It is evident that it is not enough to 
say that a transition in a net consumes a given token to infer 
what is really happening: \emph{Distinguishing between tokens 
becomes important}, since different tokens will be processed 
differently by transitions.
\begin{example}[Different histories give different results]
  Suppose that all the places of the net in 
  Figure~\ref{fig: petri evolution} hold resources of 
  type $\Int$, viz. integer numbers. Interpret the transitions 
  of the net as functions from integers to integers:
	\begin{equation*}
		t_1(x) = x + 1 \qquad t_2(x) = 2x \qquad t_3(x) = x
	\end{equation*}
  Consider the tokens in Figure~\ref{fig: petri evolution 1} to both 
  represent the number $2$. It is evident that the token processed 
  by $t_1$ has value $3$, while the token processed by $t_2$ has 
  value $4$. Since $t_3$ is the identity on integers, applying $t_3$ 
  to one token or the other will produce different results, namely:
	\begin{equation*}
		t_3(t_1(2)) = 3 \neq 4 = t_3(t_2(2))
	\end{equation*}
\end{example}
\begin{remark}[Petri nets are inadequate for active design]
  Petri nets have been explicitly designed not to distinguish 
  between tokens. This non-deterministic behavior -- i.e. not 
  knowing which token a transition is processing -- is intended, 
  since the formalism is concerned only with studying structural 
  properties of distributed, concurrent systems, and abstracting 
  from such details comes in handy. But this is unsuitable if one 
  wants to use Petri nets to actively design complex infrastructure. 
  It is evident that we need a practical way to
  distinguish between tokens, similar to pushing a switch ``on and off'' depending on the task 
  at hand. 
\end{remark}
Now that we convinced ourselves that distinguishing between 
tokens is important, we still have to figure out how to do it.
\begin{remark}[Distinguishing between tokens]
  To distinguish between tokens, the most appropriate criterion that 
  we can think of is that \emph{tokens are considered to represent 
  the same resource if they have the same history}, meaning that 
  they have been processed by the same transitions. This indeed 
  makes sense, since it is the only way we have to distinguish 
  between tokens \emph{within} the net: Consider again 
  Figure~\ref{fig: petri evolution 1}. Here we have two tokens 
  in $p_1$, but we do not know anything else about them: Surely, 
  they may represent different resources but we have no way to 
  infer this from the behavior of the net, since the tokens ``were 
  already there'' before we started executing it. From within the 
  net, these tokens may then be considered equal, since any 
  additional information is not accessible.
\end{remark}
Now that we understood what it means to consider two tokens 
equal or not, we want to formalize this mathematically. We will 
proceed as follows:
\begin{itemize}
	\item We will organize Petri nets in a category, called $\Petri$;
  \item To each Petri net $N$ we will associate a category, denoted 
  with $\Fold{N}$, representing \emph{all the possible histories 
  of all the possible tokens in the net};
  \item We sketch how this correspondence works, linking nets 
  with executions in a reversible fashion.
\end{itemize}\label{item: executions plan}
Notice how our modus operandi represents very well the philosophy 
behind category theory that we highlighted in the previous Chapter: 
Category theory is the study of patterns, and we want to create a 
correspondence between Petri nets and their possible evolutions in a 
way that is compatible with the way nets interact: Patterns have to be 
preserved, hence we want a functor.
\begin{remark}[Different approaches]
  The plan highlighted in~\ref{item: executions plan} has been declined 
  in many different ways throughout the years, by many different authors. 
  All such approaches are somewhat conceptually similar, but differ greatly on details which are crucial when it comes 
  to implementation. 

  We redirect the reader seriously interested in knowing more about 
  this to~\cite{Meseguer1990, Baldan2003, Sassone1995, Baez2018}. 
  In particular, \cite{Master2019} provides a nice overview and generalization 
  of the problem.

  All the approaches listed above focus on showing that the category of 
  Petri nets and the category of Petri nets executions are equivalent. 
  They do this by focusing on adjunctions, a fundamental concept in category 
  theory which we did not define, but that can be found 
  in~\cite[Ch. 4]{MacLane1978}. 
  
  We as well pursued this approach, even generalizing 
  it to different Petri net flavors~\cite{Genovese2018}, but we soon realized that 
  chasing equivalences, albeit categorically satisfying, was not the right strategy 
  to obtain a feasible implementation. In this Chapter, then, we will follow 
  our own approach to the problem, of which details can be found 
  in~\cite{Genovese2019}. 
\end{remark}
\section{The category \texorpdfstring{$\Petri$}{Petri}}
Our first task is to organize Petri nets into a category. 
There are many different ways to do this, all useful. 
For now, we proceed by requiring Petri nets to be the 
objects of the category we want to define. If nets are objects, 
then we need a suitable notion of morphism between nets. 
To do this, we recall the definition of Petri net:
\begingroup %
\def\thetheoremUnified{\ref{def: Petri net}} %
  \begin{definition}[Petri net]
    A \emph{Petri net} is a quadruple
    \begin{equation*}
      N := \Net{N}
    \end{equation*}
    Where:
    \begin{itemize}
      \item $\Pl{N}$ is a \emph{finite} set, representing places;
      \item $\Tr{N}$ is a \emph{finite} set, representing transitions;
      \item $\Pl{N}$ and $\Tr{N}$ are disjoint: Nothing can be a 
      transition and a place at the same time;
      \item $\Pin{-}{N}:\Tr{N} \to \Msets{\Pl{N}}$ is a function 
      assigning to each transition the multiset of $\Pl{N}$ 
      representing its \emph{input} places;
      \item $\Pout{-}{N}:\Tr{N} \to \Msets{\Pl{N}}$ is a function 
      assigning to each transition the multiset of $\Pl{N}$ 
      representing its \emph{output} places.
    \end{itemize}
    We will often denote with $\Tr{N}, \Pl{N}, \Pin{-}{N}, \Pout{-}{N}$ 
    the set of places, transitions and input/output functions 
    of the net $N$, respectively.
  \end{definition}
\addtocounter{theoremUnified}{-1} %
\endgroup
\noindent
So we see that what we have are places, transitions and 
input/output functions. A suitable notion of morphism 
between nets will have to involve at least some of these objects. 
Consider nets $\Net{N}$ and $\Net{M}$. One of the most 
na\"ive things to do is to send transitions to transitions, 
defining a function $f:\Tr{N} \to \Tr{M}$ representing a 
correspondence between processes of $N$ and processes of $M$. 
Similarly, it makes sense to send places of $N$ to places of $M$, 
that is, to define a function $g:\Pl{N} \to \Pl{M}$. This means 
that the resource types in $N$ will correspond to resource types 
in $M$. 

Notice, though, that since processes consume and produce 
resources in places, $f$ and $g$ have to be somehow connected: 
If $t \in \Tr{N}$ is sent to $f(t) \in \Tr{M}$, then it must be that 
if $p \in \Pin{t}{N}$ (respectively, $p \in \Pout{t}{N}$), then 
$g(p) \in \Pin{g(t)}{M}$ (respectively, $f(p) \in \Pout{g(t)}{M}$), 
otherwise our correspondence will make no sense. This suggests 
that if we define $g$ to be a function $\Msets{\Pl{N}} \to \Msets{\Pl{M}}$ 
then we are able to use $g$ to express the compatibility 
conditions for input/output functions stated above. Moreover, 
we will show how each function  $\Pl{N} \to \Pl{M}$
can be canonically extended to a function $\Msets{\Pl{N}} \to \Msets{\Pl{M}}$, 
proving how this new definition of $g$ generalizes the na\"ive one.

\begin{figure}[!ht]
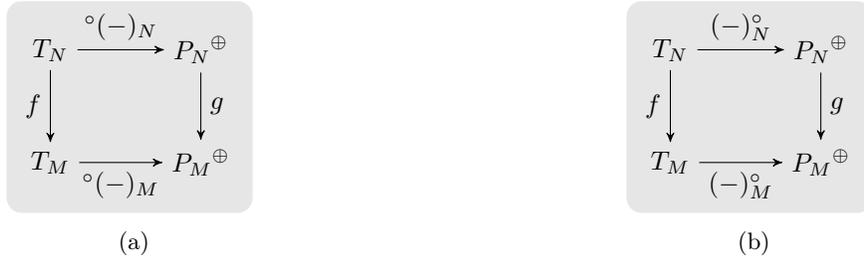

	\centering
	\begin{subfigure}[t]{0.45\textwidth}\centering
		\input{Pictures/ExecutionsOfPetriNets/01-MorphismInput.tex}
		\caption{}
		\label{fig: petri morphism input}
	\end{subfigure}
	\hfill
	\begin{subfigure}[t]{0.45\textwidth}\centering
		\input{Pictures/ExecutionsOfPetriNets/01-MorphismOutput.tex}
		\caption{}
		\label{fig: petri morphism output}
	\end{subfigure}
	\caption{Properties of net morphisms.}
	\label{fig: petri morphism diagrams}
\end{figure}

But is a function $\Msets{\Pl{N}} \to \Msets{\Pl{M}}$ enough 
to get a suitable notion of morphism between nets? Not quite. 
Multisets are not just sets, and the firing rule for multisets 
(recall Definition~\ref{def: Petri firing policy}) is defined in 
terms of multiset difference. This forces us to require additional 
properties if we do not want our correspondence to misbehave 
with respect to the firing rules of $\Net{N}$ and $\Net{M}$. 
The definition we need is the one of \emph{multiset homomorphism}, 
and it is stated below:
\begin{definition}[Multiset homomorphism]
  Consider $\Msets{P}$ and $\Msets{P'}$, the sets of multisets 
  on $P$ and $P'$, respectively. A \emph{multiset homomorphism} 
  is a function $g:\Msets{P} \to \Msets{P'}$ such that
	\begin{equation*}
    g(\Zeromset{P}) = \Zeromset{P'} \qquad g(\Mset{P}_1 \cup \Mset{P}_2) 
      = g(\Mset{P}_1) \cup g(\Mset{P}_2)
	\end{equation*}
  for each $\Mset{P}_1, \Mset{P}_2 \in \Msets{P}$, that is, 
  a multiset homomorphism is a function 
  $g:\Msets{P} \to \Msets{P'}$ that carries the zero multiset 
  to the zero multiset and respects multiset unions.
\end{definition}
\begin{remarkhard}[Multisets homomorphisms are free monoid homomorphisms]
  The reader fluent in algebra, recalling 
  Remark~\ref{rem: multisets are free commutative monoids}, will 
  have noticed that a multiset homomorphism is just a homomorphism 
  of free commutative monoids.
\end{remarkhard}
As we promised, we now show how to lift a function between base sets to a 
multiset homomorphism. Proving that the resulting function sends zero multisets 
to zero multisets and respects multiset union is a straightforward check, 
which we leave as an exercise to the reader.
\begin{proposition}[Extending functions to multiset homomorphisms]
  Let $P, P'$ be sets, and let $g:P \to P'$ be a function. $g$ can be extended to 
  a multiset homomorphism $\bar{g}: \Msets{P} \to \Msets{P'}$ by setting,
  for all $p' \in P'$ and $\Mset{P} \in \Msets{P}$:
  \begin{equation*}
    \bar{g}(\Mset{P}) (p') = \sum_{p\mid g(p) = p'} \Mset{P}(p)
  \end{equation*}
\end{proposition}
\begin{definition}[Grounded homomorphisms]
  If $g$ is a multiset homomorphism coming from a function, meaning 
  that exists a function $h$ such that $g = \bar{h}$, then we say that 
  $g$ is \emph{grounded}.
\end{definition}
The definition of multiset homomorphism allows us to carry 
the input (output, respectively) function  of a net $N$ to the 
input (output, respectively) function of a net $M$ in 
a way that respects the firing rules of both nets. We are ready to give the 
definition we were seeking:
\begin{definition}[Petri net morphisms]\label{def: Petri morphism}
  Consider the Petri nets $\Net{N}$ and $\Net{M}$. A 
  \emph{morphism of Petri nets} $M \to N$ is specified by a 
  pair $\langle f,g \rangle$ where:
	\begin{itemize}
    \item $f$ is a function $\Tr{N} \to \Tr{M}$;
    \item $g$ is a multiset homomorphism 
    $\Msets{\Pl{N}} \to \Msets{\Pl{M}}$;
    \item Diagrams in Figure~\ref{fig: petri morphism diagrams} commute.
	\end{itemize}
\end{definition}
This definition neatly packs all the discussion above, 
and the diagrams in Figure~\ref{fig: petri morphism diagrams} 
represent the idea that transitions in $\Net{N}$ and transitions 
in $\Net{M}$ are organized in a compatible way.
\begin{remark}[Net morphisms are simulations]
  One way to interpret a morphism of nets is in 
  terms of \emph{simulations}. Since a morphism 
  of nets $N \to M$ is made of a couple of \emph{functions}, 
  multiple transitions (multisets, respectively) of $N$ can correspond 
  to the same transition (multiset, respectively) of $M$. This means 
  that transitions and places of $M$ hit by the morphism act as placeholders 
  for transitions and places of $N$, and we can interpret this as if the 
  process represented by $M$ contains a subprocess simulating the one 
  represented by $N$.
\end{remark}
It is easy to see that, for each net $N$, there is a pair 
$\langle \Id{\Tr{N}}, \Id{\Msets{\Pl{N}}} \rangle$ that sends everything 
to itself. Similarly, if $\langle f,g \rangle: N \to M$ and 
$\langle f',g' \rangle : M \to L$ are net homomorphisms, then 
$\langle f;f', g;g' \rangle$ is a net homomorphism $N \to L$, and morphism 
composition is associative. Then we can define:
\begin{definition}[The category $\Petri$]
  We define the category $\Petri$ as having Petri nets as objects 
  and morphisms between them as morphisms.
\end{definition}
We can refine the category $\Petri$ further: Noticing that if $\bar{g}, \bar{h}$ 
are grounded multiset homomorphisms so is their composition 
$\bar{g};\bar{h} = \overline{g;h}$, and that the identity homomorphism is 
always grounded, we can give the following definition:
\begin{definition}[The category $\PetriGrounded$]
  \label{def: the category petrigrounded}
  We define the category $\PetriGrounded$ as having Petri nets as objects 
  and morphisms between them of the form $\langle f, \bar{g} \rangle$.

  $\Petri$ and $\PetriGrounded$ have the same objects. 
  Since every grounded multiset homomorphism is obviously a 
  multiset homomorphism, every net morphism in $\PetriGrounded$ 
  is also a morphism in $\Petri$, but the opposite is not true. 
  We say that $\PetriGrounded$ is a \emph{subcategory} of $\Petri$.
\end{definition}
Now that we managed to organize our nets into a category, it is time 
to get to the next step.
\section{The Execution of a net}\label{sec: the execution of a net}
Let us focus on the transitions of a net. There are fundamentally 
two ways in which transitions can interact:
\begin{itemize}
  \item The firing of one transition is independent from the firing 
  of the other (e.g. transitions $t_1, t_2$ in Figure~\ref{fig: petri evolution});
  \item The firing of a transition depends on the firing of the 
  other (e.g. transitions $t_1, t_3$ in Figure~\ref{fig: petri evolution}).
\end{itemize}
This should clearly suggest that a \emph{monoidal category} 
(recall Definition~\ref{def: monoidal category}) is the structure we 
want to represent transition firings, since it comes with a notion of 
sequential and parallel composition, representing presence or 
absence of interaction between transitions. Let us see how we can 
use symmetric monoidal categories to represent an execution.

Consider a net $N$, and a monoidal category $\CategoryC$ such 
that each place of $N$ corresponds to an object in $\CategoryC$. To 
avoid clutter, we will denote the places and the objects they correspond 
to in the same way. Consider then a place $p \in \Pl{N}$. The object 
$p$ can be thought of as representing a token in $p$. Using the tensor 
product, we can iterate this: $p \Tensor p$ represents two tokens in 
$p$; $p \Tensor p \Tensor p$ represents three tokens in $p$, and so on. 
Similarly, if we have two places $p,q \in \Pl{N}$, then $p \Tensor q$ 
stands for one token in $p$ and one in $q$.

\begin{figure}[!hb]
	\centering
	\input{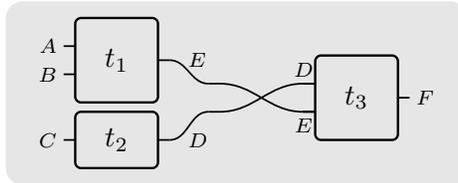}
	\caption{Swaps are necessary to do the required bookkeeping.}
	\label{fig: petri production switch example}
\end{figure}

Notice that, according to this idea, $p \Tensor q \Tensor p$ and 
$p \Tensor p \Tensor q$ both represent having two tokens in $p$ 
and one in $q$: Tokens are just being considered in a different order 
and then we should have a way to go from one object to the other. 
This means that we want our category to be \emph{symmetric} (recall 
Definition~\ref{def: symmetric monoidal category}).
\begin{remarkhard}[Frictions between nets and categories]
  The fact that monoidal products of objects in a symmetric monoidal category 
  are not in general commutative, while multisets 
  (used to express input, output and markings of Petri nets) are, 
  is the main point of friction in defining the correspondence 
  between nets and categories. This point has been tackled in many different ways 
  in the literature, for instance by imposing commutativity of monoidal products 
  -- as in~\cite{Meseguer1990,Baez2018}, by defining the monoidal products 
  as not commutative and interlinking them via natural 
  transformations -- as in~\cite{Sassone1995}, or by weakening the definition of 
  Petri net -- as in~\cite{Baldan2003}. Our own approach -- defined
   in~\cite{Genovese2019}, is mainly concerned with defining something 
   which is sensible, easy to implement and computationally efficient. We will 
   follow~\cite{Genovese2019} in the remainder of the Chapter.
\end{remarkhard}
If we fully embrace the idea that transitions are processes 
carrying resources into other resources, it is natural to say that \emph{a transition 
$t \in \Tr{N}$ corresponds to a morphism $t:\Pin{t}{N} \to \Pout{t}{N}$ 
in $\CategoryC$.} But this means that sequences of transitions are now just 
string diagrams! This has huge benefits, since using a string diagram we 
can represent which transition is consuming which tokens, and observing 
the wiring we can reconstruct how a single token is processed. Since we do 
not care about how we bracket parallel composition, it is clear that we want 
our category to be also strict (recall 
Definition~\ref{def: strict monoidal category}).
\begin{figure}[ht!]
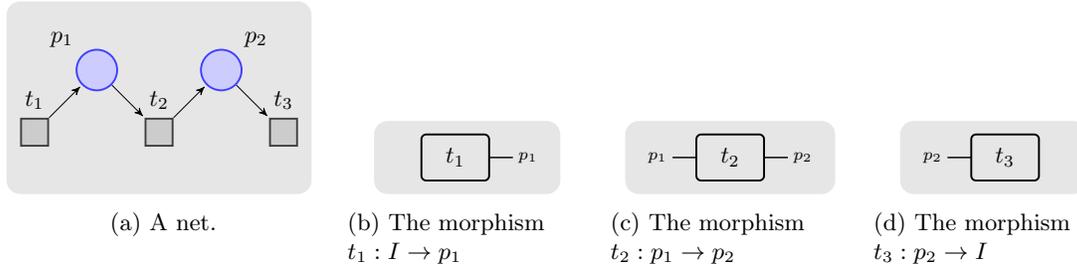

	\centering
	\begin{subfigure}[t]{0.30\textwidth}\centering
		\input{Pictures/ExecutionsOfPetriNets/02-ProductionExample.tex}
		\caption{A net.}
		\label{fig: petri production example net}
	\end{subfigure}
	\hfill
	\begin{subfigure}[t]{0.22\textwidth}\centering
		\input{Pictures/ExecutionsOfPetriNets/02-ProductionExampleT1.tex}
		\caption{The morphism\\ $t_1:\TensorUnit \to p_1$}
		\label{fig: petri production example t1}
	\end{subfigure}
	\hfill
	\begin{subfigure}[t]{0.22\textwidth}\centering
		\input{Pictures/ExecutionsOfPetriNets/02-ProductionExampleT2.tex}
		\caption{The morphism\\ $t_2:p_1 \to p_2$}
		\label{fig: petri production example t2}
	\end{subfigure}
	\hfill
	\begin{subfigure}[t]{0.22\textwidth}\centering
		\input{Pictures/ExecutionsOfPetriNets/02-ProductionExampleT3.tex}
		\caption{The morphism\\ $t_3:p_2 \to \TensorUnit$}
		\label{fig: petri production example t3}
	\end{subfigure}
	\caption{A net and its morphisms.}
	\label{fig: petri production example}
\end{figure}
\begin{example}
  Consider the net in Figure~\ref{fig: petri production example net}. 
  Its transitions can be represented as the morphisms of a strict symmetric 
  monoidal category as in Figures~\ref{fig: petri production example t1},
  \ref{fig: petri production example t2} 
  and~\ref{fig: petri production example t3} and all the possible string 
  diagrams, as, for instance, the ones in 
  Figure~\ref{fig: petri production example possible firings}, represent 
  sequences of transition firings. We also see how the monoidal unit, that 
  is not drawn in the pictures according to our convention 
  (see Section~\ref{sec: string diagrams}), is useful to represent transitions 
  with no inputs and/or no outputs. Each set of vertically aligned wires in 
  the diagram represents a state of the net, and transitions carry states into 
  states. Note how this allows us to completely disambiguate the problem of 
  distinguishing tokens. For instance, in 
  Figure~\ref{fig: petri production example possible firings 2}, $t_1$ 
  produces a token in $p_1$, and $t_2$ consumes a token from $p_1$ as 
  well. But the diagram states clearly how these tokens are not the same: 
  $t_2$ is consuming a token that was already present in $p_1$ before $t_1$ fired.
\end{example}
\begin{figure}[!ht]
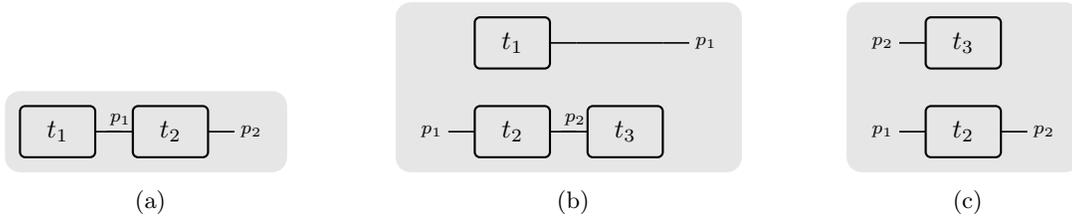

	\centering
	\begin{subfigure}[t]{0.30\textwidth}\centering
		\input{Pictures/ExecutionsOfPetriNets/03-ProductionExampleString1.tex}
		\caption{}
	\end{subfigure}
	\hfill
	\begin{subfigure}[t]{0.4\textwidth}\centering
		\input{Pictures/ExecutionsOfPetriNets/03-ProductionExampleString2.tex}
		\caption{}\label{fig: petri production example possible firings 2}
	\end{subfigure}
	\hfill
	\begin{subfigure}[t]{0.25\textwidth}\centering
		\input{Pictures/ExecutionsOfPetriNets/03-ProductionExampleString3.tex}
		\caption{}
	\end{subfigure}
  \caption{Some of the possible sequences of firings for the 
    net in Figure~\ref{fig: petri production example}}
	\label{fig: petri production example possible firings}
\end{figure}
\begin{remark}[Categories must be strict symmetric monoidal]
  We realize quite quickly why we need our category 
  to be strict (Def.~\ref{def: strict monoidal category}) 
  symmetric (Def.~\ref{def: symmetric monoidal category}) 
  monoidal (Def.~\ref{def: monoidal category}): Observe 
  Figure~\ref{fig: petri production switch example}. It is not 
  important, at this stage, to know to which net this diagram 
  may be referring to. We start from the state $A \Tensor B \Tensor C$, 
  meaning that we have one token in $A$, one token in $B$ and one 
  token in $C$. Then transitions $t_1$ and $t_2$ fire, and we can 
  represent this event as simultaneous since the two transitions have 
  nothing to do with each other, hence firing priority doesn't matter 
  in this situation. What matters is that the state produced is 
  $E \Tensor D$, so one token in $D$ and one in $E$. Now, 
  transition $t_3$ has to fire, but it is expecting a state $D \Tensor E$: 
  This is morally the same thing, one token in $D$ and one token 
  in $E$, but since category theory distinguishes between these two 
  objects, we have to introduce a swapping morphism to make this 
  composition possible.
\end{remark}
\begin{remark}[Firing sequences cannot 
  correspond to string diagrams uniquely]
  \label{rem: firing sequences cannot 
  correspond to string diagrams uniquely}
  Looking at our string diagrams closely, we get quickly 
  aware of the fact that \emph{firing sequences cannot 
  correspond to string diagrams uniquely}: If the category 
  describes all the possible ways to execute the net, it is clear 
  that the same firing sequence can correspond to different 
  diagrams. This is consistent with the idea that the category 
  provides additional information that the net cannot capture, 
  namely token histories. To see this, consider 
  Figures~\ref{fig: petri evolution}. The string diagrams in 
  Figure~\ref{fig: petri production example different executions} are 
  all legitimate executions describing the same sequence of firings.
\end{remark}
\begin{figure}[ht!]
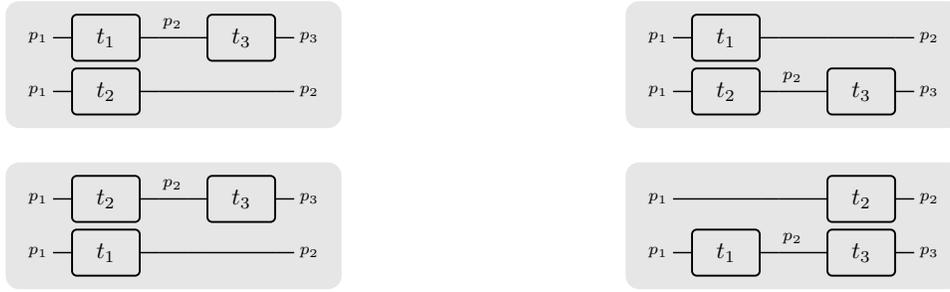

	\centering
	\begin{subfigure}[t]{0.45\textwidth}\centering
		\input{Pictures/ExecutionsOfPetriNets/05-EvolutionDiagram1.tex}
	\end{subfigure}
	\hfill
	\begin{subfigure}[t]{0.45\textwidth}\centering
		\input{Pictures/ExecutionsOfPetriNets/05-EvolutionDiagram2.tex}
	\end{subfigure}
	\par\bigskip
	\begin{subfigure}[t]{0.45\textwidth}\centering
		\input{Pictures/ExecutionsOfPetriNets/05-EvolutionDiagram3.tex}
	\end{subfigure}
	\hfill
	\begin{subfigure}[t]{0.45\textwidth}\centering
		\input{Pictures/ExecutionsOfPetriNets/05-EvolutionDiagram4.tex}
	\end{subfigure}
  \caption{Some of the possible sequences of firings for the net 
    evolution in Figures~\ref{fig: petri production example}}
	\label{fig: petri production example different executions}
\end{figure}
Now that we grasped how to represent computations of a net 
categorically, at least at an intuitive level, we are ready to 
formalize this. Up to now, in fact, we just stated how the category 
of executions should look like, but we did not build it. There are many 
categories which may be good candidates to represent a net execution, 
so how do we pick one? We will find out in the next Section, while we 
conclude the present one with a summary of what we learned so far.
{\small
	\begin{center}
		\begin{tabular}{ c | c | c }
      \hline
      Places & correspond to & Objects \\
      States & correspond to & Monoidal products of objects (places) \\
      Transitions & correspond to & Morphisms \\
			Firing sequences & correspond to & String diagrams \\
			\hline
		\end{tabular}
	\end{center}
}
\section{Free strict symmetric monoidal categories (FSSMCs)}
  \label{sec: free strict symmetric monoidal categories}
Given a net $N$, we want to generate a category 
representing all the possible ways to execute $N$ and 
we know, thanks to the previous Section, that this category 
has to be strict symmetric monoidal to allow us to do sequential, 
parallel composition and all the necessary bookkeeping given by 
swapping tokens around. But there are many different strict, symmetric 
monoidal categories out there, so which one do we need? The notion 
we need is the one of \emph{free, strict symmetric monoidal category}, 
and we will dedicate this section to defining it.
\begin{definition}[Strings generated by a set]
  Let $S$ be a set. We denote the \emph{set of strings of finite length 
  of elements in $S$} as $\Strings{S}$:
	\begin{equation*}
    \Strings{S} := \Suchthat{s_1 s_2 \dots s_n}{n \in \Naturals 
      \wedge \forall i (s_i \in S)}
	\end{equation*}
\end{definition}
We can think about $\Strings{S}$ also as 
\emph{the set of all possible monoidal products of elements in $S$}: 
If $S$ denotes a set of objects in a monoidal category, 
then we can interpret a string $s_1\dots s_n$ as 
$s_1 \Tensor \dots \Tensor s_n$: String concatenation 
stands for monoidal product, and the empty string stands 
for the monoidal unit, which is consistent with the idea 
that $s \Tensor \TensorUnit = s$ for each object $s$. From 
this we can infer a very important fact, that will be useful 
later on:
\begin{remark}[$S \in \Obj{\CategoryC}$ implies 
  $\Strings{S} \in \Obj{\CategoryC}$]
    \label{rem: mapping generators maps strings}
  If $S$ is a set, and we map each element of $S$ to an 
  object in a strict symmetric monoidal category, then each 
  element of $\Strings{S}$ can be mapped in that category 
  too. This is obvious, since a monoidal category is closed 
  for monoidal products.
\end{remark}
\begin{remarkhard}[$\Strings{S}$ is free]
    \label{rem: strings are free monoids}
    Remark~\ref{rem: mapping generators maps strings} can 
    be stated in a bit more high-level 
    way by saying that $\Strings{S}$ is the free \emph{non-commutative} 
    monoid generated by $S$ (compare this with 
    Remark~\ref{rem: multisets are free commutative monoids}): Since 
    the monoid of objects of any monoidal category is a monoid in the 
    classical sense, Remark~\ref{rem: mapping generators maps strings} 
    just states the free property of $\Strings{S}$.
\end{remarkhard}
If we think of $S$ as the set of places of a net, and we want 
to define a category $\CategoryC$ representing all the possible 
ways to execute it, then it makes sense to require that $S$, 
and all the possible monoidal products of elements in $S$, 
are objects of $\CategoryC$. This is because we know, from 
Section~\ref{sec: the execution of a net},  
that we are going to model states of a 
net as monoidal products of tokens, represented by the place 
they live in. From this we realize that $\Strings{S}$ seems to 
be a good candidate to define $\Obj{\CategoryC}$: In this case, 
the objects of $\CategoryC$ are \emph{just} the states or, to be 
precise, all the possible ways to enumerate states.

What else do we need to get a strict symmetric monoidal 
category? Not much: Clearly we need \emph{identites and symmetries}: 
Identity morphisms are necessary if we want to define a category, 
because the axioms require it. Similarly, the presence of symmetries 
is required by axioms for symmetric monoidal categories. 
These and all their possible compositions are the only morphisms 
that are ``obligatory'', according to the axioms.

But this is not the end of the story: If we think in terms of string diagrams 
up to this point the only thing we have are wires, tangled in any 
possible way we can think about. But where are the boxes?

Boxes can be thought of as just morphisms which are neither 
identities nor symmetries. To get them into the picture, we need to 
specify some \emph{generating morphisms}.

 We can give, then, the following definition:
\begin{definition}[Free strict symmetric monoidal category]
  \label{def: free strict symmetric monoidal category}
  Let $S$ be a set, and let $T$ be a set of triples 
  $(\alpha, r, s)$ with $r, s \in \Strings{S}$. 
  
  A \emph{free strict symmetric monoidal category (abbreviated 
  FSSMC) generated by $S$ and $T$} is a symmetric monoidal category 
  whose monoid of objects is $\Strings{S}$, and whose morphisms 
  are generated by the following introduction rules:
  \begin{gather}
		\frac{s \in \Strings{S}}{\Id{s}: s \to s}
      \qquad 
      \frac{r,s \in \Strings{S}}{\sigma_{r,s}: {r \Tensor s} \to {s \Tensor r}} 
      \qquad
      \frac{(\alpha, r,s) \in T}{\alpha: r \to s}
        \label{eq: fssmc generators rules} \\
    	\nonumber\\
    \frac{\alpha: A \to B, \,\, \alpha': A' \to B'}
      {\alpha \Tensor \alpha':  {A \Tensor A'} \to {B \Tensor B'}} 
      \qquad 		
      \frac{\alpha: A \to B, \,\, \beta: B \to C}{\alpha;\beta:  A \to C} 
        \label{eq: fssmc composition rules}
	\end{gather}
  Morphisms are quotiented by the following equations, 
  for $\alpha: A \to B$, $\alpha': A' \to B'$, 
  $\alpha'': A'' \to B''$, $\beta: B \to C$, $\beta':B' \to C'$, 
  $\gamma: C \to D$:
	\begin{align}
		\alpha ; \Id{B} = &\,\,\alpha = \Id{A} ; \alpha & 
      \quad  (\alpha;\beta);\gamma &= \alpha;(\beta;\gamma)
        \label{eq: fssmc axioms category}\\
    \Id{\TensorUnit} \Tensor \alpha = &\,\,\alpha = 
      \alpha \Tensor \Id{\TensorUnit}& 
      \quad  (\alpha \Tensor \alpha') \Tensor \alpha'' 
      &= \alpha \Tensor (\alpha' \Tensor \alpha'')
        \label{eq: fssmc axioms monoidal 1}\\
		\Id{A} \Tensor \Id{A'} &= \Id{A \Tensor A'} & 
      \quad  (\alpha \Tensor \alpha') ; (\beta \Tensor \beta') 
      &= (\alpha ; \beta) \Tensor (\alpha' ; \beta')
        \label{eq: fssmc axioms monoidal 2}\\
    \sigma_{A, A' \Tensor A''} = (\sigma_{A,A'} 
      &\Tensor \Id{A''}); (\Id{A'} \Tensor \sigma_{A,A''}) &
      \quad  \sigma_{A,A'};\sigma_{A',A} &= \Id{A \Tensor A'}
        \label{eq: fssmc axioms symmetric 1}\\
    \sigma_{A,A'};(\alpha' \Tensor \alpha) 
      &= (\alpha \Tensor \alpha');\sigma_{B,B'} &
      \quad  \sigma_{A, I} = &\,\,\Id{A} = \sigma_{I, A}
      \label{eq: fssmc axioms symmetric 2}
	\end{align}
  We say that a category $\CategoryC$ is a FSSMC when it is 
  generated by some $S$ and $T$.
\end{definition}
This is a big, meaty definition, so let us try to unpack it. Objects, 
as we said, correspond to all the possible ways to represent 
a state, following the intuition from 
Section~\ref{sec: the execution of a net}. Again, we 
define $\Tensor$ as string concatenation on the objects. 
Now, the morphisms: Rules are read top to bottom: Every 
time a condition expressed above the line is realized, the 
condition expressed below it is inferred. 
Rules~\ref{eq: fssmc generators rules} tell us that for each 
object $s$ we get a morphism $\Id{s}: s \to s$ that 
will -- unsurprisingly -- be our identity morphism on $s$.
Similarly, for each couple of objects $r,s$, we 
get a morphism $\sigma_{r,s}: r \Tensor s \to s \Tensor r$, that 
will be our symmetry. Then, there is the rightmost rule 
in~\ref{eq: fssmc generators rules}, 
which is perhaps the most mysterious one: The point here is that 
each triple in $T$ can be interpreted as consisting of a \emph{label} -- 
denoted with a Greek letter and representing a generating morphism -- 
and of a couple of objects, representing the morphism source and target.
Our ``mysterious'' rule just realizes this interpretation, by saying that 
for each triple $(\alpha, r, s)$ in $T$ we need to introduce a morphism 
$\alpha: r \to s$ in the category.

The rules in~\ref{eq: fssmc composition rules} 
deal with populating our category with sequential and monoidal 
compositions. As things stand now, we know that, say, generating 
morphisms are part of our category, but we never said anything about 
their compositions: This has to be, in fact, stated explicitly. 
The rules in~\ref{eq: fssmc composition rules} just say that 
whenever we have two morphisms, 
their sequential (when the morphisms are compatible) 
and parallel compositions must be part of the category as well.
Starting from generating morphisms, identities and symmetries and
iterating these rules one quickly becomes aware of how any morphism 
in our category is just a big composition -- parallel and 
sequential -- of identities, symmetries and generators.

To understand the second part of the definition, notice this: 
Up to now, we get a morphism $\Id{s}$ for each object $s$, 
but nothing assures us that this morphism behaves as an identity: 
We have $\Id{s}$ because we defined a rule that formally introduces 
it, but the rule does not say anything about its behavior. The behavior 
has to be formally imposed by identifying morphisms with each other. 
Axioms in~\ref{eq: fssmc axioms category} are the necessary 
identification to obtain a category, since they entail identity and 
associativity laws to hold. Axioms in~\ref{eq: fssmc axioms monoidal 1} 
and~\ref{eq: fssmc axioms monoidal 2} 
entail that our category is monoidal, while 
axioms~\ref{eq: fssmc axioms symmetric 1} 
and~\ref{eq: fssmc axioms symmetric 2} imply that
it is moreover symmetric.
\begin{definition}[The category of symmetries]
  We denote with $\Sym{S}$ the FSSMC generated by a set of 
  generating objects $S$, while the set of generating morphisms $T$ 
  is empty.
\end{definition}
\begin{remark}[Axioms and rules are all necessary]
  Note that all the axioms and rules in 
  Definition~\ref{def: free strict symmetric monoidal category} are 
  necessary: If we strip out only one of these ingredients then it 
  is not possible anymore to prove that our categories are strict 
  symmetric monoidal.
\end{remark}
\begin{remark}[$\Sym{S}$ is special among FSSMCs]
  Elaborating further on the last Remark, $\Sym{S}$ is the most 
  general free strict monoidal category containing $S$ among 
  its objects, since it has no superfluous components of any kind. 
  This generality is also proved by the following property, 
  called \emph{freeness}.
\end{remark}
\begin{remark}[Freeness]\label{rem: freeness}
  The reason why we call FSSMCs ``free'' is because they satisfy 
  only the bare minimum amount of equations to be a strict symmetric 
  monoidal category.

  In an arbitrary strict symmetric monoidal category, 
  for instance, there could be some other equations that are satisfied, 
  e.g. it could be that $f;g = h$ for some morphisms $f,g,h$. This is not 
  the case for FSSMCs.

  If $\CategoryC$ is a strict symmetric monoidal category, 
  then for each function $f: S \to \Obj{\CategoryC}$ \emph{there 
  is a unique strict symmetric monoidal functor 
  $F:\Sym{S} \to \CategoryC$ extending $f$}, meaning that, 
  for each $s \in S$, $Fs = f(s)$. Observe how this amounts to 
  a generalization of Remark~\ref{rem: mapping generators maps strings}
  from objects to the entire category.

  Similarly, given a strict symmetric monoidal category 
  $\CategoryC$ and a FSSMC generated by $S$ and $T$, 
  specifying a mapping between elements of $S$ and objects of 
  $\CategoryC$ and a mapping from generating morphisms in $T$ and 
  morphisms of $\CategoryC$ is enough to extend such correspondence 
  into a functor.
\end{remark}
Freeness is a great property, because it says that every time we map 
elements of $S$ and $T$ to the objects and morphisms of a strict 
symmetric monoidal category, all the structure 
of the FSSMC ``follows along'' by 
means of a uniquely determined functor. In this sense, the strict 
symmetric monoidal structure of any FSSMC is completely determined 
by $S$ and $T$.
\subsection{The category \texorpdfstring{$\FSSMC$}{FSSMC}}
  \label{subsec: the category fssmc}
Now that we defined what free strict symmetric monoidal categories 
are, we proceed by doing what any category theorist would do: We 
organize these categories into a category!

This seems difficult and counterintuitive, but it is not: We already know 
that functors can be thought of as ``morphisms between categories'', 
so our natural choice would be to define a category where FSSMCs are
the objects, and functors between them are the morphisms. 

Since our categories are strict symmetric monoidal, it makes sense to 
ask morphisms between them to preserve this structure, that is, to restrict 
to strict, symmetric monoidal functors 
(Definition~\ref{def: symmetric strong strict monoidal functors}).

Alas, this is still not enough: We know that all FSSMCs are also free, 
and we would like our morphisms to ``preserve'' this as 
well, whatever this means. To cast a 
decent definition, we first have to isolate peculiar morphisms 
existing in any FSMMC.
\begin{definition}[Symmetries]
  \label{rem: symmetries}
  Given a FSSMC $\CategoryC$, we will abuse notation and call 
  \emph{symmetry} any morphism 
  obtained by composing sequentially and monoidally identities and 
  symmetries: A symmetry is just a tangle of wires, 
  see Figure~\ref{fig: symmetry example}.
\end{definition}
\begin{figure}[!ht]
	\centering
	\input{Pictures/ExecutionsOfPetriNets/06-Symmetry.tex}
	\caption{A symmetry in a FSSMC: $(
    \sigma_{A,B} \Tensor \Id{C} ); 
    (\Id{B} \Tensor \sigma_{A,C});
    (\sigma_{B,C} \Tensor \Id{A} )
     $.}
	\label{fig: symmetry example}
\end{figure}
\begin{remark}[$\Sym{S}$ is just wires!]
  Considering our last definition, we get quickly aware of how 
  $\Sym{S}$ is special among all the FSSMCs generated by $S$: 
  It has no boxes, and so we can think of it as a category made 
  entirely of wires!
\end{remark}
\begin{remarkhard}[Symmetries are images of $\Sym{S}$]
  Given a FSSMC $\CategoryC$ generated by $S$ and $T$, we know 
  from Remark~\ref{rem: freeness} that freeness defines a functor 
  $F: \Sym{S} \to \CategoryC$ from the identity function $S \to S$.
  Symmetries in the sense of Remark~\ref{rem: symmetries} are just 
  all the morphisms of $\CategoryC$ which are images of morphisms in 
  $\Sym{S}$ through $F$.
\end{remarkhard}
In our definition of FSSMC it is clear that what really 
matters are the generating morphisms, which are the ``boxes'' of our 
category, whereas symmetries are always there ``by default''. A nice 
definition for a functor between FSSMCs then is the following:
\begin{definition}[Generator-preserving functors]
  Let $\CategoryC$ be a FSSMC generated by $S_\CategoryC$, 
  $T_\CategoryC$, and $\CategoryD$ be a FSSMC generated by 
  $S_\CategoryD$ and $T_\CategoryD$. A \emph{generator-preserving 
  functor between $\CategoryC$ and $\CategoryD$} is a strict 
  symmetric monoidal functor $F:\CategoryC \to \CategoryD$ such that 
  each generating morphism $\alpha$ of $\CategoryC$ is mapped to 
  $\sigma;\beta;\sigma'$, with $\beta$ generating morphism of 
  $\CategoryD$ and $\sigma, \sigma'$ symmetries in $\CategoryD$.
\end{definition}
As usual, let us unpack this definition. A generator-preserving functor 
just maps generating morphisms to generating morphisms, but it is also 
allowed to ``scramble their inputs and outputs a bit'' by pre- and 
post-composing with symmetries. This definition is well behaved, and 
in fact:
\begin{lemma}[Generator-preserving functors are well-behaved]
  The result of the composition of generator-preserving functors is a 
  generator-preserving functor. Given a FSSMC $\CategoryC$, the 
  identity functor $\Id{\CategoryC}$ is generator-preserving.
  The proof of these statements can be found 
  in~\cite[Prop.2]{Genovese2019}.
\end{lemma}
This is enough to prove that FSSMCs and generator-preserving 
functors between them form a category, and we have:
\begin{definition}[The category $\FSSMC$]
  We denote with $\FSSMC$ the category having free strict 
  symmetric monoidal categories as objects and generator-preserving 
  functors as morphisms.
\end{definition}
Finally, as we did in Definition~\ref{def: the category petrigrounded} 
we can specialise $\FSSMC$ further. In fact, we did not impose any 
requirement on how a generator-preserving functor has to behave 
on objects: If $\CategoryC$ and $\CategoryD$ are FSSMCs and 
$F$ is a functor between them, then any object of $\CategoryC$ can 
be mapped to any object of $\CategoryD$ as long as monoidal 
composition is preserved. Since we know that generating objects are 
somehow special among objects in a FSSMC, we can use them 
to refine our definition further obtaining the following statement, 
the proof of which can again be found in~\cite[Prop.2]{Genovese2019}.
\begin{lemma}[The category $\FSSMCGrounded$]
  \label{lem: the category fssmcgrounded}
  A generator-preserving functor is called \emph{grounded} if it maps 
  generating objects to generating objects. Composition of grounded 
  functors is grounded, and the identity functor on any FSSMC is 
  grounded. Hence FSSMCs and grounded functors between them 
  form a subcategory of $\FSSMC$, denoted with $\FSSMCGrounded$.
\end{lemma}
\section{The categories \texorpdfstring{$\Fold{N}$}{Fold(N)} 
  and \texorpdfstring{$\UnFold{\CategoryC}$}{UnFold(C)}}
  \label{sec: the category foldn}
We finally built all the theory needed to link Petri nets and FSSMCs. 
The only bits missing are provided by the following definitions.
\begin{definition}[Multiplicity]
  Let $S$ be a set. There is an obvious 
  mapping $\Multiplicity{S}: \Strings{S} \to \Msets{S}$, called 
  \emph{multiplicity}, that 
  associates to each string $str \in \Strings{S}$ a multiset 
  $S \to \Naturals$ by ``counting occurrencies'':
  \begin{equation*}
		\Multiplicity{S}(str)(s) := \text{ Occurrences of $s$ in $str$ }
	\end{equation*}
\end{definition}
\begin{remarkhard}[$\Multiplicity{S}$ ia a monoid homomorphism]
  For each set $S$, $\Multiplicity{S}$ is a homomorphism of 
  monoids.
\end{remarkhard}
What $\Multiplicity{S}$ does is very simple:
Given a string on $S$ and an element in $S$,  it counts 
how many times the element occurs in the string.  Multiplicity is 
instrumental in defining string ordering:
\begin{definition}[Ordering]
  \label{def: ordering}
  Given a set $S$, an \emph{ordering function on $S$} is a function
  $\Ordering{S}: \Msets{S} \to \Strings{S}$ such that 
  $\Ordering{S};\Multiplicity{S} = \Id{\Msets{S}}$.
\end{definition}
As we know, multisets are just sets with repetition, and do not come 
endowed with any notion of ordering. On the contrary, strings are 
sensitive to element positioning. This information can be ``canonically 
forgotten'', meaning that there is essentially just one way to 
map a string to a multiset. This is what $\Multiplicity{S}$ does.
On the other hand, there are many different ways to ``linearize'' a 
multiset into a string, hence many different choices of 
$\Ordering{S}$.
\begin{examplehard} [Orderings are not canonical]
  Note that the ``lack of canonicity'' of $\Ordering{S}$ is also 
  reflected in the fact that, whereas $\Multiplicity{S}$ is a monoid 
  morphism, $\Ordering{S}$ is not. To see this, assume $S$ is the set 
  of Latin letters, and define $\Ordering{S}$ as the function that maps 
  any set with repetition of Latin letters to the string where they are 
  alphabetically ordered. In this case, multisets  $\{c,b,c,b\}$ and 
  $\{a,b,b\}$ get mapped to strings $bbcc$ and $abb$, respectively, 
  but the union $\{c,b,c,b\} \cup \{a,b,b\}$ is mapped to the 
  string $abbbbcc$, which is not the concatenation of $bbcc$ 
  with $abb$.
\end{examplehard}
We are now ready to put Definition~\ref{def: ordering} to good use, 
finally formalizing what we are interested in, but with just one caveat:
As we said, there are many different ordering functions we can choose 
on a given base set. From now on, we will postulate that each Petri net 
comes endowed with an ordering function on its set of places, that is, 
\begin{remark}[Assumption: Nets are ordered]
  \label{rem: assumption nets are ordered}
  From now on, we will assume that for each Petri net 
  $\Net{N}$ there is a fixed function $\Ordering{\Pl{N}}$.
\end{remark}
\begin{remark}[Ordered nets do not blow up our theory]
  It is crucial to stress that the assumption in 
  Remark~\ref{rem: assumption nets are ordered} does not require to 
  change the theory developed insofar in any way. In fact, we can 
  give a formal definition of "ordered net" and prove that ordered nets 
  form a category which is equivalent to $\Petri$. Details can be found 
  in~\cite[Appendix]{Genovese2019}.
\end{remark}
\begin{remarkhard}[Ordered nets are computationally friendly]
  Even more importantly, the requirement in 
  Remark~\ref{rem: assumption nets are ordered} does not cause any 
  implementation problem: In a functional programming environment, 
  a Petri net can be implemented by giving a \texttt{place} type, 
  a \texttt{transition} type and a couple of functions defining inputs and 
  outputs respectively (these can be terms of some \texttt{input} and 
  \texttt{output} types, respectively, if one prefers). All such types are 
  then tied together in a structure called \emph{record}.
  In such setting, the only change to make to implement our assumption 
  is that the \texttt{place} type is orderable. Having done this, there is a 
  canonical procedure to define an ordering function on the set of places.
  Details are again to be found in~\cite[Appendix]{Genovese2019}.
\end{remarkhard}
\begin{definition}[The category $\Fold{N}$]
    \label{def: the category foldn}
  Let $N := \Net{N} \in \Obj{\Petri}$. We define $\Fold{N}$ 
  (called \emph{the category of executions of $N$}) to be the 
  FSSMC generated as in 
  Definition~\ref{def: free strict symmetric monoidal category}, with
  $\Pl{N}$ as the set of generating places and 
  \begin{equation*}
    T := 
      \Suchthat{(t, 
                         \Ordering{\Pl{N}}(\Pin{t}{N}),
                         \Ordering{\Pl{N}}(\Pout{t}{N})
                        )}{t \in \Tr{N}}
  \end{equation*}
\end{definition}
This time, many parts of the definition are familiar. As we already 
sketched in Section~\ref{sec: the execution of a net}, we use the places 
of the net to generate the objects of an FSSMC. On morphisms, we 
use the ordering function $\Ordering{}$ to sort the input and output 
places of each transition, and use these as generating morphisms.
We call the category $\Fold{N}$ the category of executions of $N$, 
and its morphisms \emph{executions} or \emph{histories} of $N$ 
because, unsurprisingly, $\Fold{N}$ realizes our desiderata sketched 
out in Section~\ref{sec: the execution of a net}.
To further ensure that our definition is a sound one, a nice thing to 
have would be the possibility to ``go back'' from executions to the nets 
they represent. This should in principle be possible, since different nets 
will surely have different categories of executions, so no information 
should be lost when using $\Fold{-}$. This should guarantee that 
such information can be recovered when going in the opposite direction.
This is indeed the case, and Definition~\ref{def: the category foldn} 
is invertible. In fact:
\begin{definition}[The category $\UnFold{\CategoryC}$]
  \label{def: the category unfoldn}
  Let $\CategoryC$ be a FSSMC generated by $S$ and $T$. Define the 
  Petri net $\UnFold{\CategoryC}$ as follows:
  \begin{itemize}
  \item $\Pl{\UnFold{\CategoryC}} = S$;
   \item $\alpha \in \Tr{\UnFold{\CategoryC}}$ 
    if and only if $(\alpha, r,s) \in T$ for some $r, s$;
   \item  If $\alpha \in \Tr{\UnFold{\CategoryC}}$
    and so $(\alpha, r,s) \in T$ for some $r,s$, then 
    we set $\Pin{\alpha}{\UnFold{\CategoryC}} = \Multiplicity{S}(r)$;
  \item If $\alpha \in \Tr{\UnFold{\CategoryC}}$
   and so $(\alpha, r,s) \in T$ for some $r,s$, then 
   we set $\Pout{\alpha}{\UnFold{\CategoryC}} = \Multiplicity{S}(s)$.
  \end{itemize}
\end{definition}
Here, we use generating objects 
and morphisms of a FSSMC to define places and transitions of a Petri 
net. Then we use source and target of each generating morphism to 
define transition inputs and outputs, respectively. For this last step we 
need to use multiplicities, because we need to convert strings -- 
source and target of a morphism -- to multisets -- input and output of 
a transition.

Finally, notice that the two mappings defined in this section are 
somehow one the inverse of the other, as we wanted. 
In fact we have the following result, the proof of which 
follows easily from the definitions:
\begin{lemma}
  For any Petri net $N$, $\UnFold{\Fold{N}}$ and $N$ are isomorphic. 
  For any FSSMC $\CategoryC$, $\Fold{\UnFold{\CategoryC}}$ and 
  $\CategoryC$ are isomorphic.
\end{lemma}
This last result is important, because it ultimately allows us to go back 
and forth between nets and FSSMCs.
\section{Functors between executions}
  \label{sec: functors between executions}
In the last Section, we worked out correspondences 
that assign, to each $N$, a FSSMC $\Fold{N}$, and 
vice-versa. We also know that Petri nets are the 
objects of a category $\Petri$, while FSSMCs form a 
category $\FSSMC$, so we ask: Is it possible to 
extend such correspondences to functors? 

The answer is ``yes and no''. As we will see shortly, going 
from $\FSSMC$ to $\Petri$ does not create any issue, whereas 
doing the opposite is problematic. This is again related to 
the fact that a Petri net carries \emph{less information} that 
its corresponding $\FSSMC$, since source and target of 
generating morphisms are ordered, while inputs and outputs of 
transitions are not -- see Remark~\ref{rem: firing sequences cannot 
  correspond to string diagrams uniquely}.

We will start by going from $\FSSMC$ to $\Petri$. The first thing we 
notice is that a mapping between strings can be converted to 
a mapping between multisets using the multiplicity function:
\begin{proposition}[From strings to multisets]
  \label{prop: from strings to multisets}
  Suppose to have a mapping $f:\Strings{S} \to \Strings{S'}$
  that respects string concatenation (hence such that $f(r s) = f(r)f(s)$).
  From Remark~\ref{rem: strings are free monoids} it can be proven that 
  $f$ is completely determined by where it sends elements in $S$ -- 
  the generators of $\Strings{S}$.
  We can use this fact to turn $f$ into a multiset homomorphism 
  $\bar{f}: \Msets{S} \to \Msets{S'}$, by setting, for each multiset 
  $\MsetBase{X}{S} \in \Msets{S}$ and $s \in S$,
  \begin{equation*}
   \bar{f}(\MsetBase{X}{S})(s) = \Multiplicity{S'}(f(s))
  \end{equation*}
\end{proposition}
Proposition~\ref{prop: from strings to multisets} is just a technicality, but it is instrumental 
in turning a generator-preserving functor between FSSMCs into a 
net morphism. In fact,
\begin{definition}[The functor $\UnFold{F}$]
  Suppose we have FSSMCs $\CategoryC$, generated by 
  $S_\CategoryC, T_\CategoryC$ and $\CategoryD$, generated by 
  $S_\CategoryD, T_\CategoryD$. If $F: \CategoryC \to \CategoryD$ 
  is a generator-preserving functor sending the generating morphism 
  $t_\CategoryC$ to $\sigma; t_\CategoryD; \sigma'$, then we define 
  \begin{align*}
    \UnFold{F}:=\UnFold{\CategoryC} &\to \UnFold{\CategoryD}\\
    p \in \Pl{\UnFold{\CategoryC}} &\mapsto \Multiplicity{S_\CategoryD}(Fp)\\
    t_\CategoryC \in \Tr{\UnFold{\CategoryC}} &\mapsto t_\CategoryD
  \end{align*}
\end{definition}
Now we need to check that, for each generator-preserving functor $F$, 
$\UnFold{F}$ is a morphism of Petri nets. For sure, thanks to 
Proposition~\ref{prop: from strings to multisets} the Definition 
above defines a multiset homomorphism $\UnFold{F}_{Pl}$ between 
$\Pl{\UnFold{\CategoryC}}$ and $\Pl{\UnFold{\CategoryD}}$. Also, 
it defines a function $\UnFold{F}_{Tr}$ 
between $\Tr{\UnFold{\CategoryC}}$ and 
$\Tr{\UnFold{\CategoryD}}$. The last thing we need to check is that 
the commutative squares in the definition of Petri net morphism 
(Definition~\ref{def: Petri morphism}) indeed commute, which is left 
as an exercise.

Moreover, it is not difficult to convince ourselves that, denoting 
with $\Id{\CategoryC}$ the identity functor on $\CategoryC$, we have 
$\UnFold{\Id{\CategoryC}} = \Id{\UnFold{\CategoryC}}$.
Similarly, we can prove that $\UnFold{F;G} = \UnFold{F};\UnFold{G}$, 
where the composition on the left-hand side is functor composition 
in $\FSSMC$ and composition on the right-hand side is composition 
of Petri net morphisms.

All in all, this proves that $\UnFold{-}:\FSSMC \to \Petri$ is a functor 
mapping free strict symmetric monoidal categories to Petri nets, and 
functors between them to Petri net morphsims. Moreover, recalling 
the definitions of grounded Petri morphism and grounded functor, giving 
rise respectively to the grounded version of $\Petri$ and $\FSSMC$ ( see 
Definitions~\ref{def: the category petrigrounded} 
and~\ref{lem: the category fssmcgrounded}), we can see how 
$\UnFold{-}$ plays nicely with such restrictions:
\begin{lemma}[Restricting to grounded categories]
  If $F: \CategoryC \to \CategoryD$ is a grounded functor, then 
  $\UnFold{F}: \UnFold{\CategoryC} \to \UnFold{\CategoryD}$ is a 
  grounded morphism of Petri nets, and vice-versa. So
  $\UnFold{-}$ can be restricted to a functor from $\FSSMCGrounded$ 
  to $\PetriGrounded$.
\end{lemma}

As we anticipated, things are not so easy going in the other direction.
Namely, in mapping a morphism of Petri nets to a generator-preserving 
functor we have to make some choices: We know that a
generator-preserving functor maps a generator $t_\CategoryC$ to a 
morphism $\sigma;t_\CategoryD;\sigma'$ where $\sigma, \sigma'$ are 
symmetries and we are free to choose them as we want to. The problem 
is that symmetries are just morphisms permuting objects, and if we take
multiplicities of their source and target they will obviously be the same.
This means that nets are totally blind when it comes to symmetries and 
do not provide any information about how to choose them. This is 
compatible with the idea that symmetries only deal with the 
necessaty bookkeeping to distinguish between tokens, to which nets are 
indifferent. 

So, given a net morphism $\langle f, g \rangle: N \to M$ 
which maps a transition $t$ to a transition $u$, 
if we want to lift this to a generator-preserving functor between their 
corresponding categories of executions we need to make some choices 
by ``manually specifying'' symmetries. This is not a problem per s\'e, and 
there are sensible ways to make these choices 
(see for instance~\cite[Sec.4.3]{Genovese2019}). The problem is that 
all these choices cannot be made consistent with each other, meaning 
that we have no way to prove that the functorial laws (specifically the 
one about morphism composition) hold. So, we can map nets to FSSMCs 
and their morphisms to generator-preserving functors, but not in a 
functorial way!
\section{Lack of functoriality is not the end of the world}
  \label{sec: lack of functoriality is not the end of the world}
The lack of functoriality from $\Petri$ to $\FSSMC$ has been traditionally 
considered a problem in the literature, which was focused on proving 
that these two categories (or some small modifications of them) were 
equivalent. On the contrary, in our research  we realized that leaving 
things as they are is not just enough, but actually a better solution if 
the goal at hand is to implement a programming language.

There are many reasons for this. For instance, notice how the FSSMC 
formalism is based on strings, while Petri nets need multisets. 
Manipulating strings is way easier than manipulating multisets in a 
developing environment,  because historically many more tools and 
data structures have been developed to deal with strings, 
mainly to do text manipulation. This asymmetry between strings and 
multisets is so sharp that, in practice, multisets are often dealt as they 
were strings in programming. 

\begin{figure}[!ht]
	\centering
	\includegraphics[height=100px]{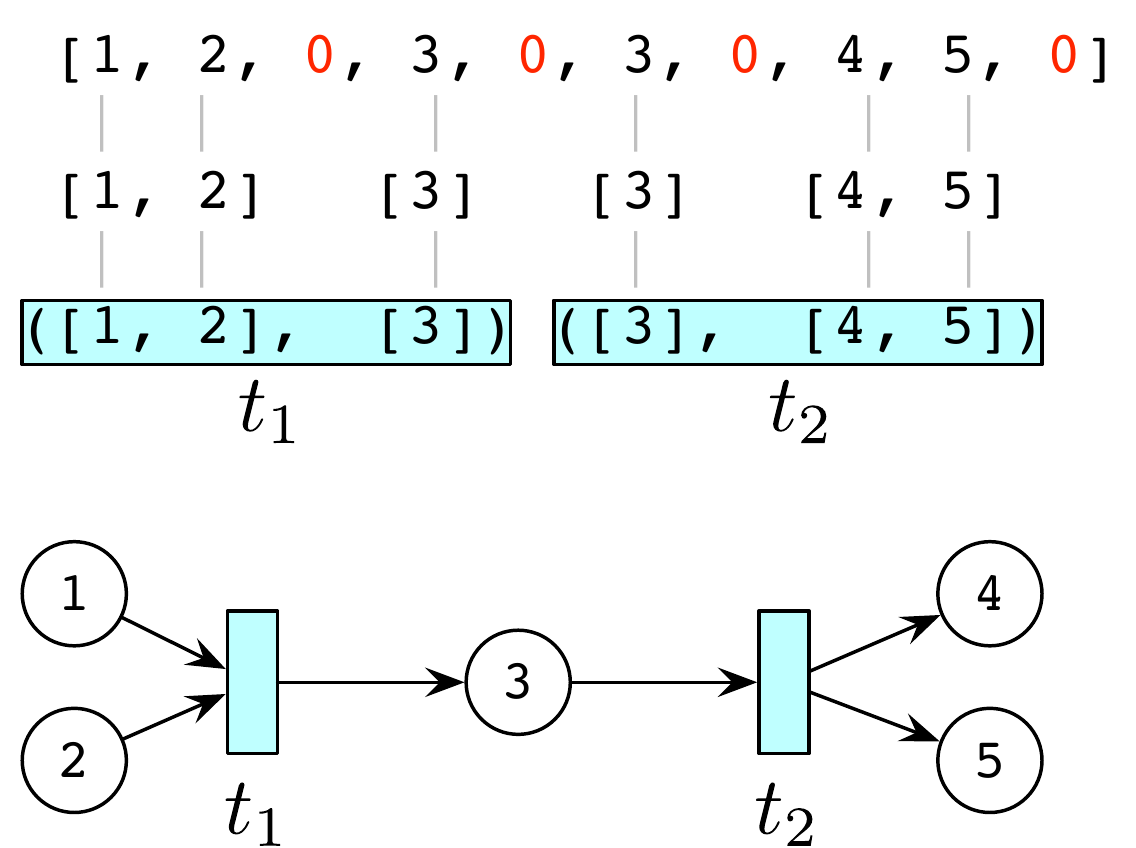}
  \caption{A way to convert a string to a Petri net, and vice-versa.}
  \label{fig: string to net again}
\end{figure}

To see this, recall the way of serializing/deserializing a Petri net,
which we introduced in Section~\ref{sec: Petri nets implementation}:
We start with a string of numbers,
where $0$ is treated as a special character. Scanning the string, we chop 
it every time we encounter a zero. What we are left with now is a bunch 
of substrings, which we sequentially group into couples. Each of these 
couples defines input and output of a transition, and as we see this is 
enough information to build a Petri net.
The serializing/deserializing procedure is again shown in 
Figure~\ref{fig: string to net again}. 

\noindent
Quite soon though we realize that since strings are ordered, 
our substrings are not really specifying input and output of transitions -- 
which are multisets -- but source and target of their corresponding 
generating morphisms! That is, the procedure in 
Figure~\ref{fig: string to net again} can be used, \emph{without any change}, 
also to pass around FSSMCs! It is clear then than even if we are 
visualizing the information in the string as a net, we are secretly dealing 
with its corresponding FSSMC. Computers simply like them more!

The way the Statebox language works, then, is the following: When a 
user draws a Petri net, its places and transitions are automatically indexed, 
and the structure is converted, under the hood, into its corresponding
 FSSMC. When transitions are fired, the user can specify which tokens 
 a firing transition has to process (if there is any choice to be made), and 
 the corresponding morphisms are composed in the FSSMC. Visualizing 
 the state of an execution simply amounts to visualize the string diagram 
 representing that history. 

 When a Petri net has to be morphed into another, a functor between 
 their corresponding FSSMCs has to be specified. The user can just
 indicate which places and transitions have to be mapped to which places 
 and transitions. This defines a morphism of Petri nets which is lifted 
 to a functor between FSSMCs (in a non functorial way, as we 
 said already) via some of the standard procedures explained 
 in~\cite[Sec.4.3]{Genovese2019}. Alternatively, the user is able to 
 define such procedures manually, de facto defining the functor between 
 FSSMCs directly.

 All in all, with this approach we use just FSSMCs, but Petri nets are 
 used both to do model checking and prove properties about the code 
 (which obviously the FSSMC preserves) or to provide enough basic 
 information to allow the computer to infer the rest. As some 
 researchers like to say, Petri nets are \emph{presentations} of FSSMCs,
 meaning that they provide the bare minimum information to build and 
 work with a given FSSMC. This is precisely the way we are using them.
\section{Beyond standard Petri nets}
  \label{sec: beyond standard Petri nets}
Up to now, we considered normal Petri nets and 
categorically described their executions. But what 
happens if we change our notion of Petri net? A nice 
change to make would be, for instance, to allow the 
net to have \emph{negative tokens}. If we represent 
a token as a black dot in a place, we can represent a 
negative token as \emph{red}; we can moreover consider 
transitions that consume/produce negative tokens 
(Figure~\ref{fig: integer petri}). We call a net that allows 
for negative tokens an \emph{integer Petri net}.
\begin{figure}[!ht]
	\centering
	\begin{subfigure}[t]{0.22\textwidth}\centering
		\input{Pictures/ExecutionsOfPetriNets/08-IntegerPetri.tex}
		\caption{Integer Petri net.}
		\label{fig: integer petri}
	\end{subfigure}
	~\hfill 
	\begin{subfigure}[t]{0.22\textwidth}\centering
		\input{Pictures/ExecutionsOfPetriNets/08-IntegerTokens.tex}
		\caption{Integer tokens.}
		\label{fig: integer tokens}
	\end{subfigure}
	~\hfill 
	\begin{subfigure}[t]{0.22\textwidth}\centering
		\input{Pictures/ExecutionsOfPetriNets/08-IntegerPreFiring.tex}
		\caption{Before firing.}
		\label{fig: integer net pre firing}
	\end{subfigure}
	~\hfill 
	\begin{subfigure}[t]{0.22\textwidth}\centering
		\input{Pictures/ExecutionsOfPetriNets/08-IntegerPostFiring.tex}
		\caption{After firing.}
		\label{fig: integer net post firing}
	\end{subfigure}
	\caption{}
\end{figure}

\noindent
If we start to explore this definition further, 
we see that very strange things can happen now. 
Since clearly a negative token and a positive one 
``annihilate'', exactly as $-1 + 1 = 0$, we can produce 
couples of tokens in any place, as in Figure~\ref{fig: integer tokens}. 
The consequence of this is that, as in 
Figures~\ref{fig: integer net pre firing} 
and~\ref{fig: integer net post firing}, now transitions can fire 
borrowing tokens from a place, and so they are always enabled!

Is there a use for this generalization? Most likely, yes. 
In fact, the study of executions of integer Petri nets is a genuine 
contribution of the Statebox team to academic research, that 
resulted in a paper~\cite{Genovese2018}, further 
generalized in~\cite{Master2019}. What motivated us 
to investigate in this direction is that integer nets can be useful 
to model conflict resolution in concurrent behaviour. Consider, 
for instance, the net in Figure~\ref{fig: conflict petri}: We know 
that transitions $t_1$ and $t_2$ have to compete for the token 
in $p_1$ and, at least in the case of standard nets, they cannot 
both fire. Now suppose that there are two users, 
say $U_1$ and $U_2$, that can operate on the net, deciding 
which transition to fire. When a user takes a decision, it is 
broadcast to the other one, and the overall state of the net is 
updated. In a realistic scenario, though, broadcasting takes 
time (imagine, for instance, that our users have bad internet 
connections): User $U_1$ could decide to fire $t_1$ and user 
$U_2$ could decide to fire $t_2$ while the broadcast choice 
of $U_1$ has still to be received, putting the overall net into 
an illegal state (Figure~\ref{fig: conflict petri not solved}).
\begin{figure}[!ht]
	\centering
	\begin{subfigure}[b]{0.25\textwidth}\centering
		\input{Pictures/ExecutionsOfPetriNets/09-ConflictPetri.tex}
		\caption{}
		\label{fig: conflict petri}
	\end{subfigure}
	\hfill
	\begin{subfigure}[b]{0.365\textwidth}\centering
		\input{Pictures/ExecutionsOfPetriNets/09-ConflictPetriNotSolved.tex}
		\caption{}
		\label{fig: conflict petri not solved}
	\end{subfigure}
	\hfill
	\begin{subfigure}[b]{0.365\textwidth}\centering
		\input{Pictures/ExecutionsOfPetriNets/09-ConflictPetriSolved.tex}
		\caption{}
		\label{fig:conflict petri solved}
	\end{subfigure}
	\caption{}
\end{figure}

\noindent
In such a situation we need a way to 
re-establish \emph{consensus}, that is, decide unambiguously 
in which legal state the net is. There are multiple ways to do 
this, but our main concern here is that the usual Petri net 
formalism does not have a way to represent illegal states, 
which is useful to attacking the problem mathematically. 
With integer Petri nets we are able to easily represent such a 
situation using negative tokens, as in 
Figure~\ref{fig:conflict petri solved}. Intuitively, we can say that 
a net is in an illegal state if the state contains a negative number of 
tokens in some place, and re-establishing consensus from an illegal 
state then amounts to getting back to one where the number of 
tokens in each place is non-negative. Clearly, the fact that any net 
can now fire just by borrowing positive tokens from its input places 
is consistent with the idea that, for whatever reason, every transition 
can put the net into an illegal state. There are, even, transitions like 
the one in Figure~\ref{fig: integer petri} that de facto ``produce illegal 
states'' out of thin air. This could be used, for instance, to represent a 
faulty component in our net architecture.

\begin{figure}[!ht]
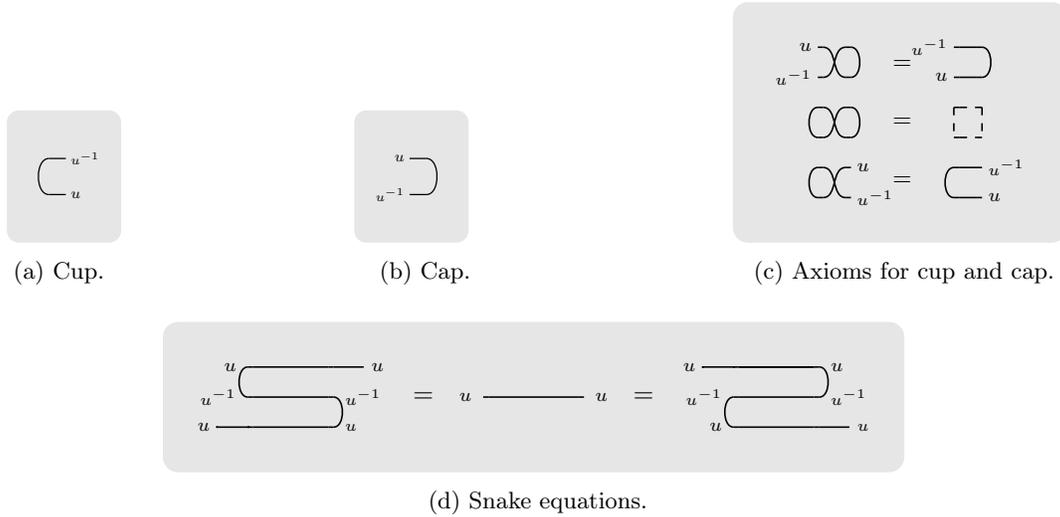

	\centering
	\begin{subfigure}[t]{0.15\textwidth}\centering
		\input{Pictures/ExecutionsOfPetriNets/10-Cup.tex}
		\caption{Cup.}
		\label{fig: cup}
	\end{subfigure}
	\hfill 
	\begin{subfigure}[t]{0.15\textwidth}\centering
		\input{Pictures/ExecutionsOfPetriNets/10-Cap.tex}
		\caption{Cap.}
		\label{fig: cap}
	\end{subfigure}
	\hfill
	\begin{subfigure}[t]{0.35\textwidth}\centering
		\input{Pictures/ExecutionsOfPetriNets/10-CupCapAxioms.tex}
		\caption{Axioms for cup and cap.}
		\label{fig: cup and cap axioms}
	\end{subfigure}
	\par\bigskip
	\begin{subfigure}[t]{0.90\textwidth}\centering
		\input{Pictures/ExecutionsOfPetriNets/10-SnakeEquations.tex}
		\caption{Snake equations.}
		\label{fig: snake equations}
	\end{subfigure}
	\caption{Additional structural morphisms and axioms for executions of integer nets.}
\end{figure}

As we saw in this Chapter, the category of executions of a net carries 
much more information than the net itself, since we can track 
precisely the history of each token in the net. When it comes to 
integer nets, it makes sense to study this category to see 
if there are na\"ive ways to resolve an illegal situation, at least in 
some cases. The category of executions of an integer net looks very 
similar to what we already saw in Section~\ref{sec: the category foldn}, 
and we do not have to change much of what we already have. 
First, we need to add a new couple of bookkeeping morphisms in 
our formalism, along with identities and symmetries. These are 
depicted as a cup (Figure~\ref{fig: cup}) and a cap 
(Figure~\ref{fig: cap}),  and represent the creation or annihilation 
of couples of negative and positive tokens in a place. These new 
morphisms have to satisfy the axioms in 
Figures~\ref{fig: cup and cap axioms} and~\ref{fig: snake equations} 
-- these last couple of axioms are called \emph{yanking} or 
\emph{snake equations}~\cite{Coecke2017}, for obvious 
visual reasons. Strict symmetric monoidal categories that have cups 
and caps and respect such axioms are called \emph{strict compact 
closed categories}~\cite{Kelly1980}.

\begin{figure}[h]
	\resizebox{\textwidth}{!}{
		\input{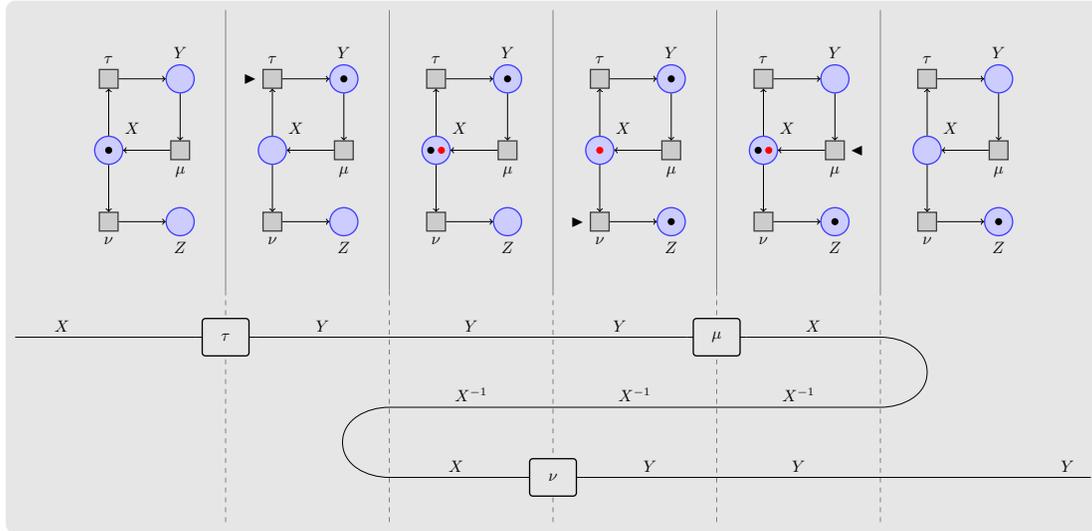}
	}
	\caption{Conflict resolution using integer Petri nets.}
	\label{fig: conflict resolution}
\end{figure}

\noindent
As we did for standard nets, we need to define the concept of 
\emph{free strict compact closed category} along with mappings 
that allow us to go from nets to categories, back and forth.
This is not the right place to dive into the technicalities of this 
construction, for which we redirect the reader to~\cite{Genovese2018}.
What is really interesting, though, is to see 
how compact closed categories solve some 
of the problems regarding nets in illegal 
states, na\"ively. For example, consider the situation in 
Figure~\ref{fig: conflict resolution}: As before, imagine that a 
user $U_1$ fires transition $\tau$, while another user fires 
transition $\nu$ putting the net into an illegal state. We can 
represent this graphically in our category of executions introducing 
a cup that produces a pair of positive and negative tokens. At 
this point we apply a morphism corresponding to $\nu$ and carry 
the ``positive part'' produced by the cup to $Y$.

Now comes the interesting part: Suppose that some other user 
fires transition $\mu$. This transition produces a token in $X$ that 
\emph{effectively cancels} the debt left in $X$ by the firing 
of $\nu$, reporting the net into an legal state. But this sequence of 
firings is still not acceptable, since the firing of $\nu$ could not have 
happened in the first place!

Nevertheless, the categorical model offers us a solution straight 
out of the box: When the token produced by $\mu$ lands in $X$, it 
annihilates the negative token left there. In the category of executions, 
this amounts to add a cap to our string diagram. But now the magic 
happens: {We can straighten the string diagram using the snake 
equations} obtaining the sequence of firings $\tau$, then $\mu$, 
then $\nu$.

\noindent
The right way to read the diagram in 
Figure~\ref{fig: conflict resolution} is as follows: The vertical 
lines divide different instants in time \emph{in the real world}. 
If we were to attach a timestamp to each transition firing, we would 
actually observe that $\tau$ has been fired, then $\nu$ has been fired, 
and finally $\mu$ has been fired. On the contrary, the wire 
represents \emph{the causal flow of the network itself}: It does not 
matter which transitions have been fired first in the real world, the 
flow represented by following the wire -- namely $\tau$, then $\mu$, 
then $\nu$ -- is the flow that does not break causality, allowing for a 
sequence of \emph{completely legal firings}. The category of 
executions then offers na\"ive solutions to re-establishing consensus by 
reshuffling the order of transition firings, cancelling out any illegal state.

If we observe the net carefully, we notice that, actually, we still need 
to establish consensus on something: We presumed that user $U_1$ 
fired $\tau$ \emph{before} user $U_2$, but obviously user $U_2$ is 
not of this opinion, otherwise he would not have fired $\nu$ in the first 
place.

This means that we need a way to establish which of the two users 
fired first, which is still a consensus problem. We argue, though, that 
this kind of consensus is much simpler to reach (for instance 
implementing a global clock) than having to reach consensus on an 
entire merging problem such as the one of re-establishing the causal 
order of transitions is. The category of executions for integer Petri 
nets takes care of this part of the problem for us, and needs only a 
very small amount of consensus to work properly.

All the details about this construction can be found 
in~\cite{Genovese2018}, where we proved results that are 
somewhat akin to the ones in 
Section~\ref{sec: functors between executions}. Namely, 
we arranged integer Petri nets in a category, called $\PetriZ$. We did 
the same for categories of executions, obtaining a category of 
categories $\FSCCC$. Finally, we produced a couple of mappings 
$\Fold{-}$ and $\UnFold{-}$, which behave exactly like their 
counterparts in Section~\ref{sec: functors between executions}.
Following the approach in~\cite{Sassone1995}, this time we were more 
interested in the mathematical results than in the implementation, so we 
massaged our categorical definitions a bit to be able to show an 
 equivalence of categories.

The study of integer Petri nets is in its infancy, and many 
questions have yet to be answered. The ``conflict resolution 
procedure'' sketched in this Section, for instance, is only of 
theoretical interest at the moment and far away from an 
industry-strength implementation but we will keep pushing 
in this direction in the hope of obtaining something that can 
eventually become a useful feature to be used in our language.

It has to be noted that integer Petri nets also possess very nice 
characteristics when one tries to model transaction flows and 
money flows in general. This is relevant for a number of applications, 
among which are Blockchain-based techniques~\cite{Nakamoto2008}, 
for which it is commonplace to represent any kind of asset -- even 
computations -- by monetizing it~\cite{Buterin2014}, and the general 
characterization of economic phenomena in terms of process theories. 
This is the object of a broader research that the Statebox team is 
carrying out along with multiple partners, and which also involves 
open games~\cite{Ghani2016}, macroeconomics~\cite{Winschel2010} 
and open systems~\cite{Sobocinski2010}.
\section{Implementation}
  \label{sec: executions implementation}
The FSSMCs formalism defines the very mathematical core of 
our way of representing net histories, hence it shouldn't be surprising 
that implementing FSSMCs is a big chunk of our coding efforts.

Since we want formal guarantees that net histories evolve without 
errors, we are implementing FSSMCs in Idris. This has the desirable 
property that if we try to compose incompatible morphisms -- e.g. 
we try to take the composition $f;g$ with $f:A \to B$ and $g: C \to D$,
the result will be a typecheck error. In simple words, \emph{Idris' typechecker
won't allow us to do any mathematically inconsistent operation}. 
This way of implementing things is clearly very powerful, 
and brings our implementation as close to the actual mathematical 
theory as possible.

Unsurprisingly, we are using the \texttt{idris-ct}~\cite{StateboxTeam2019} 
library to define what a FSSMC is, but this is not a simple task: As we can see, 
Definition~\ref{def: free strict symmetric monoidal category}
builds a FSSMC by making use of equations that identify 
morphisms inside the category. From a mathematical point of view, 
we say that we are \emph{quotienting} the morphisms. Unfortunately, 
Idris' typesystem does not really like quotients, which are indeed 
quite difficult to deal with in type theory. 

This prompted us to finding alternative, Idris-friendly ways to define 
FSSMCs. After really pushing the Idris compiler to its 
limits~\cite{Perone2019}, we got a formal,
complete definition of FSSMC.

Another implementation effort revolves around the idea of having 
the Idris core to insert symmetries in place for us during morphism 
composition. To see this, imagine to have a net history. As we 
know, we can represent this as a morphism in a FSSMC. Now
assume that this history is a morphism $m: \TensorUnit \to B \Tensor A$.
In our interpretation, this means that we started with a net having 
no tokens, and fired transitions so that we now have a token in $B$ 
and a token in $A$. Suppose that now we fire a transition corresponding 
to the generating morphism $f: A \Tensor B \to C$. We clearly can do 
this since $f$ is enabled in the net, but we quickly realize 
that we cannot take the composition $m;f$, since source and target of $f$ 
and $m$, respectively, do not match. To extend the history consistently, 
we need to do some bookkeeping, namely by inserting a symmetry in the 
composition, and taking $m;\sigma_{B, A};f$. The point is that 
in many situations -- as for instance the one we sketched -- there is only 
one way to define such symmetries. What we want, then, is a series 
of helpers and formal procedures to allow Idris to figure out these 
symmetries automatically whenever possible, to make the transition 
from Petri nets to FSSMCs as smooth as possible for the end user.
\section{Why is this useful?}
The answer to this question should be pretty clear: 
Executions allow us to track which transitions process which tokens, 
and to formalize the idea of ``history of a net''. Being able to 
represent the causal relationships between firings precisely and 
reliably is fundamental to concatenate processes in a meaningful 
way, and categories of executions, serving exactly this purpose, will 
function as a bridge to consistently link nets, seen as abstract design 
tools for complex systems, to the actual implementation -- we will start 
developing this point of view in Chapter~\ref{ch: folds}. Note how 
leveraging this formal bridge is exactly what makes Statebox 
different from any other project based on Petri nets. Petri nets have, 
in fact, been used as design tools for software many times in the past, 
but the general modus operandi was as follows:
\begin{itemize}
  \item The programmer would draft how the software about to be 
  written was supposed to work in the abstract, using a Petri net. 
  The properties of the net would be formally studied to ensure 
  some pre-set performance standards;
  \item Afterwards, code would be produced, using the net 
  implementation as a guide. This passage would be totally 
  handmade and there would be no formal link between the net 
  and the actual codebase. All things considered, the formal 
  relationship between nets and code would amount to zero, and 
  using nets to design it was not much different than sketching flow 
  diagrams on a piece of paper: Helpful, but needing a lot of common 
  sense to be implemented properly;
  \item As a consequence, it would happen that the software 
  implementation could not properly reflect the net topology 
  due to human error, and performing even small modifications 
  in the net layout would result in huge code refactoring.
\end{itemize}
Category theory, on the other hand, completely automates all 
these steps giving us a neat way to represent net executions.

\newtoggle{toggleFolds}\toggletrue{toggleFolds}
  \chapter{Folds}\label{ch: folds}
In this Chapter we will reap what we sow up to now. 
Many of the concepts presented in the previous Chapters, purely 
theoretical on their own, will start displaying evident applicative 
potential when put together. Here we will sketch the actual 
plan to turn Petri nets and category theory into a useful software 
development toolkit.

We devoted Chapter~\ref{ch: executions} to the endeavor of building 
categories associated with Petri nets. We pointed out how the main 
reason to do this was to be able to describe the net behavior in a 
completely deterministic way, keeping track of the history of any token. 
In truth, we can do much more than this: If for each net $N$ we have 
a category $\Fold{N}$, we can use our categorical intuition and do the 
most sensible thing when you have a category, namely mapping it 
to somewhere else by means of a functor.
\section{Problem Overview}
A perfectly legitimate question, at this stage, is: Why should 
we map executions to other categories? Note how, at the moment, 
both Petri nets and their executions cannot do much. We are able to 
draw a net and, as we said, we interpret its transitions as processes that, 
when firing, consume and produce resources, but where is this 
information stored? Clearly nowhere, at least for the moment. This 
interpretation exists only in our mind and is not backed up by any 
meaningful mathematics. Similarly, in defining executions, we said 
that we associate, to each transition, a family of morphisms representing 
the processes actually performed by the transition during firing. But 
again, this is an interpretation, since the actual definition of such 
processes is lacking in our category of executions. Recalling 
Definition~\ref{def: the category foldn}, we obtain a 
morphism $t: \Ordering{\Pl{N}}(\Pin{t}{N}) 
\to \Ordering{\Pl{N}}(\Pout{t}{N})$ for each transition 
$t$ by means of an inference 
rule, but that is pretty much it. Our 
processes lack any sort of actual specification.
\begin{figure}[h!]
	\centering
	\begin{subfigure}[t]{0.45\textwidth}\centering
		\input{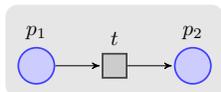}
		\caption{A net representing quicksort.}
		\label{fig: quicksort example}
	\end{subfigure}
	~\hfill %
	\begin{subfigure}[t]{0.45\textwidth}\centering
		\input{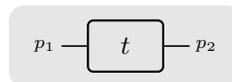}
		\caption{Morphism associated to Figure~\ref{fig: quicksort example}.}
		\label{fig: quicksort execution}
	\end{subfigure}
	\caption{Quicksort and its execution.}
\end{figure}
\begin{example}[Quicksort]\label{ex: quicksort}
  Consider the net in Figure~\ref{fig: quicksort example}. 
  We interpret the places as holding resources of type $\ListInt$, 
  that is, a token in a place represents a list of integers. 
  Transition $t$ represents an application of the 
  \emph{quicksort algorithm}~\cite{Hoare1961}, that sorts the list. 
  This information is clearly not captured by our net, which just 
  describes how the transition turns one resource into another. Also 
  our idea of tokens being of type $\ListInt$ is overimposed, since 
  the behavior of this data structure is not represented by the 
  net (for instance, we cannot concatenate tokens or perform any 
  sort of list operation on them). Similarly, in 
  Figure~\ref{fig: quicksort execution} we represent the morphism 
  associated to $t$ in its category of executions. Again, in this 
  setting, $t$ is just ``a box'', and the information describing its 
  behavior (namely, quicksort) is nowhere to be found.
\end{example}
\begin{example}[Is quicksort being done right?]
  An obvious counterargument to the reasoning in 
  Example~\ref{ex: quicksort} could be that the net in 
  Figure~\ref{fig: quicksort example} does not capture the meaning of 
  the quicksort algorithm because it is not the right model for it: It is not 
  that nets are bad at representing such a thing but that we have not used them properly.
  Up to some extent, this is actually the case, since 
  we can definitely try to model sorting algorithms in a much more 
  convincing way using nets. It is worth stressing that 
  this is often \emph{not the right way of thinking about Petri nets}. 
  The application of a mathematical gadget in computer science should, 
  nearly always, serve the purpose of stripping away complexity and 
  making things easier, possibly without giving up formal correctness 
  and consistency of our methods. Does it make sense, then, to spend 
  time to define something that already has fully debugged and 
  efficient implementations, spanning just a few lines of code? The 
  answer is clearly no, since what we would get, at best, is something that 
  needs a considerable amount of time and thought to get the level 
  of performance found in existing solutions. Petri nets should make 
  our life easier, and we would like to leverage already existing 
  implementations of algorithms if we have them, as in the case 
  of quicksort.
\end{example}
\section{Mapping executions}
What Example~\ref{ex: quicksort} entails could look like a huge 
downside: Our math is good for nothing, and our nets cannot do 
anything interesting without becoming really complicated. Luckily 
enough, this is not the case. What we obtained is, instead, far more 
valuable, and akin to what a logician would call 
\emph{the separation between syntax and semantics}. We obtained a 
model of how our Petri nets behave without having to refer to any 
particular detail which complements the declarative functional 
approach we take in the implementation of our language. We can 
talk about quicksort, as in Example~\ref{ex: quicksort}, without 
giving any specification of what quicksort does, aside of how it fits 
in the infrastructure we are designing, represented by the Petri net. 
The actual specification of quicksort (that is, its \emph{semantics}) 
can be modeled separately in another category, and then be targeted 
appropriately by mapping the category of executions of the net into it. 
The fact that this mapping is functorial guarantees that syntax 
and semantics are being glued consistently together.
\begin{example}[Design advantages]
  The clear utility of this separation is that we can undertake our 
  design efforts in stages. For instance, imagine that we are automating 
  the infrastructure of an entire company. We first talk with people 
  from various departments, asking them about their daily routines 
  and needs. We then draft a Petri net describing how these processes 
  interact with each other in real time, and since the Petri net formalism 
  is completely graphical, people giving us this information even help 
  us to visually debug it, pointing out where the diagram representing a 
  process in their workflow is incorrect. After having done this, we 
  apply tools to study reachability problems on the net, verifying that 
  it has the properties that we desire (for instance absence of deadlocks 
  or illegal states, recall Definition~\ref{def: deadlock}). If these 
  requirements are not met, then we can reshape the net until we get 
  what we want. At this point -- and only at this point -- we can start 
  writing down the code (typically in a lower level language) for the 
  programs associated to each transition, and the functorial mapping 
  from the net execution to the actual code takes care of putting 
  everything together.
\end{example}
Having intuitively described the essence of folds, let us try to fix 
the concept with a definition.
\begin{definition}[Folds]\label{def: folds}
  Given a Petri net $N$ and a symmetric monoidal 
  category $\Semantics$, a \emph{fold for $N$} is a 
  symmetric lax monoidal functor (recall 
  Definition~\ref{def: lax monoidal functor}) 
  $\Fold{N} \to \Semantics$.
\end{definition}
The reason we are choosing a Lax monoidal functor is 
because it is the weakest requirement we can think of. It is 
always a good practice to state something in the greatest level of 
generality possible, and strengthen the requirements only if needed.

Before we start digging into the real stuff, notice that there is an 
obvious fold that we can take:
\begin{remark}[Trivial Fold]
  The identity functor $\Fold{N} \to \Fold{N}$ is trivially symmetric 
  lax monoidal, hence it generates a \emph{trivial fold} for $N$. Note 
  that this remark is what prompted for the notation $\Fold{-}$ to define 
  the category of executions of a net. Executions are, to some extent, the 
  simplest fold possible, where the meaning we attach to any execution 
  of the net is the execution itself.
\end{remark}
To create more complicated instances of folds, we need to create 
semantic categories to which it makes sense to map executions. A 
good starting point is to recall Example~\ref{rem: Haskell category}, 
and use algorithms written in a functional programming language 
(Haskell, in our case) as semantics.
\begin{definition}[Haskell, again]\label{def: Haskell again}
	The category $\Hask$ is defined as follows:
	\begin{itemize}
    \item Objects are data types. A data type is a way a computer uses 
    to represent a certain kind of information. Common data types 
    are $\Bool$, consisting of the booleans $\True$ and $\False$; $\Int$, 
    consisting of integer numbers, $\ListInt$, consisting of finite lists 
    of integer numbers, and many other;
    \item Morphisms are \emph{terminating haskell algorithms}, that is, 
    algorithms that take terms of some data type as inputs , apply a 
    sequence of operations that at some point terminates, and output a 
    term of some data type as a result.
	\end{itemize}
  The category $\Hask$ can be made into a symmetric monoidal 
  category using the natural cartesian product structure that data 
  types and morphisms admit (a product of types $A, B$ is just the 
  type of couples $(a,b)$ where $a$ has type $A$ and $b$ has type $B$).
\end{definition}

\begin{remarkhard}[Is $\Hask$ a category?]
  \label{rem: Haskell not a category}
  The reader with experience in the abstract theory of programming 
  languages will have risen an eyebrow reading 
  Example~\ref{def: Haskell again}. In fact, the matter of defining 
  the category $\Hask$ is quite a can of worms. In our case we are 
  considering the strict symmetric monoidal category 
  equivalent -- via Remark~\ref{rem: monoidal equivalence} -- to what 
  is known in the functional programming folklore as the 
  \emph{platonic Haskell category}\cite{HaskellWiki}, where types 
  do not have bottom values, valid morphisms are just terminating 
  algorithms, algorithms are considered equal if they agree on all 
  inputs and the use of  \texttt{seq} is very limited. In general, the 
  question of casting a category out of the Haskell programming 
  language is still very debated, but we want to stress how this is not 
  fundamental with respect to what we are going to do here. The point 
  is that the fold from $\Fold{N}$ to $\Hask$ is implementable, and 
  offers a consistent way to map transitions into pieces of software.
\end{remarkhard}
\begin{example}[The fold to $\Hask$, in practice]\label{ex: fold to hask}
  It is interesting to see how the mapping $\Fold{N} \to \Hask$ works 
  in practice. Let us build a strict symmetric monoidal 
  functor $F:\Fold{N} \to \Hask$: Each place in $\Pl{N}$ is also an 
  object of $\Fold{N}$, and as a consequence will be mapped by $F$ to 
  a particular Haskell data type. We have complete freedom in defining 
  this mapping as we please. The mapping on monoidal products of 
  objects will then have to follow since we set, by 
  definition, $F(u \otimes v) = (Fu,Fv)$.

  Next, the bookkeeping morphisms. Identities on a object $u$ will be 
  mapped to the algorithm $Fu \to Fu$ that takes any term of 
  type $Fu$ in input and outputs the term itself without changing it. 
  Symmetries $\sigma_{u,v}: u \otimes v \to v \otimes u$ will be 
  mapped to the algorithm that takes tuples $(x,y)$ with $x$ of type 
  $Fu$ and $y$ of type $Fv$ and outputs $(y,x)$.
	
  For each transition $t \in \Tr{N}$ we get, from 
  Definition~\ref{def: the category foldn}, a  morphism
  $t_{u,v}$ such that $\Multiplicity{\Pl{N}}(u) = \Pin{t}{N}$ and 
  $\Multiplicity{\Pl{N}}(v) = \Pout{t}{N}$. Each one of these morphisms 
  will be mapped to a Haskell algorithm $Ft_{u,v}: Fu \to Fv$. 
\end{example}
\begin{remark}[Strictness makes life easier]
  Note that, in Example~\ref{ex: fold to hask}, requiring $F$ to be strict 
  monoidal is what saved the situation, allowing us to define it only on 
  places and transitions and leveraging strictness to extend it to all 
  objects and morphisms. If we require $F$ to be just lax, then we 
  have much more choice to define it, which is a good thing on the one 
  hand, giving implementational freedom, but bad on the other, since 
  we have to specify more things ``manually'' to make it work. The 
  appropriate choice clearly depends on context.
\end{remark}
\begin{example}[Quicksort, continued]
  We now turn our na\"ive interpretation of Example~\ref{ex: quicksort} 
  to something formal. Call $N$ the net in 
  Figure~\ref{fig: quicksort example}. We define the 
  fold $\Fold{N} \to \Hask$ mapping $p_1$ and $p_2$ to $\ListInt$, and 
  $t$ to the code:
  \lstset{numbers=left, 
                numberstyle=\tiny, 
                breaklines=true,
                backgroundcolor=\color{\backgrnd},
                numbersep=5pt,
                xleftmargin=.25in,
                xrightmargin=.25in} 
	\begin{lstlisting}[language=Haskell]
	quicksort :: [Int] -> [Int]
	quicksort [] = []
	quicksort (p:xs) = (quicksort lesser) ++ [p] ++ (quicksort greater)
	where
	lesser  = filter (< p) xs
	greater = filter (>= p) xs
	\end{lstlisting}
	Finally, $t$ is now formally identified with quicksort!
\end{example}
\begin{remark}[Terminating algorithms work better]
  \label{rem: terminating algorithms work better}
  In Definition~\ref{def: Haskell again} we explicitly required 
  morphisms of $\Hask$ to be \emph{terminating algorithms}. 
  It is worth to spend a few words on this: The interpretation of a 
  fold is that transitions of a net get mapped to algorithms. Folds 
  tell us which algorithm to run on which data when a transition fires. 
  Clearly, in case the algorithm is not terminating, things break down: 
  The transition in the net fires, but no tokens can be ever produced 
  since the algorithm will hang forever. We get rapidly aware of how 
  mapping transitions to algorithms that are not guaranteed to terminate 
  \emph{is not, in general, good practice}, since our formalism is no 
  longer able to ensure consistency. This means that such ``unsafe'' 
  mappings should be used only in very restricted contexts, when it is 
  absolutely necessary, and in a very localized way, so that we are always 
  able to keep track of which transitions in the net can exhibit a 
  pathological behavior. Keeping the problem circumscribed is the 
  easiest way to fix problems should they arise.
	
  With respect to this, some functional programming languages such as 
  Idris~\cite{Brady2013, Brady2017} offer the useful functionality 
  of a \emph{totality checker} already embedded in their compiler. 
  What this means is that, for a restricted class of algorithms, the 
  compiler is able to tell us if our algorithm will terminate on every 
  input. This feature is incredible in the context of Folds, since it will 
  allow us to map transitions to code that we know to be well behaved.
\end{remark}
\section{Different folds, shared types}
In reviewing the material covered in the last Section, we become 
aware that there is nothing special about using $\Hask$ as our semantics, 
and that in fact every functional programming language -- or in 
general every language in which types can be defined -- does, more 
or less, the job. One of the first things that comes to mind is: What 
happens if given a net $N$ we chose $\Fold{M}$, for some other 
net $M$, to define the semantics of a fold? We are not requiring, here, 
for this functor to be generator-preserving, as 
we did in Chapter~\ref{ch: executions}. The idea is that we could 
map a single transition of $N$ to an entire sequence of firings in $M$. 
This sort of ``net inception'' concept is very powerful, and backs up the 
intuition of transitions in a net triggering the execution of other nets as 
subprocesses, but needs far more work to be used properly.

What we can already do with the tools developed so far is sketching 
how different folds relate to each other:
\begin{definition}[Morphisms of folds]\label{def: morphisms of folds}
  Given folds $F_1: \Fold{N} \to \Semantics_1$, $F_2: 
  \Fold{N} \to \Semantics_2$, a \emph{morphism of folds} is a 
  symmetric lax monoidal functor $G:\Semantics_1 \to \Semantics_2$ 
  such that $F_1;G = F_2$. Folds and their morphisms form a category.
\end{definition}
\begin{remarkhard}
  The category of folds and their morphisms can be seen as the 
  \emph{co-slice category of strict symmetric monoidal categories 
  and symmetric lax monoidal functors over $\Fold{N}$}. $\Id{\Fold{N}}$ 
  is clearly initial in this category.
\end{remarkhard}
Embracing the interpretation of semantics in terms of data types and 
algorithms, a morphism of folds can be seen as a ``translation'' from 
one programming language to another, that allows us to rewrite our 
mapping altogether. This is not very useful in practice: such 
translation exists very rarely since different programming languages 
have different properties.

What would be very useful, on the contrary, would be to have 
\emph{all the programming languages modeled in the same category}. 
In fact, up to now, we are mapping executions into one programming 
language at a time, but this is not always what we would like to have. 
If our Petri nets represent a complex system, then transitions can 
represent processes radically different in nature, that would be better 
implemented in different programming languages, or, most likely, for 
which efficient implementations already exist in different languages.

With respect to this we want to be resourceful, and be able to use as much 
preexisting stuff as we can. Experienced programmers know, in fact, 
that one of the biggest barriers in the adoption of a new programming 
language is having to rewrite entire libraries from scratch: This is not 
only time-consuming, forcing developers to spend many hours of good 
work on just preparing the software instead of using it to solve the 
problems at hand, but also very inefficient, since rewriting complex 
code is a tedious process that needs a lot of further testing. In real-life 
applications, there is virtually no porting of industry-strength products 
that works out of the box, and in translating libraries from one language 
to another one is almost always guaranteed sub-optimal 
performance -- both in terms of time/space efficiency and presence of 
bugs and errors -- for a big portion of the development stage. The ideal 
semantic category we aim at, then, looks like this:
\begin{definition}[Generalized semantics for folds]\label{def: pathological semantics}
  Let $\mathcal{L}_1, \dots, \mathcal{L}_n$ denote programming 
  languages. We define a category having:
	\begin{itemize}
		\item As objects, data types of $\mathcal{L}_1, \dots, \mathcal{L}_n$;
    \item As morphisms, terminating algorithms between data 
    types in the languages $\mathcal{L}_1, \dots, \mathcal{L}_n$.
	\end{itemize}
\end{definition}
\begin{figure}[h!]
	\centering
	\input{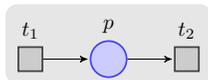}
  \caption{A net showing how 
    Definition~\ref{def: pathological semantics} is pathological.}
	\label{fig: pathological semantics}
\end{figure}
This definition is obviously pathological, and will never serve 
any real purpose. We can see it directly considering the net in 
Figure~\ref{fig: pathological semantics}: Suppose that 
transition $t_1$ has to correspond to some very efficient algorithm 
we want to use, written in Haskell. On the other hand, the best 
choice for $t_2$ would be to map it to some code written in 
Elm~\cite{Czaplicki2012, Czaplicki2018}. This implies that the 
place $p$ has to correspond to a data type that is shared by Haskell 
and Elm, which is basically impossible since different languages 
implement data types differently.

For the same reason, defining a symmetric monoidal structure on the 
category in Definition~\ref{def: pathological semantics} is next to 
impossible, since we do not even know what it means to take tuples of 
data types defined using different specifications. Luckily enough, 
there is a solution to this problem, that relies precisely on 
\emph{defining a shared data type specification for different languages}. 
It is clear that such shared data types will have to be somehow limited, 
since types have different capabilities in different languages 
(for the experienced readers, notice how Idris has dependent types 
while Haskell does not, so they cannot be part of our shared type 
structure). This is something to which the Statebox team is devoting 
a lot of work, and its specification, called 
\emph{Typedefs}~\cite{StateboxTeam2018a}, is an open, 
ongoing project. Typedefs will make 
it easier to mix programming languages and will 
empower developers to use 
Petri nets to design their software without having to give up on the 
tools they already created, which is a very desirable feature.
\section{Why is this useful?}
The usefulness of folds is self-evident: They 
allow for neat compartmentalization of development stages and, by 
giving the programmer freedom to chose the semantic category that 
suits their needs best, ensure full backwards compatibility. With 
Typedefs, finally, this compatibility can be seamlessly extended 
across different languages, finally allowing for a consistent linking 
between different layers of a complex system. From the point of 
view of industry-strength coding, especially in fail-sensitive 
applications, such features are simply invaluable, and put Statebox 
into a unique status among the myriad of programming languages 
out there. \emph{As category theory is the glue of mathematics, 
Statebox is the glue of programming}.

\chapter*{List of Symbols}
  \addcontentsline{toc}{chapter}{List of Symbols}
Please note that some of the symbols we have used are overloaded (as for $\to$ to 
denote morphisms, functors and reachability). Their disambiguation depends on 
the context.
\section*{Sets}
\begin{tabularx}{\linewidth}{@{}S{p{5cm}}X@{}}
  $S, T, \dots$ &
    Generic set names, when usable \\
  $\cup$ & 
    Set union \\
  $\cap$ &
    Set intersection \\
  $\sqcup$ &
    Set disjoint union \\
  $\Naturals$ &
    Set of natural numbers \\
  $\Integers$ &
    Set of integer numbers \\
\end{tabularx}
\section*{Multisets}
\begin{tabularx}{\linewidth}{@{}S{p{5cm}}X@{}}
  $\MsetBase{X}{S}$ &
    Finite multiset $\MsetBase{X}{S}: S \to \Naturals$ \\
  $\Mset{X}$ &
    Finite multiset $\MsetBase{X}{S}: S \to \Naturals$, base implicit  \\
  $\MsetBase{X}{S}, \MsetBase{Y}{S}, \MsetBase{Z}{S}, \dots$ &
    Generic finite multiset names, when usable \\
  $\Mset{X}, \Mset{Y}, \Mset{Z}, \dots$ &
    Generic finite multiset names (base implicit), when usable \\
 $\Msets{S}$ &
    Set of finite multisets over $S$ \\
  $\subseteq$ &
    Multiset inclusion \\
  $\cup$ &
    Multiset union \\
 $ - $ &
    Multiset difference, defined when the second argument is included in the first \\
  $\cdot$ &
    Scalar multiplication of multisets \\
  $\Disjoint$ &
    Disjoint union of multisets \\
  $\Zeromset{S}$ &
    Zero multiset over $S$ \\
  $\hookrightarrow$ &
    Multiset injection \\
 $\left| \quad \right|$ &
    Cardinality of multisets \\
  $g: \Msets{S} \to \Msets{S'}$ & 
    Multiset homomorphism \\
  $\bar{g}: \Msets{S} \to \Msets{S'}$ & 
    Multiset homomorphism coming from function $g: S \to S'$ \\
\end{tabularx}
\section*{Strings}
\begin{tabularx}{\linewidth}{@{}S{p{5cm}}X@{}}
  $r, s, \dots$  &
    Generic names for strings over $S$, when usable \\
  $\Strings{S}$ &
    Set of strings of finite length over $S$ \\
  $\Multiplicity{S}$ &
    Multiplicity function from $\Strings{S}$ to $\Msets{S}$ \\
  $\Ordering{S}$ &
    Ordering function from $\Msets{S}$ to $\Strings{S}$ \\
\end{tabularx}
\section*{Petri nets}
\begin{tabularx}{\linewidth}{@{}S{p{5cm}}X@{}}
  $N,M,L, \dots$ &
    Generic net names, when usable \\
  $p, q, \dots$ &
    Generic place names, when usable \\
  $t, u, v, \dots$ &
    Generic transition names, when usable \\
  $\Pl{N}$ &
    Places of $N$ \\
  $\Tr{N}$ &
    Transitions of $N$ \\
  $\Pin{t}{N}$ &
    Input of transition $t$ of $N$ \\
  $\Pout{t}{N}$ &
    Output of transition $t$ of $N$ \\
  $\Marking{X}$ &
    Marking of a net \\
  $N_\Marking{X}$ &
    Net $N$ along with its marking $\Marking{X}$ \\
  $N_\Marking{X} \xrightarrow{t} N_\Marking{Y}, \quad
    \Marking{X} \xrightarrow{t} \Marking{Y}$ &
    Firing of $t$ in $N$, reachability of $\Marking{Y}$ from $\Marking{X}$ \\
  $\langle f, g \rangle$ &
    Morphism of Petri nets \\
\end{tabularx}
\section*{Category theory}
\subsection*{Category names}
\begin{tabularx}{\linewidth}{@{}S{p{5cm}}X@{}}
  $\FSCCC$ &
    Category of free compact closed categories and generator preserving functors \\
  $\FSSMC$ &
    Category of free strict symmetric monoidal categories and generator preserving functors \\
  $\FSSMCGrounded$ &
    Category of free strict symmetric monoidal categories and grounded generator preserving functors \\
  $\Group$ &
    Category of groups and homomorphisms \\
  $\Hask$ &
    Platonic Haskell category of datatypes and functions \\
  $\hTop$ &
    Category of pointed topological spaces and homotopy classes of continuous functions \\
  $\Petri$ &
    Category of Petri nets and morphisms between them \\
  $\PetriGrounded$ &
    Category of Petri nets and grounded morphisms between them \\
  $\PetriZ$ &
    Category of integer Petri nets and morphisms between them \\
  $\Set$ &
    Category of sets and functions \\
  $\Top$ &
    Category of topological spaces and continuous functions \\
\end{tabularx}
\subsection*{Basic notions}
\begin{tabularx}{\linewidth}{@{}S{p{5cm}}X@{}}
  $\CategoryC$ &
  A category \\
  $\CategoryC, \CategoryD, \CategoryE, \dots$ &
    Generic category names, when usable \\
  $\Obj{\CategoryC}$ &
    Objects of $\CategoryC$ \\
  $\Homtotal{\CategoryC}$ &
    Morphisms of $\CategoryC$ \\
    $A, B, C, \dots$ &
    Generic object names, when usable \\
  $f: A \to B, \quad A \xrightarrow{f} B$ &
    Morphism from $A$ to $B$ \\
  $f, g, h, \dots$ &
    Generic morphism names, when usable \\
  $\Source{f}$ &
    Source (or domain) of morphism $f$ \\
  $\Target{f}$ &
    Target (or codomain) of morphism $f$ \\
  $f;g, \quad A \xrightarrow{f} B \xrightarrow{g} C$ &
    Composition of $f$ and $g$ \\
  $f^{-1}$ &
    Inverse of morphism $f$ (when it exists) \\
  $F: \CategoryC \to \CategoryD, \quad \CategoryC \xrightarrow{F} \CategoryD$ &
    Functor from $\CategoryC$ to $\CategoryD$ \\
  $F, G, H, \dots$ &
    Generic functor names, when usable \\
  $FA$ &
    Application of functor $F$ to object $A$ \\
  $Ff$ &
    Application of functor $F$ to morphism $f$ \\
  $\mathfrak{D}$ & 
    A commutative diagram in some category \\
  $\eta: F \to G$ & 
    Natural transformation between $F$ and $G$ \\
  $\eta, \tau, \dots$ & 
    Generic natural transformation names, when usable \\
  $\eta_A$ &
    Component of natural transformation $\eta$ on object $A$ \\
\end{tabularx}
\subsection*{Monoidal Categories}
\begin{tabularx}{\linewidth}{@{}S{p{5cm}}X@{}}
  $(\CategoryC, \Tensor, \TensorUnit)$ &
    Verbose notation for monoidal categories \\
  $A \Tensor B$ &
    Monoidal product of objects $A$ and $B$ \\
  $f \Tensor B$ & 
   Monoidal product of morphisms $f$ and $g$ \\
  $\TensorUnit$ &
    Monoidal unit \\
  $\alpha_{A, B, C}$ &
    Associator component on $A, B, C$\\
  $\lambda_{A}$ &
    Left unitor component on $A$ \\
  $\rho_{A}$ & 
    Right unitor component on $A$ \\
  $\sigma_{A,B}$ &
    Symmetry on $A, B$ \\ 
  $\epsilon:I' \to FI$ &
    Unit morphism for a lax monoidal functor \\
  $\phi_{A,B}$ &
    Composition component on $A,B$ for a lax monoidal functor \\
\end{tabularx}
\subsection*{Limits, colimits}
\begin{tabularx}{\linewidth}{@{}S{p{5cm}}X@{}}
  $\times$ &
    Product (of sets, of categories, ...)\\
  $\pi_1, \quad \pi_2$ &
    Product projections \\
  $\langle f, g \rangle$ &
    Universal morphism of products applied to $f, g$ \\
  $\sqcup$ &
    Coproduct (of sets, of categories, ...); pushout (when subscripted, as in $\sqcup_B$)\\
  $\iota_1, \quad \iota_2$ &
    Coproduct injections; pushout injections (when superscripted, as in $\iota_1^B$) \\
  $[f,g]$ &
    Universal morphism of coproducts applied to $f,g$; universal morphism of pushouts applied on $f,g$ (when subscripted, as in $[f,g]_B$) \\
\end{tabularx}
\section*{Executions}
\begin{tabularx}{\linewidth}{@{}S{p{5cm}}X@{}}
  FSSMC &
    Free strict symmetric monoidal category \\
  $(\alpha, r, s)$ &
      Generating morphism for a FSSMC \\
    $\Sym{S}$ & 
      Category of symmetries generated by a set $S$ \\
    $\Fold{N}$ &
      Category of executions of a net $N$\\
    $\UnFold{-}$ &
     Functor from $\FSSMC$ to $\Petri$ \newline
     Functor from $\FSSMCGrounded$ to $\PetriGrounded$ \\
\end{tabularx}
\section*{Folds}
\begin{tabularx}{\linewidth}{@{}S{p{5cm}}X@{}}
  $\Semantics$ & 
    Generic name for categories serving as semantics \\
  $\mathcal{L}$ &
    Generic name to denote a programming language \\
\end{tabularx}
\newpage
\printbibliography[heading=bibintoc]
\end{document}